\documentclass[twocolumn,prr]{revtex4-2}	
\usepackage{graphicx}
\usepackage{amsmath, amsfonts,amssymb, ulem}
\usepackage{color}
\usepackage{outlines}
\usepackage{hyperref}

\usepackage{xcolor}

\definecolor{imcolor}{rgb}{0.5,0.,0.5}
\definecolor{fncolor}{rgb}{0.,0.,0.9}				

\newcommand{\rv}{\vec r}

\newcommand{\be}{\begin{equation}}
\newcommand{\ee}{\end{equation}}
\newcommand{\kint}{\int\!\!\frac{{\rm d}^3\kv}{(2\pi)^3}}

\newcommand{\bs}{\boldsymbol}

\newcommand{\addFN}[1]{{{#1}}}
\definecolor{MyColor}{RGB}{0,0,240}

\newcommand{\sgn}{{\rm sgn}}

\renewcommand{\Re}{{\rm Re}\,}
\renewcommand{\vec}[1]{{\bf #1}}
\newcommand{\Tr}{{\hspace{0.5pt} \rm Tr}}

\newcommand{\norm}[1]{\lVert#1 \rVert}
\newcommand{\kv}{\vec k}

\begin{document}

\title{Topological frequency conversion in Weyl semimetals}
\date{\today}
\author{
    Frederik Nathan$^{1,2}$, 
    Ivar Martin$^3$, 
    Gil Refael$^1$
}
\affiliation{
    $^1$Department of Physics and Institute for Quantum Information and Matter, California Institute of Technology, Pasadena, California 91125, USA
    \\ 
    $^2$Center for Quantum Devices, Niels Bohr Institute, University of Copenhagen, 2100 Copenhagen, Denmark
    \\
    $^3$Material Science Division, Argonne National Laboratory, Argonne, IL 60439, USA
}
\begin{abstract}
We theoretically predict  a new working principle for optical amplification, based on  Weyl semimetals:
when a Weyl semimetal is suitably irradiated at two frequencies,  electrons close to the Weyl points   convert energy  between the frequencies    through the mechanism of topological frequency conversion from [Martin {\it et al}, PRX {\bf 7} 041008 (2017)].
Each electron converts energy at a quantized rate  given by  an integer  multiple  of Planck's constant multiplied by the product of the  two  frequencies.
In simulations, we show that optimal, but feasible band structures can support topological frequency conversion in the  ``THz gap'' at intensities  down to $ 2{\rm W}/{\rm mm^2}$; the gain from the effect can exceed the dissipative loss when the frequencies are larger than the  relaxation  time of the system. 
Topological frequency conversion provides a new paradigm for optical amplification, and 
further extends Weyl semimetals'  promise for  technological applications.
\end{abstract}	
\maketitle

Weyl semimetals are at the center of topological materials research thanks to their rich phenomenology~\cite{Nielsen_1981_a,Nielsen_1981_b,Wan_2011,Son_2013,Potter_2014,Baum_2015,Huang_2015,Lv_2015,Weng_2015,Xu_2015_1,Yan_2017,Armitage_2018,Garcia_2020} 
and  promising technological applications~\cite{Parkin_2003,Shekha_2015,Osterhoudt_2019}.
They %se novel materials 
are  characterized by topologically protected nodes in the band structure near the Fermi surface that give rise to (pseudo)spin-momentum locked low-energy excitations with linear dispersion.
% \addFN{The nontrivial band topology induced by these %
Being surrounded by very high Berry curvature, these
nodal points, or Weyl  points, lead to unusual linear and nonlinear optical properties which make Weyl semimetals promising platforms  for, e.g., photovoltaics and high-harmonic generation~\cite{Wu_2016,Morimoto_2016,Chan_2017,de_Juan_2017,Osterhoudt_2019,Sirica_2019,Oka_2019,Takasan_2020,Dantas_2020,Yang_2021,Shao_2021}.

In recent years, it was also appreciated that the interplay between external driving and band topology can give rise to a rich variety of exotic phenomena~\cite{Oka_2009,Kitagawa2010,Jiang2011,Lindner_2011,Dehghani_2015,Nathan_2015,Khemani_2016,Else_2016b,Potter_2016,Po_2016,Titum_2016,Nathan_2017,Bauer_2019,Nathan_2019b,McIver_2019,Sato_2019,Long_2020,Nathan_2021,Yates_2021,Topp_2022}. 
Particularly relevant for our work,  % showed that 
{\it bichromatic} driving (i.e., simultaneous driving at two distinct frequencies) 
\addFN{has emerged as a versatile  tool for control of  matter~\cite{Neufeld_2019,Galan_2020,Else_2020,Neufeld_2021,Long_2022_1}, that can even    induce its own unique} topological phenomena~\cite{Martin_2017,Kolodrubetz_2018,Nathan_2020c,Long_2020}: Ref.~\cite{Martin_2017} showed that   
a spin  driven by two oscillating magnetic fields with incommensurate frequencies $f_1$ and $f_2$ %, t it 
can enter a regime where it  transfers energy between the  driving modes at an average rate  given by  the universal ``energy transfer quantum'', $hf_1f_2$, where $h$ denotes Planck's constant\cite{Kolodrubetz_2018,Nathan_2020c,Long_2020}.  
This effect was termed topological frequency conversion.
% \addFN{Related phenomena with the same universal rate of energy transfer were also discussed in Refs.~.}

While the model from Ref.~\cite{Martin_2017} has been experimentally implemented and studied~\cite{Boyers_2020,Malz_2021},  actual observation of topological frequency conversion is still lacking. %yet to be  achieved. % lacking. 
The reasons are two fold: 
First, in the magnetic realm,   topological frequency conversion  in the  desirable frequency regime of THz and above requires extremely high amplitudes of the oscillating magnetic field (of about 1 Tesla and above, corresponding to radiation intensities of more than $240 \,{\rm MW}/{\rm mm}^2$). 
Even then,  measurable -- and especially {\it useful} --  conversion rates would require many spins acting synchronously.

\begin{figure}[h!]
\includegraphics[width=0.99\columnwidth]{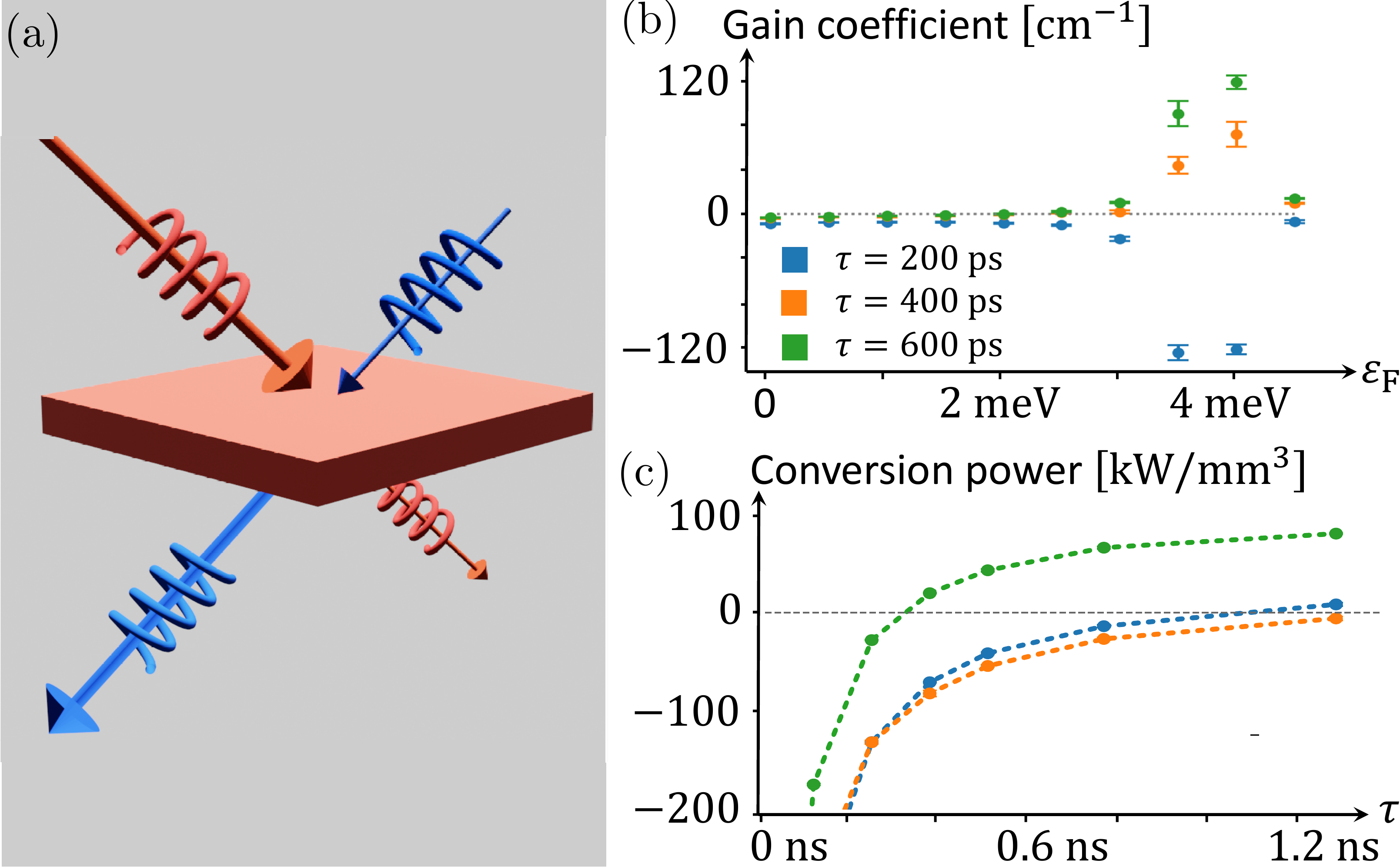}
\caption{
(a) Illustration of main result: %topological frequency conversion. % topological frequency conversion
    a Weyl semimetal irradiated by distinct   frequencies can transfer   energy between the modes through a novel mechanism known as topological frequency conversion.
(b)
    \addFN{Gain coefficient in an inversion-symmetric Weyl semimetal with a  Fermi surface consisting of two  Weyl points with Fermi velocities $3.87\cdot 10^5\,{\rm m}/{\rm s}$,   as a function of the Fermi energy $\varepsilon_{\rm F}$.
    % The coefficient is computed for a small spherical grain, taking into account  plasma oscillations.
    The ``pump'' and ``signal'' modes have frequencies $0.82\,{\rm THz}$ and $1.23 \,{\rm THz}$ and  amplitudes $50\,{\rm kV}/{\rm m}$ and $100\,{\rm kV}/{\rm m}$ inside the material, respectively. 
    Near $\varepsilon_{\rm F}= 4\,{\rm meV}$, these values correspond to radiation intensities of  $0.5\,{\rm W}/{\rm mm}^2$, and  $ 2\, {\rm W}/{\rm mm}^2$, respectively. 
    Blue, orange and green data result from relaxation times $200$, $400$, and $600\,{\rm ps}$, respectively.}
(c) 
    Energy transfer  per unit volume, as a function of  relaxation time $\tau$, for an isolated Weyl node with Fermi velocity $3.87\cdot 10^5 {\rm m}/{\rm s}$. 
    %irradiated by  two modes  with .
    \addFN{Modes $1$ and $2$ have  amplitudes  $900 \, {\rm kV}/{\rm m}$ and   $1800 \, {\rm kV}/{\rm m}$} inside the material, respectively.
    Mode $2$ has  frequency $f_2 = 1.23\, {\rm THz}$, and mode $1$ has  frequency $f_1= ({\sqrt{5}-1})f_2/{2}$, (blue),  $f_1 = {2}f_2/{3}$ (green), and  $f_1 ={2}f_2/({3+0.001\pi}) $ (orange).
\addFN{See Secs.~\ref{sec:plasmonnumerics}~and~\ref{sec:numerics}  for further details of the calculation used for panels (b) and (c), respectively.}
}
\label{fig:front_page_figure}
\end{figure}

In this work  we propose a Weyl semimetal as the medium of choice %capable of 
for realizing topological frequency conversion at high frequencies and with large conversion rates. 
For that we consider a Weyl semimetal, subjected to incoming radiation at two incommensurate frequencies, as depicted in Fig.~\ref{fig:front_page_figure}(a). 
Under appropriate driving, individual electrons near the Weyl nodes act as an ensemble of topological frequency converters (as in Ref.~\onlinecite{Martin_2017}), with the (pseudo-)spin of each electron playing the role of the spin, and  the vector potential potential \addFN{inside the material playing the role of the  magnetic field (the  ``transduction" being provided by the Fermi velocity of the Weyl point).
As a result,  the system hosts an ensemble of electrons that each convert energy from mode $2$ to mode $1$ at the quantized rate $\pm hf_1f_2$ per electron; the number of %of electrons
active  frequency converters is controlled by the magnitude of the vector potential.  % Notably, the the strong coupling needed to achieve topological frequency conversion  does not that
\addFN{Importantly, topological frequency conversion can be realized in Weyl semimetals  at relatively modest radiation intensities. 
This is because the effective spins interact directly with the (strongly coupled) electric field of the radiation %through the vector potential,
rather the than the magnetic field.
As another benefit, Weyl semimetals host a macroscopic number of active frequency converters, giving rise to very large conversion rates.
 }

\addFN{As  a bulk response, topological frequency conversion  is   unique to Weyl semimetals, and constitutes a   fundamentally new mechanism for optical amplification. The phenomenon has novel features of intrinsic interest: first, it is  a 2-wave mixing effect that does not require an idler beam or phase matching. 
Secondly, it is in essence a nonperturbative  effect, beyond the regime of standard  ``$\chi_n$'' responses: in the ideal, fully adiabatic, limit, we show that the rate of topological frequency conversion is non-analytic as a function of the driving amplitude, and hence cannot be captured through a standard Taylor expansion. Away from this limit (i.e., in the presence of finite driving frequency and relaxation), the nonperturbativeness  persists in the form of a highly nonlinear amplitude-dependence.}}

\addFN{The novel features above, along with the modest radiation intensities required and the macroscopic number of active frequency converters give Weyl semimetals a significant potential for optical amplification.
This  is   demonstrated in  Fig.~\ref{fig:front_page_figure}(b): here we plot  the gain coefficient (i.e., the  exponential rate at which the intensity of the  amplified mode increases  inside the material), obtained from simulations with a somewhat optimized, but feasible % Weyl points with realistic
band structure of a Weyl semimetal. The material is irradiated at   frequencies in the   ``THz gap,''  where new effective amplifiers are in high demand, due to a lack of powerful  coherent  radiation  sources. 
%Exploiting a plasma resonance, and a
Assuming  sufficiently slow relaxation, 
%(see below and Sec.~\ref{sec:plasmonnumerics} for details), 
our simulations indicate % THz conversion with
gain coefficients of order $100\, {\rm cm}^{-1}$ %\im{should we give a number in inverse cm here too?} 
can be achieved at intensities of order $1\,{\rm W}/{\rm m}^2$.
% of coherent  radiation in the ``THz gap'' between $0.5\,{\rm THz}$ and $1.5\,{\rm THz}$; in this frequency range  sources~\cite{Bachmann2015,Homann2022}. 
%In this way, topological frequency conversion offers an 
This value is comparable with  current methods such  quantum cascade lasers~\cite{Jukam2009,Kao2017,Snively2019,Boubanga2020,Homann2022}, 
%plasmons in two-dimensional superconductors~\cite{} or graphene heterostructures~\cite{}, or free electron lasers~\cite{}, 
which report gain coefficients, $20-50{\rm cm}^{-1}$ range~\cite{Jukam2009,Kao2017}.
We emphasize it may  be possible to realize significantly larger gain coefficients than $\mathcal O(100 \,{\rm cm}^{-1})$ in other parameter ranges; e.g., with stronger intensities. 
}

% \addFN{The above feature highlights that topological frequency conversion . Along with its nonperturbativity, and significant there  are other des
% Secondly, }

\addFN{There still are challenges  that need to be overcome before optical amplification can become reality: being a conductor,  Weyl semimetal respond with plasma oscillations to radiation which  renormalize the vector potential inside the material. 
%(relative to that of the plane waves irradiating the material). 
It is therefore necessary to drive the system above its plasma frequency  to allow the vector potential enter  the material. 
The plasma oscillations on the other hand  provides an opportunity: driving the material close to resonance with the plasma frequency amplifies the internal vector potential, thus significantly {\it enhancing} the rate of energy conversion.
Indeed, we exploit this resonance effect to achieve the simulated gain coefficients of $\sim 100\,{\rm cm}^{-1}$ for the data depicted in Fig.~\ref{fig:front_page_figure}(b).}

% \addFN{Topological frequency conversion persists even after increasing intensity by multiple orders of magnitude.} 

%
    %FN: there was a factor of 8pi^3 wrong on the y-axis in the figure. I have found the issue, and create a new figure. The new (rescaled) conversion rates correspond much better with our estimate.
%\im{should we drop this para? it conflictrs with the last sentece of the previous para}
%\comment{I think the paragraphs are consistent -- note the different units of conversion rate and intensity}

\addFN{Another, more serious, challenge is  electronic relaxation processes. These counteract the frequency conversion by providing a channel for trivial energy dissipation -- material heating.  }
For the parameters considered in Fig.~\ref{fig:front_page_figure}(b), net energy gain of the pumped mode becomes possible for a characteristic relaxation time of order $300$ picosecond at THz frequencies.
Such relaxation times are longer than the relaxation times that have been mostly reported experimentally to date, which range from  $0.25\,{\rm ps}-3\,{\rm ps}$~\cite{Dai_2015,Weber_2015,Weber_2017,Cheng_2021} to $40\,{\rm ps}$~\cite{Jadidi_2017}.
The nature and timescales for scattering processes in Weyl semimetals is an interesting subject on its own which is still being explored, however: some experiments report signatures with much longer lifetimes~\cite{Wu_2016,Ishida_2016,Liu_2020} that can even exceed  $1000\,{\rm ps}$~\cite{Jadidi_2020}. 
\addFN{
In addition experiments and theoretical studies indicate regimes dominated by non-standard, momentum-conserving channels of dissipation, resulting in  hydrodynamical behavior~\cite{Coulter_2018,Osterhoudt_2020}.
}

\addFN{We speculate that   slower relaxation rates can be achieved, e.g. through improvement of materials quality and bath/substrate engineering.
As another example, we show that dissipation is significantly reduced at commensurate frequencies,  without affecting the energy transfer from topological frequency conversion [see Fig.~\ref{fig:front_page_figure}(c)]. 
%We term this regime {\it Lissajous} conversion, due to the driving-induced electric-field forming a 3-dimensional Lissajous figure.
 %\fn{Are we sure? Is there something we can do to suppress phonon scattering?}
%ould \addFN{result in}. % practically feasible. 
Excessive heating can  be countered through pulsed driving, by allowing the system to dissipate away heat between the pulses.
If sufficiently slow relaxation  can be reached through such or similar incremental improvements, there is a potential for significant benefits in the form of a new and powerful mechanism for optical amplifcation. 
}

The rest of this paper is structured as follows: in Sec.~\ref{sec:model} we review the characteristic properties of Weyl semimetals, which forms the basis for our discussion. %which allow them to act as frequency converters. %introduce the model we study. 
In Sec.~\ref{sec:tfc}, we present the mechanism for frequency conversion from a single-particle perspective.
%The rest of the paper is devoted to showing how topological frequency conversion arises in a realistic many-body system: 
 Sec.~\ref{sec:many_body_tfc}  shows  
how topological frequency conversion arises in a realistic many-body system,  taking into account the effects of finite frequency and dissipation. % and  identify an energy-conserving and dissipative component of this current density.  
% In Sec.~\ref{sec:non_dissipative} we show that the energy-conserving component of the current causes a  transfer of energy between the modes due to the mechanism of topological frequency conversion. 
% %providing complementary perspectives on the effect. Furthermore, we analyze the impact of relaxation processes on the effect. 
% In Sec.~\ref{sec:finite_dissipation} we  estimate the rate of dissipative energy loss in the system.
In Sec.~\ref{sec:numerics}, we support our conclusions with  numerical simulations.
%that directly capture the effects of finite frequency and dissipation.
In Sec.~\ref{sec:conditions}, we summarize the conditions that a Weyl semimetal and  driving modes must satisfy to allow for topological frequency conversion.
\addFN{In Sec.~\ref{sec:phased_array}, we incorporate the effects of plasmons on the single-grain frequency converter, calculate the work in the context of Maxwell equations for the problem, and propose a practical implementation of an amplifier based on this effect using a ``phase array'' of Weyl grains.}
We conclude with a general discussion in Sec.~\ref{sec:discussion}. %, we conclude with a discussion.
Details of derivations %and additional simulations 
are provided in Appendices.

\section{Review of Weyl semimetals}
\begin{figure*}[t]
\includegraphics[width=1.99\columnwidth]{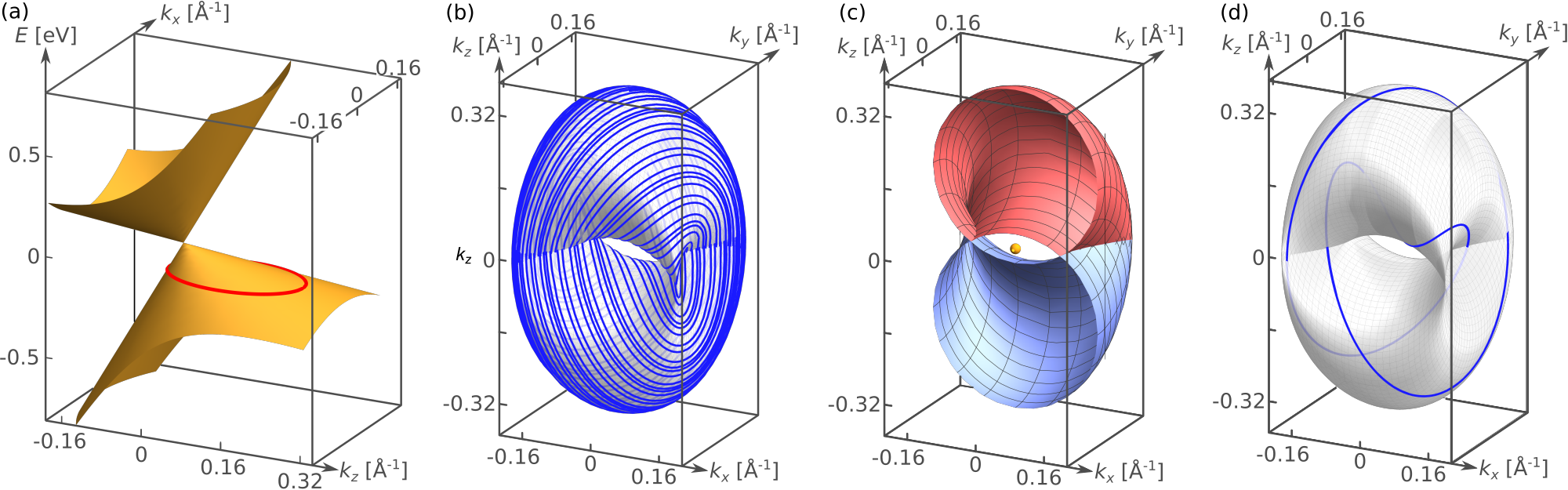}
\caption{
(a): Energy bands of the linearized Hamiltonian  in Eq.~\eqref{eq:weyl_hamiltonian}  in the plane $k_y = 0$, with $\varepsilon_0=0$, $R_{ij}=\delta_{ij}3.87\cdot 10^5\,{\rm m/\rm s}$, and $\vec v= (0,0,3.1\cdot 10^5{\rm m}/{\rm s})$.
%The plot shows the energy bands.
The red line shows  an example of the Fermi  surface (with  Fermi energy $-115 \,{\rm meV}$) when projected into the same plane. %($k_y'=0$).
(b): Trajectory of $e \vec A(t)/\hbar$ resulting from  two modes with circular polarization in the $xz$ and $yz$ planes with amplitudes $\mathcal E_1 = 740\,{\rm kV}/{\rm m}$, $\mathcal E_2 = 1000 \,{\rm kV}/{\rm m}$ and  frequencies, $f_2=1,{\rm THz}$, $f_1=\frac{\sqrt{5}-1}{2}f_2$ (blue). 
Also shown is  the surface $\mathcal B_0$ (gray). 
See main text for further details.
(c):  Cross-section of $\mathcal B_0$ for the same parameters as in (b). 
%The volumes   enclosed volumes by
Within the red and blue sub-surfaces  $W(\kv)$ takes value $1$ and $-1$, respectively, while $W(\kv)=0$ outside the surface.  
(d): Trajectory of $e \vec A(t)/\hbar$   for  the same values of $\mathcal E_2$ and $f_2$ as in (b), and with $E_1=720 \,{\rm kV}/{\rm m}$,  $f_1 = \frac{2}{3}f_2$ (resulting in a commenusrate frequency ratio). 
The different value of $\mathcal E_1$ is chosen to ensure that  the vector potential of mode $1$ has the same amplitude in  panels (b) and (d), such that the topological phase boundary $\mathcal B_0$ is the same for panels (b-d).
%of the vector potential in the same }%\fn{Double check in code that this is what we plot. Add units to axes.}}
}
\label{fig:weyl_surfaces}
\end{figure*}

\label{sec:model}
\label{sec:weyl}

We begin by reviewing the  characteristic properties of Weyl semimetals.
%(see e.g. also Refs. recent reviews of this topic).
This review forms the basis for our subsequent discussion.

Weyl semimetals  are  3-dimensional materials  in which two adjacent  energy bands touch at  isolated points in the Brillouin zone~\cite{Yan_2017,Armitage_2018}, as depicted  in  Fig.~\ref{fig:weyl_surfaces}(a).
These band-touching points are known as Weyl points.
To understand Weyl points better, we consider   the Bloch Hamiltonian of the system, $H(\kv)$,   near one such Weyl point, which we (without loss of generality) take to be located at wave vector $\kv=0$. 
When restricted to the subspace spanned by the two touching bands, and linearized in   $\kv$ around $\kv=0$, $H(\kv)$ takes the following characteristic form: 
\be
H(\kv) =\varepsilon_0 + \hbar\sum_{i,j}\sigma_i R_{ij} k_j +  \hbar\sum_i V_i k_i  
+\mathcal O( \kv^2), %\cdot \kv,
\label{eq:weyl_hamiltonian}
\ee
where  $\sigma_x,\sigma_y$ and $\sigma_z$ 
denote the Pauli matrices acting on the subspace spanned by the two touching bands in some given basis, $R$ is a real-valued symmetric full-rank $3\times 3$ matrix, while $\vec V=(V_1,V_2,V_3)$ and $\varepsilon_0$ is a real-valued velocity and energy, respectively.
%\addFN{Here $R$ and $\vec V$ have units of velocity.}
%Here $H'(\kv)$ denotes the restriction of $H(\kv)$ to the subspace spanned by the two touching bands.
Evidently, the  two energy bands of $H(\kv)$ included above % the Hamiltonian above
 touch  at the Weyl point ($\kv=0$). % ($\kv'=0$).
When the touching energy bands are plotted in the plane $k_i=0$ (for $i=x,y,$ or $z$), the bands form a characteristic ``touching cones'' structure, as for example in Fig.~\ref{fig:weyl_surfaces}(a).
$\varepsilon_0$ determines the location of the touching point on the energy axis, while $\vec V$ determines the ``tilt'' of the cones.
The eigenvectors and spectrum of $R$ determines the  anisotropy  (or ``squeezing'') of the band gap around the Weyl point.

Once it is  present, a Weyl  point is a very robust feature:  as long  $R$ remains full-rank, any infinitesimal perturbation  to the system can  only  shift the location of the band-touching  point, but not eliminate it. 
This is straightforward to verify through direct calculation.
Hence,  a smooth change of  system parameters can only cause  Weyl   points to continuously move around  in the Brillouin zone~\footnote
{
    A Weyl point can only  disappear when this continuous motion causes it to  merge and annihilate with another Weyl point.
    The two annihilating Weyl points must have opposite chiralities (see below). 
    Under such an elimination, one of the eigenvalues of $R$ will  decrease to zero. 
    Such an elimination is always possible, since   the chiralities of all Weyl points in any given gap must sum to zero~\cite{Nielsen_1981_a}. 
    Conversely, Weyl points can be ``nucleate''  pairwise under a smooth change of system parameters.
}.
As a result of this robustness, Weyl semimetals   are  a generic class of materials.
Indeed, many materials have recently been shown to be Weyl semimetals~\cite{Xu_2015,Xu_2015_1,Xu_2015_2,Xu_2015_3,Huang_2015b,Yan_2017,Armitage_2018,Xu_2018,Belopolski_2019,Swekis_2021}.
 %its location. 
%Hence an isolated Weyl point can never be eliminated under a smooth  deformation of the system.

%In addition to their novel band dispersion, %an  exotic band structure,
% Weyl semimetals are characterized by 
Another novel feature of Weyl semimetals is the  nontrivial band topology associated with  the eigenstates of the Bloch Hamiltonian, $\{|\psi_\alpha(\kv)\rangle\}$.
These topological properties are captured by the Berry curvature %~\footnote{
%Specifically, for a differential  area element in the BZ, ${\rm d}^2\vec S$, the net area  in Bloch space covered by  $|\psi_n(\kv)\rangle$ is given by $\vec \Omega^{(n)}_i(\kv)\cdot {\rm d}^2\vec S$.}, 
$\vec \Omega_\alpha(\kv) = (\Omega_\alpha^1(\kv),\Omega_\alpha^2(\kv),\Omega_\alpha^3(\kv))$ where 
\be 
\Omega^{i}_\alpha(\kv) = i \sum_{j,k=1}^3\epsilon_{ijk}\langle \partial_{j}\psi_\alpha(\kv)| \partial_{k}\psi_\alpha(\kv)\rangle,
\ee
with $\epsilon_{ijk}$ denoting the Levi-Civita tensor
and $\partial_i$  the partial derivative with respect to the $i$th component of crystal {wave vector}, $k_i$ (we discuss the physical significance of the Berry curvature below).
% The nontrivial band topology of Weyl semimetals arises because 
%Specifically, *
Weyl points  act as point sources for   Berry curvature: 
for two bands, $1$ and $2$, touching at an isolated Weyl node at $\kv=\kv_i$, the  Berry curvature of the upper band, $2$, satisfies 
\be
\nabla \cdot \vec \Omega_{2}(\kv) = 2\pi\, \sgn (|R|) \delta(\kv-\kv_i),
\label{eq:isolated_wp_bc_correspondence}
\ee
were $|\cdot|$ denotes the determinant, and $\nabla$ the nabla operator in $\kv$-space.  
The sign is reversed for the the lower band. 
The relationship between Weyl points and Berry curvature  is  in exact analogy to point charges and the electric field.
In this analogy, the index $q\equiv \sgn|R|$ determines the ``charge'', or chirality, of the Weyl point~\footnote{Note that $q$  is invariant under continuous basis transformations and coordinate system rotations, and hence 
takes the same value for any choice of basis for the two touching bands, and for any coordinate system}.
The net charge of all Weyl points that appear within a given gap % connecting two given bands must sum to 
is zero~\cite{Nielsen_1981_a}; thus any gap must hold an even number of Weyl points.

For a system with many bands and multiple Weyl points, Eq.~\eqref{eq:isolated_wp_bc_correspondence} generalizes to %the Berry curvature satisfies
\be 
\nabla\cdot \vec \Omega_\alpha = 2\pi \sum_{i} q_i s_{i,\alpha} \delta(\kv-\kv_i),
\label{eq:weyl_berry}
\ee
where the sum runs over all Weyl points in the system, $q_i$ denotes the chirality of Weyl point $i$, and
$s_{i,\alpha}$ indicates  how the Weyl points of the system connect the bands: specifically $s_{i,\alpha}=1$ %takes values $1$ and $-1$ 
if  Weyl point $i$ connects band $\alpha$ with the adjacent band above, $s_{i,\alpha}=-1$ if it connects band $\alpha$ with the band below, and $s_{i,\alpha}=0$ if band $\alpha$ is not involved at Weyl point $i$. %\comment{use the term band connectivity throughout}

 Eq.~\eqref{eq:weyl_berry} can equivalently be expressed using the divergence theorem:
for a closed surface in the Brillouin zone, $\mathcal C$, the total Berry flux of band $\alpha$,  $\oint_{\mathcal C}{\rm d}^2{\vec S}\cdot \vec \Omega_\alpha(\kv)$ (which is identical to the Chern number of band $\alpha$ when constricted to the $2$-dimensional closed surface $\mathcal C$), is given by $\sum_{\kv_i \in \mathcal C} q_i s_{\alpha,i}$ where the sum runs over all Weyl points %of the system 
contained within $\mathcal C$.

%The nontrivial band topology of Weyl semimetals manifests itself through a nonzero anomalous velocity of electrons.
 Berry curvature   acts as a magnetic field in {reciprocal} space: 
an  electron  in band $\alpha$ with a relatively well-defined  position and {wavevector}, $\vec r$ and $\vec k$, acquires a transverse velocity proportional to $\vec \Omega_\alpha(\vec k)$ when subject to a weak  external force~\footnote
{
    See below for the conditions for this result, and see Appendix~\ref{app:master_equation_solution} for derivation
},  
$\vec{\dot k}$:
% for  an electron wave-packet in band $n$ whose position $\vec r$ and momentum $\vec p$are relatively well-defined, a sufficiently weak external force~\footnote{See below for conditions}, $\vec{\dot p}$, combined with  nonzero Berry curvature results in  a transverse contribution to the velocity (in addition to the  contribution from the energy band dispersion)~\cite{X}:    
\be
\vec{\dot r}_\alpha(\kv)=\frac{1}{\hbar}\nabla_\kv \vec \varepsilon_\alpha(\kv) +\dot{\vec{k}}\times   \vec \Omega_\alpha(\kv) .
\label{eq:vanom}
\ee
This second term above is known as  {\it anomalous velocity}, and can be seen as a canonically-conjugate analog to the Lorentz force: whereas  a magnetic field $\vec B$ generates  a   velocity in {reciprocal space}   perpendicular to the real-space velocity, $ \vec{\dot k}=- \frac{e}{\hbar }\vec B \times \vec{\dot r}$ (the Lorentz force),  Berry curvature generates a real-space velocity perpendicular to the {reciprocal space}    velocity, $ \vec{\dot r}=\dot{\vec{k}}\times  \vec \Omega_n $ (the anomalous velocity); here $e$ denotes the elementary charge.

Eq.~\eqref{eq:isolated_wp_bc_correspondence} implies that  the Berry curvature diverges  near  Weyl points.
Hence electrons with wavevectors near a Weyl point experience a divergent   anomalous velocity~\footnote{Not taking into account nonadiabatic effects}. 
When subject to an applied electric field, $\vec E$,  such that $\dot \kv = -e\vec E /\hbar$, Weyl semimetals can thus produce a large  current response which may be  nonlinear as a function of $\vec E$. 
% of the driving field. 
This significant   nonlinearity makes Weyl materials  particularly attractive as nonlinear optical media, with potential applications including  high-harmonic generation, frequency conversion and photovoltaics~\cite{de_Juan_2017,Osterhoudt_2019}. 

In principle, any material the band geometry that has large local Berry curvature near the Fermi level is prone to having strong nonlinear response;  Weyl semimetals are just a prominent example of those thanks to the divergent Berry curvature near the Weyl points. 
However, this is not the full story: the exotic band {\it topology}  of Weyl semimetals (i.e. the nontrivial winding of the Berry curvature around  Weyl points) in itself gives rise to unique nonlinear  response phenomena. 
The effect we explore in this paper -- topological frequency conversion -- 
is an example of such an inherently topological   response phenomenon.

\section{ frequency conversion from  a single electron}
\label{sec:tfc}
Here we show how the nontrivial band topology of Weyl semimetals allows electrons to act as topological frequency converters ~\cite{Martin_2017,Nathan_2019}. 
We consider a Weyl semimetal   irradiated by two electromagnetic waves, or ``modes'', with distinct propagation  angles and  frequencies, and with  elliptical or circular polarization.
 Fig.~\ref{fig:front_page_figure}(a)  depicts a concrete example in which  the two waves are circularly polarized in the $xz$ and $yz$ planes. 
We let $\vec E_1(t)$ and $\vec E_2(t)$ denote the electric fields resulting from mode $1$ and $2$, respectively, such that the net electric field in the Weyl semimetal at time $t$ is given by $\vec E(t)= \vec E_1 (t)+\vec E_2 (t)$.
We assume the wavelengths of the incoming waves to be much longer than the relevant length scales we consider, and hence take $\vec E_i(t)$ to be spatially uniform. 
The two modes are  oscillating with  frequencies $f_1$ and $f_2$, such that, for $i=1,2$, $\vec E_i(t) = \vec E_i(t+T_i)$, where $T_i \equiv 1/f_i$.
For simplicity, we first consider the case where $f_1$ and $f_2$ are incommensurate; we consider the case of commensurate frequencies in Sec.~\ref{sec:commensurate}. 

The coupling between the Weyl semimetal and  the electromagnetic radiation is captured by the Peierls substitution~\footnote
{
    We  could also  have formulated the problem in the ``comoving frame'', by introducing the time-dependent wavevector $\vec h(t) =  \kv+ e \vec A(t)/\hbar $.
    The   Hamiltonian that results from this transformation is not  block-diagonal in $\vec h$, but is given by $\tilde H(\vec h,t) =H(\vec h)- i \frac{e}{\hbar}\vec E(t) \cdot \nabla_{\vec h}$.
    However, the form in Eq.~\eqref{eq:drive_hamiltonian} is more convenient for our purpose.
    The transformation from $\tilde H$ to $H$ corresponds to a time-dependent gauge transformation in real space.
}, 
which causes the driven system to be governed by the time-dependent Bloch Hamiltonian  
\be 
H(\kv,t) = H(\kv + e \vec A(t)/\hbar),
\label{eq:drive_hamiltonian}
\ee
where $\vec A(t) =-\int_0^t{\rm d}s\, \vec E(s)$ denotes vector potential induced by $\vec E(t)$~\footnote{Here shifts in the   time-origin correspond to shifts of $\vec A(t)$ by some constant displacement $\vec A_0$. 
Such shifts are equivalent to gauge transformations and hence do not affect the physical behavior of the system.}.
In the following  $H(\kv)$  denotes the  Hamiltonian of the system in the absence of the driving, while $H(\kv,t)$ denotes the Hamiltonian  in the presence of the driving. 
%as given by the expression above).

%As we find below, in the limit of rapid relaxation, the steady state of the system will be stationary as a function of $\vec p$, while in the limit of slow relaxation, the steady state will be stationary as a function of $\kv$.
% 
%Below we find that, for slow relaxation. 
% steady state will be constant as a function of $\kv$.

It is useful to decompose the vector potential as  $\vec A(t) = \vec A_1(t) + \vec A_2(t)$, where $\partial_t\vec A_i(t) =\vec E_i(t)$.
Since $\vec E_i(t)$ is generated by electromagnetic radiation, its time-average vanishes; hence $\vec A_i(t)$ is also $T_i$-periodic with respect to $t$.
Without loss of generality we take  both $\vec A_1(t)$ and $\vec A_2(t)$  to have time-average zero
(recall that constant shifts in $\vec A(t)$ correspond to benign gauge transformations).
It is convenient to represent the  vector potentials $\vec A$ and $\vec A_i$ as explicit functions of the phases of the two modes , ${\bs \alpha}$ and ${\bs \alpha}_i$ (rather than single time variable). 
Specifically, ${\bs \alpha}(\phi_1,\phi_2) \equiv \vec A_1(\phi_1/\omega_1) +\vec A_2(\phi_2/\omega_2)$ and ${\bs \alpha_i(\phi_i)}\equiv \vec A_i(\phi_i/\omega_i)$, where $\omega_i \equiv 2\pi f_i$ denotes the angular frequency of mode $i$.
We similarly let   ${\bs \epsilon}(\phi_1,\phi_2)$ and ${\bs \epsilon}_i(\phi_i)$ denote the electric fields $\vec E$ and $\vec E_i$ as functions of the individual phases the two modes.

To reveal how  topological frequency conversion emerges, we consider the dynamics of a single electron  in band $\alpha$, in a wavepacket  with some relatively well-defined position, $\vec r$, and wavevector, $\kv$. 
The rate of energy transferred to mode $1$ by the wavepacket, $P_\alpha(\kv,t)$,
is given by Ohm's law~\footnote{This result can be derived from the equation of motion of the energy density in the system: $dE/dt = \kint \langle \partial_t \hat  H(\kv,t)\rangle$. 
Using that $\partial_t\hat H(\kv,t) = \frac{e}{\hbar} \vec E(t)\cdot \nabla \hat H(\kv,t)$ along with the expression for the current density, we find  
$dE/dt = -(\vec E_1 (t) \cdot \vec j(t)+\vec E_1 (t) \cdot \vec j(t))$. We thus identify $-\vec E_i (t) \cdot \vec j(t)$ as the energy flow into the system from mode $i$ per unit volume.
},
\be 
P_\alpha(\kv,t) =-e \vec E_1 (t) \cdot \vec{ \dot r}_{\alpha}(\kv,t).
\label{eq:p_alpha_def}
\ee
where $\dot{\vec r}_{\alpha}(\kv,t)$ denotes the velocity of the wavepacket in  band $\alpha$ at wavevector $\kv$, given the Hamiltonian  $H(\kv,t)$.
When $\omega_1$ and $\omega_2$ are  small enough so that the time-dependence of  $H(\kv,t)$ is adiabatic~
\footnote{
    This condition of adiabaticity  corresponds to the condition of a ``weak  force'' that we alluded to in Sec.~\ref{sec:weyl}.
},
$\vec {\dot r}_{\alpha}(\kv,t)$ is given  by Eq.~\eqref{eq:vanom}, with the instantaneous reciprocal space velocity given by $\dot{\vec k}(t)=-e\vec E(t)/\hbar$: 
\be 
\dot {\vec r}_{\alpha}(\kv,t) =\frac{1}{\hbar} \nabla _\kv\varepsilon_\alpha(\kv+e\vec A(t))-\frac{e}{\hbar} \vec E(t)\times  \vec \Omega_{\alpha}(\kv+e\vec A(t)/\hbar).
\label{eq:driven_group_velocity}
\ee 

Our goal is  to compute the time-averaged rate of energy transfer into mode $1$, 
\be
\bar P_\alpha(\kv) \equiv \lim_{t\to \infty}\frac{1}{t}\int_0^{t}\!\!{\rm d}s\, P_\alpha(\kv,s).
\label{eq:bar_p_def}
\ee
Here and throughout this work, we use the $\bar \cdot$ accent to indicate time-averaging, such that, for any function of time and, possibly, other parameters $ f(t,x)$, $\bar f(x)\equiv \lim_{t\to \infty}\frac{1}{t} \int_0^t ds f(s,x)$.

To compute $\bar P_\alpha(\kv) $, we express $\vec {\dot r}_{\alpha}(\kv,t)$ as a direct function of $\phi_1$ and $\phi_2$:
 $ 
 \vec {\dot r}_\alpha(\kv,t) = \vec v_{\alpha}(\kv;\omega_1 t,\omega_2t).
 \label{eq:current_af_fct_of_phase}
 $
Here $\vec v_{\alpha}(\kv;\phi_1,\phi_2)$ is obtained  from the expression for $\vec{\dot r}_\alpha(\kv,t)$ in Eq.~\eqref{eq:driven_group_velocity} after replacing $\vec A(t)$ and $\vec E(t)$ with ${\bs \alpha}(\phi_1,\phi_2)$ and ${\bs \epsilon}(\phi_1,\phi_2)$, respectively. 
Since we assume $\omega_1$ and $\omega_2$ to be incommensurate, the time-averaged value of $ \vec E_1(t)\cdot \vec {\dot r}_\alpha(\kv,t) $ is identical to the phase-averaged value of ${\bs \epsilon}_1(\phi_1)\cdot\vec v_{\alpha}(\kv;\phi_1,\phi_2)$.
Hence,
\be 
\bar P_\alpha(\kv) = 
\frac{-e}{4\pi^2}\int_0^{2\pi}\!\!\!{\rm d}\phi_1{\rm d}\phi_2\,  {\bs \epsilon}_1 (\phi_1) \cdot \vec v_\alpha(\kv;\phi_1,\phi_2).
\label{eq:phase_integral}
\ee
%Eq.~\eqref{eq:phase_integral} allows us to  
Using  the expression for  $\vec v$  we described above,  along with % $\tilde p_1(\phi_1,\phi_2)=\tv E_1(\phi_1)\cdot \tv j(\phi_1,\phi_2)$ along with
 ${\bs \epsilon}_i = 2\pi f_i \partial_{\phi_i} {\bs \alpha}$, we obtain
\be
\bar P_\alpha(\kv)= f_1f_2 \frac{e^2 }{\hbar}
\int_0^{2\pi}\!\!\! {\rm d}\phi_1 {\rm d}\phi_2\, (\partial_{\phi_1}{\bs \alpha}
 \times \partial_{\phi_2}{\bs \alpha} )\cdot \vec \Omega_\alpha(\kv-e{\bs \alpha}/\hbar).
\label{eq:pn_def}
\ee
See Appendix~\ref{app:p_alpha_derivation} for detailed derivation.
The integral above has a direct geometrical interpretation: 
 $\frac{e^2}{\hbar^2}{\rm d}\phi_1 {\rm d}\phi_2\, (\partial_{\phi_1}{\bs \alpha} \times \partial_{\phi_2}{\bs \alpha} )$ gives  the  differential  area element of the closed  surface defined by $e{\bs \alpha(\phi_1,\phi_2)}/\hbar $ in reciprocal space,
\be 
\mathcal B_{0 } \equiv \{e{\bs \alpha}(\phi_1,\phi_2)/\hbar\}, \quad  0\leq \phi_1,\phi_2 <2\pi,
\ee
The direction of the differential area element $(\partial_{\phi_1}{\bs \alpha} \times \partial_{\phi_2}{\bs \alpha} )$  defines the orientation of  $\mathcal B_0$.
In Fig.~\ref{fig:weyl_surfaces}(b) we depict $\mathcal B_0$ for the case where modes $1$ and $2$ are circularly polarized in the $xz$ and $yz$ planes respectively, and have electric field amplitudes $\mathcal E_2=1000 \,{\rm kV}/{\rm m}$, $\mathcal E_1 =740{\rm kV}/{\rm m}$, and frequencies $f_2 = 1\,{\rm THz}$, $f_1 = \frac{\sqrt{5}-1}{2} f_2$. 
For incommensurate frequencies, the trajectory of $e\vec A(t)/\hbar$ fills out $\mathcal B_0$ completely at long times, as also illustrated in Fig.~\ref{fig:weyl_surfaces}(b). 

With the above geometric interpretation, we find
\be 
\bar P_\alpha(\kv)=  f_1f_2\hbar
\oint_{\mathcal B_0} {\rm d}^2  \vec k'  \cdot\vec \Omega_\alpha(\kv+ \vec k'),
\ee
 where $\oint_{\mathcal B_0} {\rm d}^2\kv' $ denotes the surface integral of $ \kv'$  over the surface $\mathcal B_0$. %  $S_\kv$. %the area element above. 
From Sec.~\ref{sec:weyl} we recall that this integral  is {quantized}  as $2\pi$ times 
%z(\kv)$, where $z(\kv)$ is an integer and gives
 the net charge of Weyl points of band $\alpha$ enclosed within the surface $\mathcal B_0$ after displacing it by $\kv$ from the origin in reciprocal space, $Q_{\alpha}[\kv]$  (here the enclosed charge is weighted by the orientation of $\mathcal B_{0}$ with respect to the  volume in which the Weyl point is enclosed):
\be 
\bar P_\alpha(\kv)
    =  
    hf_1f_2 Q_\alpha[\kv]
\ee
where we used $h=2\pi \hbar$.

%We now consider the value of $ Q_\alpha[\kv]$.
For  an isolated Weyl  point with charge $+1$ located at $\kv =0$ in a two-band system,
%Here, and throughout the paper, we label the bands according to their energies, such that band $n+1$ has higher energy than band $n$. 
 $Q_\alpha[\kv]$ is given by the following for the upper band ($\alpha=2$):
\be 
Q_2[\kv] 
    =  
    -W(\kv),
\ee
where the function $W(\kv)$ is integer-valued and denotes the net winding number of  $\mathcal B_0$ around $\kv$  as a function of $\phi_1$ and $\phi_2$.
In Fig.~\ref{fig:weyl_surfaces}(c) % and \ref{fig:conv_power}(a)
we plot $W(\kv)$ for the configuration of two circularly polarized modes also considered in Fig.~\ref{fig:weyl_surfaces}(b).
The sign is reversed for mode  $2$ [i.e., % the conversion rate that follows when 
when  replacing $\vec E_1$ with $\vec E_2$ in Eq.~\eqref{eq:p_alpha_def}]. 
%\im{replacing E1 by E2 in Eq. 7} ).
Hence the electron  acts as a conversion medium that transfers energy between mode $2$ and $1$.

For a system with multiple bands, $Q_{\alpha}[\kv] = \sum_i W(\kv-\kv_i)q_is_{i,\alpha}$,
where the index $s_{i,\alpha}$ encodes how Weyl point $i$ connects the bands of the system (see Sec.~\ref{sec:weyl}).
We hence arrive at 
\be 
\bar P_\alpha(\kv)
    =  
    hf_1f_2 \sum_i W(\kv-\kv_i)q_is_{i,\alpha}.
\label{eq:tfc_result_1}
\ee
This  constitutes one of our main results.

Eq.~\eqref{eq:tfc_result_1} shows that each electron in the Weyl semimetal     transfers energy from mode $2$ to mode $1$ at a rate which is {\it quantized}, as an integer multiple of  $hf_1f_2$. 
The value of the integer depends on the location of the  electron in the Brillouin zone, $\kv$. 
%, relative to the Weyl points in the system.
Specifically, the conversion rate $\bar P_\alpha(\kv)$ is nonzero for electrons whose wavevectors $\kv$ are  located within the surface $\mathcal B_{0}$ relative to a Weyl point. 
Thus, a nonzero  conversion power can  be realized for electrons near Weyl points. 
 
The  energy conversion  predicted in Eq.~\eqref{eq:tfc_result_1}  can be seen as a realization of the topological frequency conversion  that was discovered in Ref.~\cite{Martin_2017}. Ref.~\cite{Martin_2017} showed that a $2$-level system (such as a  spin-$1/2$) initialized in its lower band and adiabatically driven by two modes with frequencies $f_1$ and $f_2$ can transfer energy between the  modes at an   average rate quantized as $hf_1f_2z$, where $z$ is an integer. 
Ref.~\cite{Martin_2017} explained this  conversion as an anomalous velocity along the synthetic dimensions that correspond to the photon numbers of the two modes.
To understand the relationship between our result and Ref.~\cite{Martin_2017}, note that for fixed $\kv$,  $H(\kv,t)$ is a Hamiltonian of a $2$-level system of the exact same form as considered Ref.~\cite{Martin_2017}, with the pseudospin of the electron playing the role of the physical spin in Ref.~\cite{Martin_2017}.
Indeed, 
 the arguments of Ref.~\cite{Martin_2017} show that for the two-level system described by $H(\kv,t)$, $z= W(\kv)$. 
In this way, each electron in a Weyl semimetal can be seen as a topological frequency converter from Ref.~\cite{Martin_2017}, with the quantized rate of conversion controlled by its location in the Brillouin zone.

\subsection{Commensurate frequencies}
\label{sec:commensurate}
The discussion above for simplicity assumed the frequencies $f_1$ and $f_2$ incommensurate. 
Here we consider  the case where the frequencies of the modes are commensurate such that $f_1/f_2 = p/q$ for some integers $p$ and $q$. 
In this case, $\vec E(t)$ and $\vec A(t)$ thus are time-periodic with the extended  period $T_{\rm ext}= pT_1 = qT_2$.
This  time-periodicity  significantly affects the  electron's trajectory in the BZ (relative to its equilibrium wavevector),  $e\vec A(t)/\hbar$.
For  incommensurate frequencies,  the trajectory  fills a closed surface, namely $\mathcal B_0$, as illustrated in  Fig.~\ref{fig:weyl_surfaces}(b).
In contrast,  commensurate frequencies causes the  trajectory   to form a  closed {\it curve}, $\mathcal C_0$, as in Fig.~\ref{fig:weyl_surfaces}(d).
The curve $\mathcal C_{0}$ is still located on the surface $\mathcal B_0$.

For commensurate frequencies, the driving  experienced by the electron depends on the initial phase difference between the modes, $\Delta \phi$; here   nonzero   $\Delta \phi$ corresponds to a shift of the phase of mode $2$ such that $\vec E(t) = \vec E_1(t) + \vec E_2(t+\Delta\phi/\omega_2)$, resulting in $\vec E(t) = {\bs \epsilon}(\omega_1t,\omega_2t+\Delta \phi)$ and   $\vec A(t)={\bs \alpha}(\omega_1t,\omega_2t+\Delta \phi)$. 
For incommensurate frequencies, different values of $\Delta \phi$ are equivalent to shifts in the  time origin and hence do not affect the long-term dynamics of the electron.
In contrast, for commensurate frequencies, each distinct value of $\Delta \phi$ results in a different closed trajectory of  $\vec A(t)$, $\mathcal C_{0}$. 
The surface $\mathcal B_0$ is recovered by combining the curves $\mathcal C_0$ for all possible values of $\Delta \phi$.

For commensurate frequencies, the quantization of $\bar P$ breaks down.
The breakdown of quantization arises because the trajectory of the modes' phases $(\phi_1(t),\phi_2(t))=(\omega_1t,\omega _2 t+\Delta \phi)$ does not cover the whole 2d phase Brillouin zone over time, $\phi_1,\phi_2 = [0,2\pi)$, thus invalidating the step leading to Eq.~\eqref{eq:phase_integral}. 
However, quantization is recovered  when averaging $\bar P$ over all possible values of $\Delta \phi$:
for commensurate frequencies, Eq.~\eqref{eq:phase_integral} remains valid for  the {\it average} value of  $\bar P$  with respect to $\Delta \phi$. 
\addFN{Thus,  for commensurate frequenices it is  possible to  {\it enhance} conversion rates relative to the quantized value by tuning the phase difference to a value where the conversion rate  exceeds its average value. 
For uncontrolled (random) phase differences, the conversion rate  remains quantized on average. }
% This phase-averaging can be justified in  realistic situations where  phase difference of the modes is  effectively random. 

\section{Frequency conversion in many-body systems}
\label{sec:many_body_tfc}

Our next goal is to show how   topological frequency conversion emerges  in a realistic Weyl semimetal  where electrons are affected by interactions, impurities and phonons. We focus on the rate of energy transfer  to mode $1$  {per unit volume for a Weyl semimetal driven by two  modes}, $\eta(t)$. 
If $\eta(t)$ is positive, there is a net flow of energy into mode $1$, implying amplification of this mode. This energy must originate from mode 2. 
The conversion rate $\eta(t)$ can be computed from the current density, $\vec j(t)$, using Ohms law:
\be 
\eta(t) = -\vec E_1(t) \cdot \vec j(t).
\label{eq:ohms_law}
\ee

To obtain the current density $\vec j(t)$ we characterize the many-body state of the Weyl semimetal in terms of the momentum resolved  density matrix, 
\be
\hat \rho(\kv,t)\equiv \Tr_{\kv'\neq \kv}[\hat \rho_{\rm F}(t)],
\label{eq:rho_def}
\ee
where $\hat \rho_{\rm F}(t)$ denotes the full density matrix of the  Weyl semimetal at time $t$, which is subject to interactions, impurities,  and phonons, while $\Tr_{\kv'\neq \kv}[\cdot]$ denotes the trace over all possible occupations of electronic states with crystal momentum other than $\kv$.
$\hat \rho(\kv,t)$ is a matrix in the $2^d$ dimensional Fock space associated with the $d$ orbitals (or bands) accessible by the electrons at wavevector $\kv$ 
\footnote{For example, this definition, Eq. \ref{eq:rho_def} implies  $\Tr(\hat \rho(\kv,t)\hat c^\dagger_i \hat c_j)=\langle \hat c^\dagger_{i,\kv}(t)\hat c_{j,\kv}(t)\rangle$ where $\hat c^\dagger_{i,\kv}$ creates a fermion in orbital $i$ with crystal momentum $\kv$.}
Below, the ``hat'' accent~$\hat{\cdot}$ indicates operators that act on many-body orbital Fock states. Operators without the accent, such as the Bloch Hamiltonian from % we considered in 
Secs.~\ref{sec:weyl}-\ref{sec:tfc}, $H(\kv,t)$, are single-particle operators. 
$\hat \rho(\kv,t)$  encodes the band occupancies alongside with inter-band  coherences and all multi-particle correlations of electrons with the same wavevector $\kv$. 
The inter-band coherences are crucial for capturing topological  energy conversion, since they give rise to the anomalous velocity in our formalism.

$\hat \rho(\kv,t)$ determines the current density in the system,  $\vec{j}(t)$, through 
\be 
\vec j(t) = -\frac{e}{\hbar}  \kint\Tr [\hat \rho(\kv,t) \nabla_{\kv} \hat H(\kv,t)],
%\vec v(\kv,t).
\label{eq:current_expr}
\ee
where momentum integrals are taken over the full  Brillouin zone, and  $ \hat H(\kv,t)$ denotes the second-quantized Bloch Hamiltonian of the system:
\be 
\hat H(\kv, t)= \sum_{ij}H_{ij}(\kv,t)\hat c^\dagger_i \hat c_j,
\label{eq:2nd_qtzd_ham}.\ee 
Here $ H_{ij}(\kv,t) \equiv \langle i|H(\kv,t)|j\rangle$, and  $|i\rangle$ denotes the $i$th orbital state in the standard Bloch space.

In the presence of driving $\hat\rho(\kv,t)$ approaches to a time-dependent steady state. 
We obtain this steady state by solving a master equation for $\hat \rho(\kv,t)$, in which the  effects of interactions, impurities and disorder are included as a dissipative term.  
The master equation and steady-state solution are summarized in Sec.~\ref{sec:steady_state_summary} below.
The calculation of the steady-state  is straightforward, but involved, and is detailed in Appendix~\ref{app:master_equation_solution}. 
A key feature of the steady-state solution is that the current response can be split into an energy-conserving ``adiabatic component'',    $\vec j_{\rm 0}(t)$, and a dissipative correction due to non-adiabaticity and scattering,  $\delta \vec j(t)$:
\be 
\vec j(t) = \vec j_0(t) + \delta \vec j(t). %+ \vec j_{\rm na}(t)
%\ee \vec j_{\rm dis}(t)
\label{eq:current_result}
\ee
This decomposition allows us to identify an   energy-conserving and dissipative component of $\eta(t)$:
\be 
\eta_0(t) = -\vec E_1(t)\cdot \vec j_0(t), \quad \eta_{\rm dis}(t) \equiv -\vec E_1(t) \cdot \delta \vec j(t).
\ee 

The component $\vec j_0(t)$ is responsible for topological frequency conversion, and we find that this term dominates in the limit of adiabatic driving and slow relaxation. 
As a central result, we find that
\be 
\vec j_0(t) \equiv -e \kint \sum_\alpha \bar \rho_\alpha(\kv) \dot{\vec r}_\alpha(\kv,t)
\label{eq:j0_def}\ee
with $\dot{\vec r}_\alpha(\kv,t)=\frac{1}{\hbar} \nabla _\kv\varepsilon_\alpha(\kv+e\vec A(t))-\frac{e}{\hbar} \vec E(t)\times  \vec \Omega_{\alpha}(\kv+e\vec A(t)/\hbar)$ denoting the wavepacket velocity in band $\alpha$; $\bar \rho_\alpha(\kv)$ is the time-averaged  occupation in the  $\alpha$th band  of the instantaneous Bloch Hamiltonian, $H(\kv,t)$.

% The energy-conserving and dissipative components of $\eta(t)$  arise from these two components, 
% \be 
% \eta_0(t) = -\vec E_1(t)\cdot \vec j_0(t), \quad \eta_{\rm dis}(t) \equiv -\vec E_1(t) \cdot \delta \vec j(t).
% \ee 
% \im{this equation already appeared as (22)! have to keep one or the other}
In what follows, we first discuss the steady state solution of the density matrix (Sec. \ref{sec:steady_state_summary}).
We then  compute the time-averaged energy pumping resulting from  the non-dissipative component of the current response, $\vec \eta_0$. 
We finally consider the dissipative component of $\eta$, $\eta_{\rm dis}$ in Sec. \ref{sec:finite_dissipation}.  
It is crucial to  estimate $\eta_{\rm dis}$, since amplification is only achieved  when $\eta_0$ exceeds $\eta_{\rm dis}$. 
%it is exceeded by $\eta_0$ does frequency conversion .

\subsection{Steady state solution}
\label{sec:steady_state_summary}
We now discuss how we obtain the steady-state of $\hat \rho(\kv,t)$. 
Details of this discussion are provided in Appendix~\ref{app:master_equation_solution}.

For a clean and non-interacting system, $\hat \rho(\kv,t)$ evolves according to the von Neumann  equation,
$ 
\partial _t\hat \rho(\kv,t)= -(i/\hbar) [\hat H(\kv,t),\hat \rho(\kv,t)].
$
Interactions, phonons, and impurities 
cause a dissipative correction to this equation. 
For sufficiently weak dissipation, this  correction can be derived approximately from first principles
and  takes the form of a   trace- and positivity-preserving linear operator  acting on $\hat \rho(\kv,t)$, $\mathcal D(\kv,t)$~\cite{Nathan_2020b}.
Thus, $\hat \rho(\kv,t)$ is governed by the following Lindblad-type quantum master equation
\be 
 \partial_t \hat \rho(\kv,t) \approx \frac{-i}{\hbar} [\hat H(\kv,t),\hat \rho(\kv,t)] + \mathcal D(\kv,t)\circ \hat \rho(\kv,t).
\label{eq:vn_modified}
\ee
 
 In Appendix.~\ref{app:master_equation_solution}, we obtain a solution to Eq.~\eqref{eq:vn_modified}.
The solution $\hat \rho(\kv,t)$ is accurate as long as the driving is adiabatic with respect to the energy gap of $H(\kv,t)$, $\delta \varepsilon$, and much faster than the the magnitude of the dissipator $\mathcal D$~\footnote{Here $\norm{\cdot}$ can for example denote the spectral norm for linear operators}:  
%regardless of the dissipators 
%\im{confusing sentence}.\comment{FN: changed the text. Is the new version ok?}
\be
\norm{\mathcal D(\kv,t)}\ll\omega_1,\,\omega_2\ll\delta \varepsilon.
\ee
We term this limit,  as the {\it coherent adiabatic regime}.

When the above conditions are satisfied, we find the-steady state value of $\hat \rho(\kv,t)$ is diagonal in the eigenbasis of the Hamiltonian, $\hat H(\kv,t)$, up to minor nonadiabatic corrections. 
The corresponding eigenvalues (which determine  the the occupations of the instantaneous bands of $\hat H(\kv,t)$) are nearly stationary, except for minor fluctuations of order $\norm{\mathcal D}/\mathcal O(\omega_1,\omega_2)$. 
These fluctuations, along with the subleading (second-order) nonadiabatic corrections to $\hat \rho(\kv,t)$ give rise to the dissipative current $\delta j(t)$.
%The anomalous velocity, which appears in 
The term $\vec j_0(t)$ results from just keeping the (dominating) time-independent component of the eigenvalues of $\hat \rho(\kv,t)$ and including leading-order nonadiabatic correction to its eigenbasis. 
Here the leading-order non-adiabatic correction to the eigenbasis is responsible for the anomalous velocity which enters in $\vec j_0(t)$. 

Our solution to Eq.~\eqref{eq:vn_modified}  applies to any dissipator $\mathcal D$, and this dissipator  can  be derived from first principles~\cite{Nathan_2020b}.
However, for illustrative purposes, we now  demonstrate our solution for  the concrete example where  $\mathcal D$ takes a particular  phenomenological  form: the ``Boltzmann'' form.
In this approximation, the dissipator uniformly relaxes electrons towards their instantaneous equilibrium state at  some given ambient temperature $1/\beta$ and chemical potential $\mu$, and   with some 
rate $1/\tau$: 
 \be 
\mathcal D_{\rm B}(\kv,t)\circ \hat \rho  =- \frac{1}{\tau}[\hat \rho-\hat \rho_{\rm eq}(\kv,t)].
\label{eq:boltzmann}
\ee
Here $\hat \rho_{\rm eq}(\kv,t)$ is the instantaneous equilibrium state described above, and is given by $ e^{-\beta [\hat H(\kv,t) - \mu \hat n]}/\Tr(e^{-\beta [\hat H(\kv,t) - \mu \hat n]})$, where $\hat n = \sum_i \hat c^\dagger_i \hat c_i $. 
We also use this  dissipator in our numerical simulations (see Sec.~\ref{sec:numerics}).

% In Sec.~\ref{sec:non_dissipative} we show that the  time-averaged value of $ \eta_0(t)$ due to the mechanism of topological frequency conversion which we explored in Sec.~\ref{sec:tfc}{\bf Incomplete sentence, no?}. Our computation in Sec.~\ref{sec:non_dissipative} does not depend on the specific details of $ \bar \rho_\alpha(\kv)$;  its time-independence and qualitative properties  are sufficient to establish  conditions for topological frequency conversion. However, we provide a prescription for computing $\bar \rho_\alpha(\kv)$ in Appendix~\ref{app:master_equation_solution}. 
The Boltzmann-form dissipator [Eq.~\eqref{eq:boltzmann}] leads to the following steady state density matrix:
\be 
\bar \rho_\alpha(\kv)\approx  \lim_{t\to \infty}\frac{1}{t}\int_0^{t} {\rm d}s\, f _\beta[\varepsilon_\alpha(\kv,t) - \mu] . %+ \mathcal O(\lambda).
\label{eq:boltzmann_result}
\ee %\rho_\alpha^{\rm eq}(\kv,s)$.
where $f_\beta(E)$ denotes the Fermi-Dirac distribution at temperature $1/\beta$.
This result (see also Appendix~\ref{app:master_equation_solution}) % for derivation). 
indicates a steady state occupation which is the average band-population on the trajectory $\kv+ e\vec A(t)/\hbar$, as if the equilibrium distribution is   ``smeared'' over a characteristic wavevector scale $eA/\hbar$, where $A$ is the drive vector potential magnitude.
This smearing is confirmed in our numerical simulations (see Sec.~\ref{sec:numerics} and Fig.~\ref{fig:conv_power} in particular). 

%We could have obtained %the expression for the current density $\vec j_0$  in  Eq.~\eqref{eq:j0_def} by naively extending the results of Secs.~\ref{sec:weyl}-\ref{sec:tfc} to the many-particle case [see Eq.~\eqref{eq:vanom}].   However, our  analysis in Appendix~\ref{app:master_equation_solution} provides the dissipative correction $\delta \vec j(t)$, along a prescription for computing the steady-state band occupancies, $\{\rho_\alpha(\kv,t)\}$, which could not have been inferred from the single-particle dynamics in Sec.~\ref{sec:tfc}. 

%%%%%%%%%%%%%%%%%%%%%%%%%%%%%%%%%%%%%%%%%%%%%%%%%%%%%%%%%%%%%%%%%%%%%%%%%%%%%%%%%%%%%%%%
% The extent to which the system satisfies condition \eqref{eq:coh_ad_regime} is quantified by the dimensionless parameter  %$\lambda$, defined as
% \be 
% \lambda \equiv \max\left(\frac{\hbar \Omega}{\delta \varepsilon},\frac{\Gamma}{ \Omega}\right).
% \label{eq:lambda_def}
% \ee
% where $\Omega$ denotes the characteristic magnitude of the frequencies $\omega_1$ and $\omega_2$. 
% The system is in the coherent adiabatic regime when $\lambda \ll 1$. 
%%%%%%%%%%%%%%%%%%%%%%%%%%%%%%%%%%%%%%%%%%%%%%%%%%%%%%%%%%%%%%%%%%%%%%%%%%%%%%%%%%%%%%%%

%
\subsection{Non-dissipative frequency conversion}
\label{sec:non_dissipative}

We first compute the average rate of energy transfer in the limit of adiabatic driving and zero dissipation. 
I.e., we compute the the time-average of the component $\eta_0(t)$, $\bar \eta_0$.
We find that $\eta_0(t)$ can have nonzero time-average because of the mechanism of topological frequency conversion that we discovered in the last section.

To compute $\bar \eta_0$ we first note  $\eta_0$ can be written
\be 
\eta_{0} (t)=  \sum_\alpha \kint \bar \rho_\alpha(\kv) P_\alpha(\kv,t)\label{eta}
\ee
where $P_\alpha(\kv,t)\equiv e\vec E(t)\cdot \dot{\vec r}_\alpha(\kv,t)$ [see also Eq.~\eqref{eq:p_alpha_def}]. 
We find the time-average of the above using the main result from Sec.~\ref{sec:non_dissipative},
$
\bar P_\alpha(\kv) 
    =
    - h f_1f_2 \sum_i  W(\kv-\kv_i)q_is_{i,\alpha}
$  [Eq.~\eqref{eq:tfc_result_1}].
Here $q_i$, $s_{i,\alpha}$, and $\kv_i$ denote the charge, band connectivity,  and wave vector of Weyl point $i$ in the system, respectively, while $W(\kv)$  measures  the net winding of the surface $ \mathcal B_{0}$ % = \{-e{\bs \alpha}(\phi_1,\phi_2)/\hbar,0\leq \phi_1,\phi_2 <2\pi\}$ 
around wavevector $\kv$ (see  Sec.~\ref{sec:finite_dissipation} for further details).
with this, Eq. (\ref{eta}) becomes:
\be 
\bar \eta_{0} 
    = 
    -hf_1f_2
    \sum_{\alpha,i}\kint 
    q_i s_{i,\alpha} \bar \rho_\alpha(\kv)W(\kv-\kv_i).
\label{eq:main_result}
\ee
Thus, each Weyl point is surrounded by a region of reciprocal space (namely the region where $W(\kv-\kv_i)\neq 0$), in which  electrons act as topological frequency converters. 
In this region, each transfers energy to mode $1$ at the quantized rate $\pm hf_1f_2$. This is a many-electron generalization of Eq. (\ref{eq:tfc_result_1}) and constitutes  another main results of this paper.
In the following we thus refer to $\bar \eta_0$ as the topological frequency conversion rate of the system, to distinguish it from the dissipation rate, which is given by the time-average of  $ \eta_{\rm dis}(t)$.

\addFN{While the conversion rate from each electron is quantized, the net number of electrons with nonzero conversion rate is not fixed, but depends on the amplitude and configuration of the driving field (through the function $W(\kv)$) and the steady-state distribution surrounding each Weyl point, $\bar \rho_\alpha(\kv)$. 
This steady-state distribution is in turn  controlled by the band structure of the system, as well as the configuration and intensity of the external driving.}

To explore how the band structure and driving configuration controls the conversion rate, we first estimate the ``gross'' rate of topological frequency conversion from a Weyl point (i.e., %the rate that conversion rate 
not taking into account cancellation between electrons that transfer energy at opposite rates). 
\addFN{Note that $W(\kv)$ is positive within volume of order  $\sim \frac{2e^3}{\hbar^3} A_1 A_2 (A_1+A_2)$ in reciprocal space, with $A_i =E_i/\omega_i$ denoting the vector potential amplitude of mode $i$. 
This volume corresponds  to an electronic density of $\sim \frac{e^3}{4\hbar^3} A_1 A_2 (A_1+A_2)$ for each Weyl point.
Since each electron contributes  $hf_1f_2$ to $\bar \eta_0$, $\bar \eta_{\rm gross}$  is of order
\be 
\eta_{\rm gross} \sim \frac{e^3 E_1E_2}{8\pi^4 \hbar^2}\left(\frac{E_1}{\omega_1}+\frac{E_2}{\omega_2}\right).
\label{eq:eta_gross}\ee 
%{\bf THIS EQUATION CON'T Be right - too many 2 pi's missing. Also below - quote the electric field, not A}
% FN: fixed it
% Here, $E$ denotes the characteristic magnitude of $\vec E(t)$, $\Omega$ denotes the characteristic scale of $\omega_1$ and $\omega_2$, and   we used $A\sim E/\Omega$.
As an example, for $\omega_i \sim 2\pi \,{\rm THz}$ and $E_i\sim 1500\,{\rm kV}/{\rm m}$. }
%(corresponding to intensities of $6{\rm kW}/{\rm mm^2}$, which we expect can be reached using pulsed lasers~\cite{French_1995}) %FN: see also https://www.americanscientist.org/article/high-power-lasers},
the above estimate yields $\eta _{\rm gross} \sim 500 \,{\rm kW}/{\rm mm}^3$. % the Fermi energy.
%\im{this estimate is different from Abstract. We should keep one or the other?} \comment{FN: I think the estimate is consistent with abstract; see below}

The  actual, {\it net}, topological conversion power, $\bar \eta_0$ is significantly smaller than the gross rate we estimated above, due to cancellation between electrons that convert energy at rates $hf_1f_2$ and $-hf_1f_2$.
Specifically, when modes $1$ and $2$ only contain a single harmonic each, the driving induced vector potential satisfies $\vec A_i(t)=-\vec A_i(t+T_i/2)$, 
%\im{what does this mean?},
implying $W(\kv)=-W(-\kv)$~
\footnote{
    When  $\vec A_i(t)$ contains higher harmonics, $W(\kv)$ can potentially break this inversion antisymmetry. 
    However, even for such cases  $\int {\rm d}^3{\kv} W(\kv)$ still vanishes implying that the regions of reciprocal space characterized by conversion rates $hf_1f_2$ and $-hf_1f_2$ have equal net volumes.
} [this symmetry is  clearly evident %for the specific example we depict 
in Fig.~\ref{fig:weyl_surfaces}(c)].
Hence the regions of reciprocal space characterized by conversion rates $hf_1f_2$ and $-hf_1f_2$ have equal net volumes.
In realistic situations, both volumes will be occupied by electrons, implying  $\bar \eta_0\ll \eta_{\rm gross}$. 
%being significantly smaller %reduced compared to
%than the estimate in Eq.~\eqref{eq:eta_gross}.
%The actual  {\it net} conversion rate will thus be a small fraction of  the ``gross'' rate estimated above, due to  the contributions from the two regions cancelling out with one another.
However, because $\eta_{\rm gross}$  can be quite large, even a small imbalance in the  filling of the two regions can lead to significant net frequency conversion.
%as we demonstrate in Sec.~\ref{sec:eta_estimate}. 
%below, where we compute $\bar \eta_0$ for the steady state obtained from the Boltzmann-type dissipator in Eq.~\eqref{eq:boltzmann_dissipator}. % taking into account the c

%The considerations above show that a Weyl point can only generate
To achieve  a nonzero  $\bar \eta_0$, the steady-state occupation of the bands, $\bar \rho_\alpha(\kv)$, must be anisotropic around the Weyl point to counteract the  antisymmetry $W(\kv)=-W(-\kv)$.
Such an anisotropy  is generally achieved when  the ``Weyl cone tilt'', $\vec V$, is nonzero, since we expect the steady-state inherits the same symmetry properties as the equilibrium state [see Eq.~\eqref{eq:weyl_hamiltonian}].
Additionally, the Weyl point must be within a distance of order $\sim eA_i/\hbar$ from the Fermi surface to ensure that $\bar \rho_\alpha(\kv)$ does not take constant value ($1$ or $0$) within $\mathcal B_0$. 
Indeed,  our numerical simulations demonstrate that nonzero $\bar \eta_0$  can arise when $\vec V\neq 0$ and the Fermi surface lies close to the Weyl point.

Topological frequency conversion is \addFN{in essence a nonperturbative} %a highly nonlinear
effect: it is controlled by the overlap of the quantized (i.e., nonanalytic) function $W(\kv)$ with the steady-state distribution. 
% In particular, it eerges only when the driving vector potential the  threshold where the surface $\mathcal B_0$ becomes large enough to enclose $\kv$-points where the gap of $H(\kv)$ is larger than $\Omega$ (see Sec.~\ref{sec:nonadiabatic} for further discussion) and also large enough so that the Fermi surface is within a distance $\sim A$ from the Weyl point). 
Hence topological frequency conversion does not have a simple  power-law dependence on $A$ in the  limit of small $A$, and is therefore beyond  standard nonlinear response theory.
\addFN{In Sec.~\ref{sec:numerics} [Fig.~\ref{fig:escaling}(b)] we provide data from numerical simulations indicating this highly nonlinear nature of the phenomenon. }% is not an efficient way to describe the effect. 
\subsection{Dissipative energy loss }
\label{sec:finite_dissipation}
\begin{figure}
    \centering
    \includegraphics[width=0.99\columnwidth]{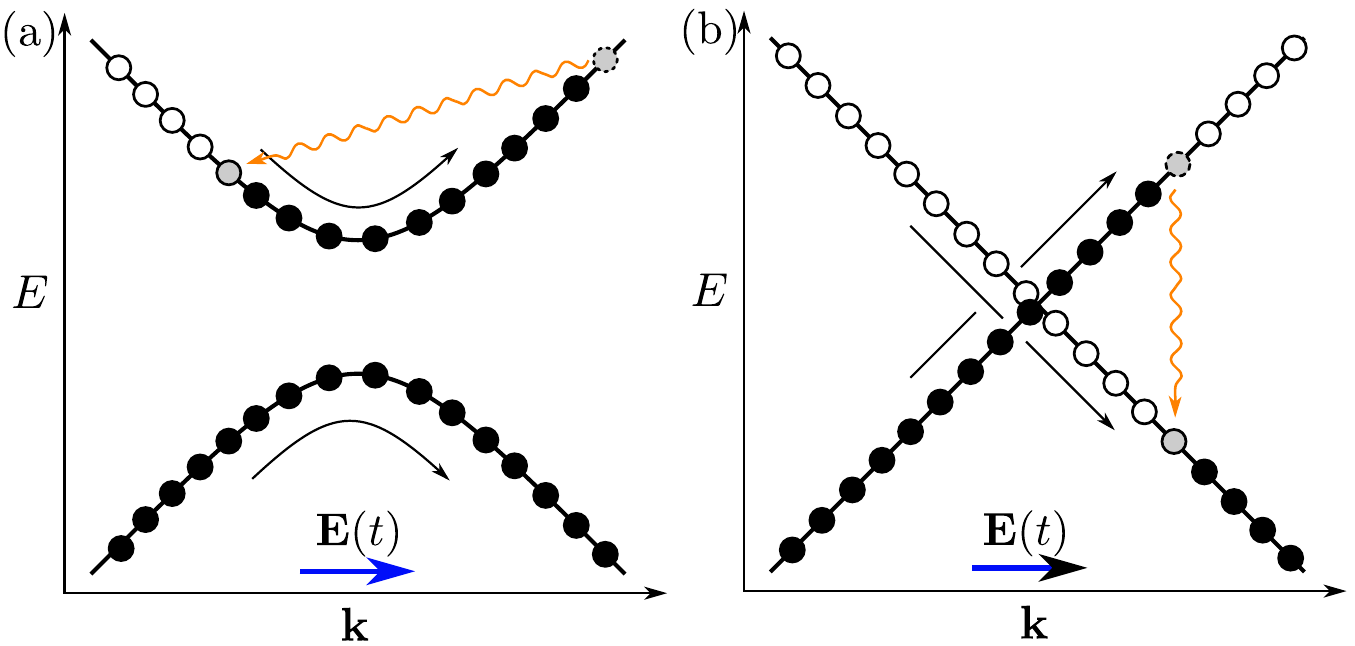}
    \caption{Schematic illustration of the two distinct dissipation mechanisms in a driven Weyl semimetal.  (a)  Momentum relaxation: as electrons are adiabatically translated in the Brillouin zone by the driving-induced electric field (black arrows), electrons shifted to higher energies (black dots) can relax by decaying to vacant states  that have become available at lower energies within the same band (white dots), causing net dissipation of energy.
    This mechanism can occur for all wavevectors near the Fermi surface, where  driving increases the energy of electrons beyond the energies of vacant states elsewhere in the same band. 
    (b): Nonadiabatic heating. 
    Electrons that are taken through or near a Weyl node by the driving-induced electric field can undergo complete or partial Landau-Zener tunneling from the valence to the conduction band. 
    This results in dissipation when the excited electrons in the conduction band relax back into vacant states in the valence band. % (orange arrow). 
    }
    \label{fig:dissipation_mechanisms}
\end{figure}

For topological frequency conversion to cause a net amplification of mode $1$, $\bar \eta_0$ must exceed the rate of energy loss due to dissipation, $\eta_{\rm dis}$. 
It is therefore crucial to estimate this dissipation rate. 
This is the goal of this subsection. 

Our solution of the master equation in Appendix~\ref{app:master_equation_solution} shows that the dissipative current response, $\delta \vec j(t)$ [see Eq. (\ref{eq:current_result})], contains two components:
\be
\delta \vec j(t)=\delta \vec j_{\rm mr}(t)+\delta \vec j_{\rm na}(t),
\ee
which we interpret as arising from momentum-relaxation  ($ \delta \vec j_{\rm mr}$) and nonadiabaticity-induced particle-hole pair creation ($ \delta \vec j_{\rm na}$); see discussion below. 
Consequently, $\eta_{\rm dis}(t)$ can also be separated into these two components:
\be 
\eta_{\rm dis}(t) = \eta_{\rm mr}(t) + \eta_{\rm na}(t).
\label{eq:eta_dis_decomp}
\ee 
where $\eta_{\rm mr}(t) \equiv \vec E_1(t) \cdot \vec j_{\rm mr}(t)$, and $\eta_{\rm na}$ is defined likewise. 

While $\eta_{\rm mr}$ and $\eta_{\rm na}$ are given in Appendix~\ref{app:master_equation_solution},  here we discuss their origin and estimate their magnitudes  based  on a phenomenological discussion. 

Energy loss due to {momentum relaxation}, $\eta_{\rm mr}$,   arises when perturbed electrons in the close vicinity the Fermi surface relax due to their displacement from instantaneous equilibrium, as schematically indicated in Fig.~\ref{fig:dissipation_mechanisms}(a). 
In contrast, $\eta_{\rm na}(t)$ arises from the particle-hole pair creation that results because the driving-induced electric field inevitably overpowers the gap sufficiently close to Weyl points. 
%due to the   violation of the adiabaticity condition   which necessarily occurs sufficiently close to  Weyl points. 
%Here, the driving-induced electric field inevitably    overpowers the gap, resulting in particle-hole pair creation. 
Equivalently, $\eta_{\rm na}$ arises because effective gap closing of $\hat H(\kv,t)$ near a  Weyl point gives rise to Landau-Zener tunneling from the conduction to the valence band upon driving.
These excited electrons dissipate energy as they relax back to the conduction band, as schematically illustrated in Fig.~\ref{fig:dissipation_mechanisms}(b).

%Interestingly,  below we find that nonadiabatic heating  is very sensitive to the commensurability of the driving frequencies, and is strongly suppressed near highly rational frequency ratios of $\omega_1/\omega_2$ (i.e., when $\omega_1/\omega_2 \approx p/q$ for small integers $p$ and $q$), as is also evidenced in our numerical simulations [see Fig.~\ref{fig:front_page_figure}(b)]. 

% To understand why $\eta_{\rm na}$ corresponds to non-adiabatic heating, note that  $\vec v _{\rm na}(\kv,t)\vec \sim v_{\rm F}\lambda^2$ is only for $\kv$ where  $\lambda \gtrsim 1$. %due to the  time-dependence of $\hat H(\kv,t)$ becoming  non-adiabtic because  $\kv$ is taken onto, or near, a Weyl point '
% From the definition of $\vec v_{\rm na}(\kv,t)$ [see text below Eq.~\eqref{eq:current_response_3}], we see that this occurs for $\kv$-points for which  the time-dependence of $ H(\kv,t)$ is non-adiabatic, as we expect for non-adiabatic heating.%

Below we estimate $\eta_{\rm mr}$ and $\eta_{\rm na}$ based on phenomenological arguments. 
For simplicity, we consider  system with two bands and an isolated Weyl point at $\kv= 0$ (which is easily generalized to multiple Weyl points). Moreover, we do not distinguish between the characteristic relaxation rates associated with momentum relaxation and particle-hole pair creation (which  may be  different in real materials), but use
\be 
\Gamma \sim 
\norm{\mathcal D(\kv,t)}
\ee 
as an estimate for both characteristic relaxation rates.

\subsubsection{Momentum relaxation}
\label{sec:momentum_relaxation}
We first estimate  the rate of energy loss arising from momentum relaxation, $\eta_{\rm mr}(t)$.
\addFN{For convenience, in the following we let  $A=E/\Omega$ denote the  characterstic magnitude of the driving-induced vector potential, $E$  the characteristic magnitude of $\vec E(t)$, and $\Omega$  the characteristic scale of $\omega_1$ and $\omega_2$.}

Effects of momentum relaxation  can only emerge within a distance $\sim eA/\hbar$ from the Fermi surface, where electronic occupation fluctuates.
Therefore, only a density of $ e S_{\rm F}  A /(2\pi)^3 \hbar $ contributes to $\delta \vec j_{\rm mr}(t)$, where $S_F$ is the area of the Fermi surface.
Electrons near the Fermi surface on average gain an energy of order $eAv_{\rm F}/2\hbar $ due to the driving (with $v_{\rm F}$ the characteristic Fermi velocity). 
Assuming their relaxation rate is of order $\Gamma$, the average rate of energy loss in the system due to momentum relaxation thus is given by 
$
  \frac{\Gamma e^2A^2 S_{\rm F}v_{\rm F}}{16\pi^3 \hbar}.
$
We estimate that half of this comes from mode $1$.
This estimate agrees well with our predictions based on the definition of $\delta \vec j_{\rm mr}(t)$ in Appendix~\ref{app:master_equation_solution}. %%%% FN: Comment on this in Appendix -- or is this sentence even necessary?
Using that $A\sim E/\Omega$, we find
%, along with 
%$E^2 \sim I/c\epsilon_0 $,  where $I$ denotes the radiation intensity, and $c$ the speed of light, we thus obtain 
\be 
\eta_{\rm mr}\sim  \frac{\Gamma e^2 \addFN{E^2}  S_{\rm F}v_{\rm F}}{ 16\pi^3     \hbar \Omega^2}.
\label{eq:momentum_relaxation_estimate}
\ee 

It is interesting to note that $\eta_{\rm mr}$  is proportional to the size of the Fermi surface, $S_{\rm F}$. 
Therefore type-I Weyl points  are  more suitable for frequency conversion than type-II Weyl semimetals, since  the surrounding Fermi surface  forms  a compact ellipsoid for the former case, in contrast to an  extended hyperboloid in the latter.  

As an example,  we estimate $\eta_{\rm mr}$ for the same parameters we used to  estimate  $\eta_{\rm gross}$ in Sec.~\ref{sec:non_dissipative} %(we consider equivalent parameters in our numerical simulations below), % in Sec.~\ref{sec:numerics}), 
i.e., $E\sim 1500\,{\rm kV}/{\rm m}$, $v_{\rm F}\sim 5\cdot 10^5{\rm m}/{\rm s}$, $\Omega = 2\pi\, {\rm THz}$.
For an ellipsoid  Fermi surface  with   principal semi axes $(1.5,1.5,2.4)eA/\hbar$ (yielding $S_{\rm F}\approx 0.06 {\rm Å}^{-2}$),
our estimate then results in $\eta_{\rm mr}\sim 5\cdot 10^{-5} {\rm J}/{\rm mm}^3\tau $, where $\tau = 1/\Gamma$.
Note that our estimated value of $\eta_{\rm mr}$ is proportional  to  the  Fermi surface area and thus can be easily adjusted to other values of this quantity. %$S_{\rm F}$. % two quantities. 
%estimate is proportional to the relaxation rate, and to the Fermi surface area. 
Recalling that $\eta_{\rm gross}\sim 500\,{\rm kW}/{\rm mm}^3$ for the same parameters, we expect net frequency conversion to only exceed momentum relaxation when $\tau \gg 100\,{\rm ps}$. 
This expectation is confirmed in our numerical simulations [see Fig.~\ref{fig:front_page_figure}(b)]. %estimate be comparable  to the {\it net} rate of topological frequency conversion in the system, $\bar \eta_0$. 

\subsubsection{Non-adiabatic heating}
\label{sec:nonadiabatic}
Non-adiabatic heating arises from electrons at 
$\kv$-points where the time-dependence of $H(\kv,t)=H(\kv+e\vec A(t)/\hbar)$ overwhelms the band gap. 
In Weyl semimetals, such $\kv$-points inevitably exist (even for arbitrarily slow driving), because the band gap of $H(\kv)$ closes   at Weyl points. % at $\kv=0$. 
%in the bands of $H(\kv)$.
For each such $\kv$-point, the band gap of $H(\kv,t)$ effectively closes at  certain times $t$, namely  when $\kv+e\vec A(t)/\hbar$ is sufficiently close to the Weyl point at $\kv=0$ (see below for more detailed conditions). 
At each such  gap-closing event,  electrons at wavevector $\kv$ will undergo partial or complete  Landau-Zener transition from the conduction to the valence band.
This mechanism   effectively heats the electrons and eventually results in dissipation once the   excited electrons relax. 
%As explained below Eq.~\eqref{eq:eta_na_def} \im{fix link}, 
The  dissipation induced by the mechanism above is captured by $\eta_{\rm na}(t)$.

To estimate $\eta_{\rm na}$ we first identify the set of $\kv$-points for  which the time-dependence  of $H(\kv,t)$ is non-adiabatic;
we term this region of reciprocal space as the ``non-adiabatic'' region and denote it by $\mathcal V_{\rm na}$.
%\addFN{The dynamics of an electron which is taken near a Weyl point has the form of the Landau-Zener problem. }
The Landau-Zener formula~\cite{Landau_1932,Zener_1932} states that  time-dependence of $H(\kv,t)$  is non-adiabatic if, for some $t$, 
\be 
\hbar \norm{\partial_t H(\kv+e\vec A(t))}
    \gtrsim
    \delta \varepsilon^2(\kv+e\vec A(t))/\hbar,
    \label{eq:adiabaticity_condition}
\ee 
where 
$
\delta \varepsilon(\kv)
    \equiv 
    \varepsilon_2(\kv)-\varepsilon_1(\kv)
$, 
and 
$\varepsilon_\alpha(\kv)$ denotes the $\alpha$th energy band of $H(\kv)$.
 Using the linearized form of $H(\kv)$ in Eq.~\eqref{eq:weyl_hamiltonian}, a straightforward derivation (see Appendix~\ref{app:d0_bound}) shows that this condition is satisfied at % $\mathcal V_{\rm na}$ consists of
$\kv$-points for which
$
\min_t |\kv+ e \vec A(t)/\hbar|
    \lesssim d _0
$
where
\be 
d_0 
    \equiv 
    \sqrt{
        \frac{eE\norm{R}}{\hbar v_0^2},
        }
\label{eq:d0_result}
\ee 
while $\norm{R}$ and  $v_0$ denotes the largest and smallest eigenvalue of the velocity matrix $R$, respectively [see Eq.~\eqref{eq:weyl_hamiltonian}]~\footnote{We do not exclude the possibility that  dynamics are still be adiabatic for some $\kv$-points where $\min_t |\kv+ e \vec A(t)/\hbar|<d_0$. 
Our condition above  may thus overestimate the extent of $\mathcal V_{\rm na}$, and hence also $\eta_{\rm na}$. }.
%. \im{should we say here that this overestimates LZ heating, since in general the nodes will be approached from other directions?}
For incommensurate frequencies, $\mathcal V_{\rm na}$ thus consists of all $\kv$-points within a distance $d_0$ from the topological phase boundary $\mathcal B_0$ by our estimate. 
For commensurate frequencies $\mathcal V_{\rm na}$ consists all $\kv$-points within  a distance $d_0$ from $\mathcal C_0$, which forms a closed curve on $\mathcal B_0$, as in  Fig.~\ref{fig:weyl_surfaces}(d).

Electrons with wave vectors $\kv$ within $\mathcal V_{\rm na}$ encounter a vanishing gap of $\hat H(\kv,t)$ at times $t$ where $| \kv + e\vec A(t)/\hbar| \lesssim d_0$.
These electrons then undergo Landau-Zener tunneling, which  effectively heats them to a high-temperature state, as explained in the beginning of this subsection. 
%due to Landau Zener  tunneling between the  conduction and valance band, as . 
These high-temperature electrons then relax back to equilibrium after a characteristic time $1/\Gamma$. $\eta_{\rm na}(t)$ is then the rate of energy loss, or heating, (per unit volume) arising from  this relaxation. % process.
We estimate %  to have magnitude 
\be 
\eta_{\rm na} \sim \Delta\varepsilon_{\rm na} n_{\rm na} \Gamma/2,
\label{eq:eta_na_estimate_general}
\ee
where $n_{\rm na}$ is the concentration of excited electrons within $\mathcal V_{\rm na}$,  and $\Delta \varepsilon _{\rm na}$ denotes the characteristic   average value of  $\delta \varepsilon(\kv+e\vec A(t))$  for $\kv$ within $\mathcal V_{\rm na}$. %in the non-adiabatic region. 
Here the factor of $2$ comes because we estimate that the other half of the dissipated energy comes from  mode $2$.

We obtain  $\Delta \varepsilon _{\rm na}$ using that  $\mathcal V_{\rm na}$ %consists of $\kv$-points 
is located a distance $\sim eA/\hbar$ from the Weyl node, such that $\Delta \varepsilon _{\rm na}\sim  e A \norm{R} $, and $\norm{R}$ is the largest velocity implied by the velocity tensor $R$.  % (where $A$ denotes the characteristic magnitude of the driving-induced vector potential). 
Using $A \sim E/\Omega$, we obtain 
\be
\Delta \varepsilon _{\rm na}\lesssim eE\norm{R}/\Omega .
\label{eq:delta_e_estimate}
\ee
%The reminder of this subsection is devoted to estimating $n_{\rm na}$.

To  estimate $n_{\rm na}$, it is crucial to know the characteristic time interval between successive gap-closing events experienced by electrons with a given wavevector within $\mathcal V_{\rm na}$, $\Delta t$. 
%We estimate this timescale below. %(Sec.~\ref{sec:incommensurate} and \ref{sec:lissajous}).
To build intuition,  let us  first consider what happens when $\Delta t \gg 1/\Gamma$, i.e., when  electrons have time to fully relax  between successive gap-closing  events~\footnote{This regime typically does not occur for the limit  we consider where $\Gamma \ll \Omega$. However, it may  arise for rapid relaxation.}.
Electrons at wavevector $\kv$ are  taken to a high-temperature state whenever $\kv$ comes within a sphere of radius $\sim d_0$ from $e\vec A(t) /\hbar$.
Electrons are in equilibrium as they ``enter'' the sphere (due to our assumption $\Delta t \gg 1/\Gamma$), and we estimate that half of them are excited to the conduction band  as they ``leave'' the sphere.  
The concentration of electrons per unit time that are heated by %to the high-temperature state 
this process is hence given by the cross-section of this sphere times $\frac{0.5}{(2\pi)^3} e|\partial_t \vec A|/\hbar$.
Therefore, we expect the concentration of electrons heated per unit time to be given by $ e  d_0^2 |\partial_t {\vec A}(t)|/16\pi^2 \hbar$.
Assuming the electrons relax with characteristic rate $\Gamma$, we estimate $n_{\rm na}$ as the fixed point of $\partial_t n_{\rm na} = e  \pi d_0^2 |\partial_t {\vec A}(t)|/16\pi^2 \hbar  - \Gamma n_{\rm na}$.
Using $\partial_t \vec A = E$, we thus find  
$
n_{\rm na} \sim { e   d_0^2 E  }/{16\pi^2 \hbar \Gamma}. % \quad \textrm{for $\Delta t \gg 1/\Gamma$}.%{\rm for} \quad \Gamma \gg \Omega.
$
%where $E$ denotes the characteristic magnitude of the driving-induced vector potential.

Next, we consider the case where   $\Delta t\lesssim 1/\Gamma$. 
In this case, a significant fraction of electrons are already in a high-temperature state when they experience a gap-closing event (i.e., when they ``enter'' the sphere with radius $d_0$ centered at $e\vec A(t)/\hbar$).
Assuming that the gap-closing event effectively randomizes the state of the electrons (i.e., the electrons are in a infinite-temperature state right after ``leaving'' the sphere, regardless of their initial state), a subsequent gap-closing event only re-heats a reduced number of electrons to a high-temperature state. 
%since a significant fraction of electrons   have already been excited to a high-temperature from a previous gap-closing event.
We estimate the fraction of pre-excited electrons to be  of order $0.5\, e^{-\Gamma \Delta t}$ right before the gap closing and $0.5$ right after; thus the heating rate is reduced by a factor $\mathcal O(1-e^{-\Gamma \Delta t})$, resulting in 
\be
n_{\rm na} \sim \frac{ e   d_0^2 E  }{16\pi^2\hbar \Gamma} (1-e^{-\Gamma \Delta t}). 
% \quad \textrm{for $\Delta t \gg 1/\Gamma$}.%{\rm for} \quad \Gamma \gg \Omega.
\ee
%Combining this result with 
Combining this result  with  Eqs.~\eqref{eq:d0_result}-\eqref{eq:delta_e_estimate}, we  obtain
\be
\eta_{\rm na} \sim \frac{  e^3 E^3 \norm{R}^2  }{32\pi^2 \hbar^2\Omega  v_0^2}(1-e^{-\Gamma \Delta t}).
\label{eq:eta_na_result}
\ee
Below we  estimate $\Delta t$ (i.e., the characteristic time between gap closing events) for the two cases incommensurate and commensurate frequencies; as we find these two situations  lead to significantly different  $\Delta t$, and hence also different values of $\eta_{\rm na}$. %where we estimated $\Omega^2 \sim 4\pi^2 f_1f_2$.

Evidently, the bound above is controlled by the ratio between the largest and smallest eigenvalues of the matrix $R$,   $\norm{R}/v_0$. 
As we argued in Sec.~\ref{sec:weyl}, this number quantifies the anisotropy of the band gap around the Weyl point.

Note that the first factor in Eq.~\eqref{eq:eta_na_result} is {larger} than the {\it gross} rate of topological frequency conversion in Eq.~\eqref{eq:eta_gross}  (this follows from $ \norm{R}/v_0\geq 1$, and $\pi^2>8$).
Thus,  $\Delta t$  needs to be much shorter than  $\Gamma^{-1}$ for nonadiabatic heating not to overwhelm the net rate of topological frequency conversion. 
In particular, since $\Delta t$ is at least $2\pi/\Omega$, we expect $\Gamma\ll \Omega$ to be a necessary condition for topological frequency conversion. 

To illustrate the above result, we estimate $\eta_{\rm na}$ for the same  parameters as gave us  the estimates  $\eta_{\rm gross}= 500\,{\rm kW}/{\rm mm}^3$  and $\eta_{\rm mr}\sim 100\,{\rm kW}/{\rm mm}^3$, namely,   $E\sim 1500\,{\rm kV}/{\rm m}$, $\Omega = 2\pi\, {\rm THz}$, $\norm{R}\sim v_{0}\sim 5\cdot 10^5{\rm m}/{\rm s}$ [see text below  Eqs.~\eqref{eq:eta_gross} and Eq.~\eqref{eq:momentum_relaxation_estimate}]. 
With these parameters Eq.~\eqref{eq:eta_na_result} yields  $\eta_{\rm na} \sim 650{\rm kW}/{\rm mm^3}(1-e^{-\Gamma \Delta t})$. 
In the case of fast relaxation $\Gamma$,   $\eta_{\rm na}$ is clearly the dominant heating mechanism.
%For comparison, recall that $\eta_{\rm gross} \sim 500\,{\rm kW}/{\rm mm}^3$ for the same parameters.

\subsubsection{Nonadiabatic heating at incommensurate frequencies}

For incommensurate frequencies, we estimate $\Delta t$ as the  time-window  over which   trajectory of $e\vec A(t)/\hbar$ 
has length $|\mathcal B_0|/d_0$, where $|\mathcal B_0|$ denotes the area of the surface on which $e\vec A(t)/\hbar $ is confined to at all times, $\mathcal B_0 \equiv\{e{\bs \alpha}(\phi_1,\phi_2),\; 0<\phi_i<2\pi\}$ [see Sec.~\ref{sec:tfc} and Fig.~\ref{fig:weyl_surfaces}(b)] 
Since $\partial_t \vec A(t) =\vec E(t)$, the trajectory of $e\vec A(t)/\hbar$ over the time-window $\Delta t$ has length $ eE\Delta t/\hbar$.
Estimating $|\mathcal B_0|\sim 4\pi^2 e^2 A^2/\hbar^2$ and using $d_0 =  \sqrt{ {eE }\norm{R}/{\hbar v^2_0}}$ along with $A\sim E/\Omega$, we hence obtain
\be 
\Delta t  \sim \frac{4\pi^2}{\Omega^2} \sqrt{\frac{eEv_0^2}{\hbar\norm{R}}}%\sqrt{\frac{eE}{\hbar\norm{R}}}
\quad \textrm{for irrational $\omega_1/\omega_2$}
\ee
% Using this in Eq.~\eqref{eq:eta_na_result}, along with $1-e^{-\Gamma \Delta t}\approx \Gamma \Delta t$, we thus find % we find 
% \be
% \eta_{\rm na} \lesssim  \frac{\Gamma\norm{R}}{2\pi^2} \sqrt{\frac{eE\hbar}{ v_0}} \left(\frac{eE}{\hbar\Omega}\right)^3 %e^{7/2} E^{7/2}\norm{R}^{3/2}    \Gamma  }{  \hbar ^{5/2}\Omega^3  v_0}.
% \label{eq:p_rec_incomm}
% \ee
% \addFN{Note that the rate scales as $I^{7/4}$ but saturates to a $I^{3/2}$ scaling. }

For the parameters we used to estimate $\eta_{\rm na}$ above [$E\sim 1500\,{\rm kV}/{\rm m}$, $v_{\rm F}\sim 5\cdot 10^5{\rm m}/{\rm s}$, and $\Omega = 2\pi\, {\rm THz}$] this estimate yields 
$\Delta t \sim 30\,{\rm ps}$. 
To achieve topological frequency conversion at incommensurate frequencies with these parameters, the characteristic relaxation time $\tau = \Gamma^{-1}$ must be much longer than this timescale.
In this limit (i.e., $\tau \gg \Delta t$), we find  $\eta_{\rm na}\sim 5 \cdot 10^{-8}{\rm kJ}/{\rm mm}^3\tau $, which  is smaller than $\eta_{\rm gross}$ %the {gross} rate of topological frequency conversion
when $\tau \gtrsim 100\,{\rm ps}$. %$\bar \eta_0$. 

\subsection{Lissajous Conversion}
\label{sec:lissajous}

We now consider the case of commensurate frequencies, which can give a marked reduction of the non-adiabatic losses.

 Eq.~\eqref{eq:eta_na_result} shows that
 small time-intervals between subsequent gap-closing events, $\Delta t$, leads to suppression of nonadiabatic heating, $\eta_{\rm na}$. 
An important consequence of this  is that   $\eta_{\rm na}$ is strongly suppressed for commensurate frequencies, i.e., when 
\be 
f_1/f_2= q/p
\ee
 for some  integers $p$ and $q$ (which we, without loss of generality, take to have no common divisor).  
In this case, $e\vec A(t)/\hbar$  forms a 3-dimensional Lissajous figure in reciprocal space, as in Fig.~\ref{fig:weyl_surfaces}(d), and is time-periodic with period 
$
qT_1 =pT_2 .
$ 
Thus, a given $\kv$-point within $\mathcal V_{\rm na}$ experiences  gap-closing events with periodicity~\footnote{In principle, gap-closing events may occur more often if  $e\vec A(t)/\hbar$ returns to    within a distance $d_0$ from the same point over a time-interval shorter than $\Delta t$ (as occurs for near commensurate-frequencies, or if $p/q$ is close to some rational fraction which involves smaller integers). In any case, $  qT_1$ provides an upper bound for $\Delta t$.}
\be
\Delta t = qT_1=pT_2.
\label{eq:delta_t_lissajous}% \quad \textrm{for  $\omega_1/\omega_2 = q/p$}
\ee
% Since we assume $\Gamma \ll \omega_1,\omega_2$, we may hence approximate $1-e^{-\Gamma\Delta t}\approx \Gamma \Delta t$. 
% Using $\omega_1,\omega_2 \sim \Omega$, this leads us to  
% \be
% \eta_{\rm na} \sim p \frac{e^3 E^3\norm{R}    \Gamma  }{ 4\pi  \hbar ^2\Omega^2v_0}.
% \label{eq:rational_eta_na_result}
% \ee
% The freedom in choosing $p$ or $q$ arises because we assume $\omega_1$ and $\omega_2$, and hence also $p$ and $q$, to have the same magnitude. 
% We expect the more general case can also be analyzed using the approach above. 
As a consequence, $\eta_{\rm na}$ 
is strongly  suppressed for highly rational frequency ratios, i.e., when $p$ and $q$ are small. 

The  suppression of $\eta_{\rm na}$  means that net amplification from topological frequency conversion is significantly enhanced at highly  commensurate frequencies.
We term this mechanism of topological frequency conversion at commensurate frequencies   {\it Lissajous conversion}.
The dramatic suppression of nonadiabatic heating results in an enhanced  net frequency conversion rate in  the Lissajous conversion regime, as     is evident in  our numerical simulations (see Fig.~\ref{fig:front_page_figure}(b), and Sec.~\ref{sec:numerics}).

As an example, we consider  Lissajous conversion at frequencies $\omega_1 = 2\pi\, {\rm THz}$ $\omega_2 = \frac{3}{2}\omega_1$.
These parameters result in  $\Delta t \sim 3 {\rm ps}$.
To compare, recall that $\Delta t $ was estimated to $30\,{\rm ps}$ in the same frequency range. 
Using Eq.~\eqref{eq:eta_na_estimate_general}, we estimate the nonadiabatic heating rate to be given by $\eta_{\rm na}\sim 3.5 \cdot 10^{-9}{\rm kJ}/{\rm mm}^3\tau $, which  we expect can be smaller than $\eta_{\rm gross}$ when $\tau \gg 10\,{\rm ps}$. 
In contrast, recall that our estimated nonadiabatic heating rate at incommensurate frequencies in the same frequency range is   given by $5\cdot 10^{-8}{\rm kJ}/{\rm mm}^3\tau$, and thus more than $10$ times larger.

%We expect Lissajous conversion to also occur for {near-commensurate} frequencies. 
%Specifically, we expect Eq.~\eqref{eq:delta_t_lissajous} also holds when $\omega_1/\omega_2$ is sufficiently close to $q/p$ such that $e|\vec A(t)  -\vec A(t+2\pi q/\omega_1)|/\hbar \ll d_0  $ for all $t$.
In the limit of large $p$, $q$, we expect that our estimate for $\Delta t$ saturates at the expression we obtain for incommensurate frequencies. % below [Eq.~\eqref{eq:p_rec_incomm}]. 
%In this regime, we consider $\omega_1$ and $\omega_2$ to be effectively incommensurate.

\section{Numerical simulations of frequency conversion}
\label{sec:numerics}
  
\begin{figure}[t]
\includegraphics[width=0.99\columnwidth]{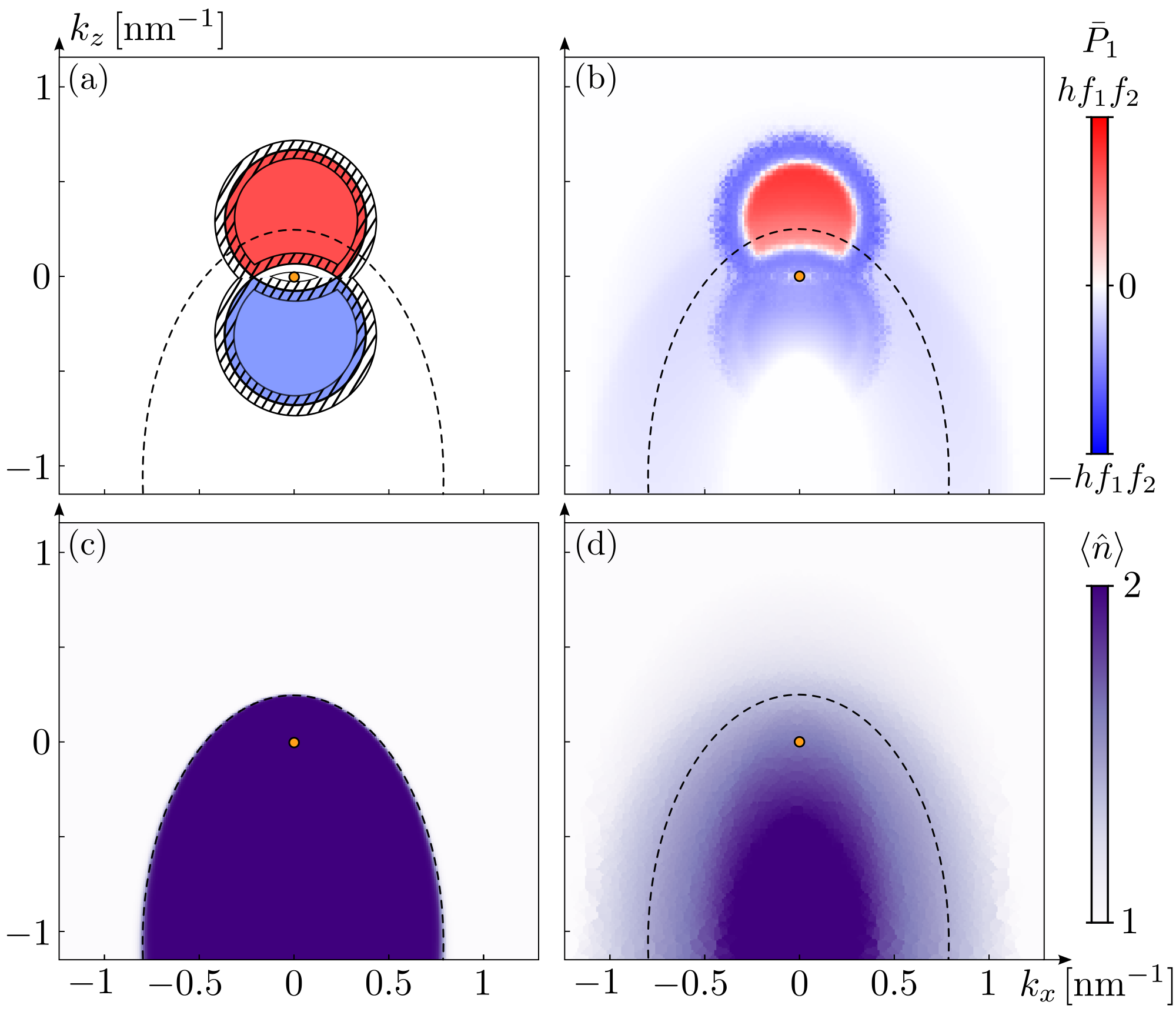}
\caption{
Energy conversion in  the model we study in Sec.~\ref{sec:numerics}. 
%See main text for  model details and parameters. % and parameters. 
(a) Plot of  $W(\kv)$ in the plane $k_y=0$, with red, blue, and white  indicating  values $-1$, $1$, and $0$, respectively.  
Black circle indicates  $\mathcal B_0$ and shaded region schematically indicates  the non-adiabatic region
 $\mathcal V_{\rm na}$. 
 Orange circle  and dashed line indicate the location of the Weyl point and equilibrium Fermi surface.
(b): net rate of energy transfer to  mode 1 from   electrons  with a given wavevector $\kv$, $\bar P(\kv)$, as a function of $\kv$ in the plane $k_y=0$, for same parameters as in Fig.~\ref{fig:conv_power}(a).
(c) occupation number in equilibrium of electronic modes with   wavevector $\kv$, as a function of $\kv$ in the plane $k_y=0$, for the same system as in panels (a-b). 
(d): net  time-averaged occupation number of electronic modes with   wavevector $\kv$, $\langle \hat n(\kv)\rangle$ as a function of $\kv$ in the plane $k_y=0$ for the system  depicted in panel (a-c). 
% when in equilibrium.
%See main text for  details and parameters.
}
%\comment{In panel a, use uniform, opaque gray shading}}
\label{fig:conv_power}
\label{fig:tfc}
\end{figure}

We now support our theoretical predictions by data from numerical simulations. 

In our simulations, we consider  the dynamics of  electrons  near a single Weyl node in a Weyl semimetal with $2$ bands.
The electrons are subject to the  linearized Bloch Hamiltonian
\be 
H(\kv) = \hbar  v \vec k\cdot {\bs \sigma} +   \hbar \vec k \cdot {\vec V}.
\ee 
We also introduce two  electromagnetic modes that are circularly-polarizedthat propagate in the $yz$- and $xz$-planes, respectively.
For $i=1,2$, mode $i$ has angular frequency  $\omega_i $ and electric field amplitude  $\mathcal E_i$ inside the material.
It thus induces the time-dependent electric field $\vec E_i(t)$, where
\begin{align}
\vec E_1(t) &=\mathcal  E_1 (\cos \omega_1 t ,0,\sin \omega_1 t),\label{eq:e1}\\ \vec E_2(t)& = \mathcal E_2 (0,\cos   \omega_2t ,\sin  \omega_2 t).\label{eq:e2}
\end{align}
The irradiated electrons are governed by the  time-dependent Bloch  Hamiltonian $H(\kv,t)= H(\kv+e\vec A(t)/\hbar)$, where $\vec A(t)$ denotes the driving-induced vector potential and is  defined through   $\partial_t \vec A(t) = \vec E_1(t) + \vec E_2(t)$ (see also Sec.~\ref{sec:tfc}).
As in the previous sections, we work in a gauge where $\vec A(t)$ has vanishing time-average.

We numerically obtain the evolution of  the momentum-resolved density matrix of the system, $\hat \rho(\kv,t)$ (see Sec.~\ref{sec:many_body_tfc} for definition), using the master equation in Eq.~\eqref{eq:vn_modified}.
We take the dissipator   $\mathcal D$ to be    given by  the Boltzmann form  [Eq.~\eqref{eq:boltzmann}]:
$
\mathcal D(\kv,t)\circ \hat \rho  =- \frac{1}{\tau}[\hat \rho-\hat \rho_{\rm eq}(\kv,t)].
$
Here $\hat \rho_{\rm eq}(\kv,t)$ denotes the instantaneous equilibrium state of electrons with crystal momentum $\kv$ at time $t$  at some given temperature $T$ and chemical potential $\mu$ [see text below Eq.~\eqref{eq:boltzmann} for explicit definition]. %which we take to be positive in our simulations.
Since Eq.~\eqref{eq:vn_modified} describes evolution in the $4$-dimensional second-quantized Bloch space of the system,  its  numerical solution is relatively inexpensive. % for a given  value of $\kv$. 

For each $\vec k$, we numerically solve Eq.~\eqref{eq:vn_modified} to obtain the  steady-state evolution of $\hat \rho(\kv,t)$. 
From this steady state we extract the  quantity
\be 
\bar P(\kv)  \equiv \lim_{t\to \infty}\frac{1}{t}\int_0^{t}\!\! {\rm d}s \frac{e}{\hbar}  \vec E_1(s)\cdot \Tr[\nabla \hat H(\kv,s) \hat \rho(\kv,s)]. %\vec v(\kv,s).
\ee
$\bar P(\kv)$ gives the time-averaged    {\it total}  rate of energy transfer  to mode $1$ from  electrons with wavevector $\kv$. % (i.e., both including dissipation and topological frequency conversion). 
The total time-averaged rate of energy transferred to mode $1$ per unit volume of the whole system, $\bar \eta$,  % Weyl semimetal 
is obtained by integrating $P(\kv)$ over all wavevectors: 
\be 
\bar \eta = \kint \bar P(\kv).
\label{eq:eta_numeric_result}
\ee
In our simulation, we evaluate the $\kv$-integral above by sampling $\bar P(\kv)$ over a large number of uniformly distributed values of $\kv$~\footnote{We estimate the accuracy of our computed value of $\bar \eta$   using the standard deviation of the values of $\bar \eta$  that result from restricting the integral random subsets of the sampled $\kv$-points.}.

We solve the master equation for $\hat \rho(\kv,t)$ through  direct  integration, not making use of any of the  approximations  of Sec.~\ref{sec:finite_dissipation}.  
In particular, our simulation does not distinguish between coherent and incoherent dynamics, and our obtained value for $\bar \eta$  thus includes both  contributions both from topological frequency conversion {and} dissipation. 
Hence our simulation  can be used to  test the conclusions in Sec.~\ref{sec:many_body_tfc}.

We probe different values of $f_1$ and $\tau$, while keeping all other parameters fixed at values $f_2 = 1.23\,{\rm THz}$, $v=3.87\cdot 10^5\, {\rm m}/{\rm s}$, $\vec V = (0,0,3.1\cdot 10^5\, {\rm m}/{\rm s})$,  $\mu = 115 \,{\rm meV}$,  $T = 20\,{\rm K}$, $\mathcal E_1 =0.9\, {\rm M V}/{\rm m}$, and $\mathcal E_2 = 1800\,{\rm kV}/{\rm m}$. 
% Our chosen  values of $\mathcal E_1$ and $\mathcal E_2$ correspond to intensities $2.14\,{\rm kW}/{\rm mm}^2$ and $8.57\,{\rm kW}/{\rm mm}^2$, respectively, which are experimentally accessible with pulsed lasers in some frequency ranges~\cite{French_1995}.
%\im{we should check achievable E with THz; if this ref is for lasers, it is at wrong frequency. Should we ask Armitage?}\fn{good point. Let's discuss and decide what to do}
Our chosen values of $v$ and $\vec V$ have magnitudes  comparable  to those in real materials~\cite{Arnold_2016,Ramshaw_2018}.
The  values of $\mu$ and  $\vec V$ are chosen to maximize the imbalance between the number of electrons acting as frequency converters at rates $hf_1f_2 $ and $-hf_1 f_2$, as discussed in Sec.~\ref{sec:non_dissipative} (see also Sec.~\ref{sec:origin_of_ec} below). %\im{need to fix link} below). 

\subsection{Identification of amplification regime}
In Sec.~\ref{sec:many_body_tfc}, we showed that the time-averaged rate of energy transfer to mode $1$ can be decomposed as $\bar \eta=\bar \eta_0+\bar \eta_{\rm dis}$. 
Here $\bar \eta_0$ can be positive due to topological frequency conversion, %and and denotes the time-averaged rate  of  topological energy conversion, while
while $\bar \eta_{\rm dis}$ is negative and measures the time-averaged rate of energy dissipated from mode $1$ due to heating in the system. 
We expect $|\bar \eta_{\rm dis}|$  to decrease with increasing relaxation time $\tau$, while $\bar \eta_0$   remains constant. 
Thus, $\bar \eta$ should increase with $\tau$.
There should also exist  a critical value of $\tau$ for which $\bar \eta=0$. 
When  $\tau$ is larger than  this ``amplification threshold'', the system will amplify mode $1$  ($\bar \eta>0$). %, corresponding to net amplification  of mode $1$. 
We expect the amplification threshold  to  be significantly lower in the Lissajous conversion regime (i.e., at rational frequency ratios) than for irrational frequency ratios due to the suppression of nonadiabatic heating in the former case (see  Sec.~\ref{sec:lissajous_numerics} and Sec.~\ref{sec:lissajous_numerics} below).

To identify the amplification threshold for the system, we computed   $\bar \eta$  as a function of   $\tau$ for three representative choices of $f_1/f_2$; namely, irrational $f_1=\frac{1}{\varphi} f_2 $, rational $f_1=\frac{2}{3}f_2$, and  nearly-rational $f_1 =\frac{2}{3+\epsilon}f_2$, where $\varphi$ is given by the ``golden mean'', $\frac{1}{2}(1+\sqrt{5})$, and $\epsilon=\pi/1000$. 
We keep all other parameters fixed at the values we specified earlier.  
% These two values are thus  chosen to reveal the dependence of the amplification threshold on the commensurability of $f_1/f_2$ (see below). 
The two latter values  of $f_1$ are chosen to demonstrate the mechanism of Lissajous conversion: 
Whereas  $f_1=\frac{2}{3}f_2$ is commensurate with $f_2$, $f_1 =\frac{2}{3+\epsilon}f_2$ is not, and hence  the former value of $f_1$ is expected to yield more efficient---Lissajous---conversion.

In Fig.~\ref{fig:front_page_figure}(c) we plot $\bar \eta$ as a function of $\tau$ for the three values of $f_1$ above.
As we expect, $\bar \eta$ increases as a function of $\tau$ for all  choices of $f_1$,  and attains positive value for sufficiently large $\tau$. 
For the irrational frequency ratio $f_ 1= f_2 /{\varphi}$, the amplification threshold is reached  at $\tau\approx 1000\,{\rm ps}$,  for $f_1=2f_2/3$  at $\tau \approx 300\,{\rm ps}$ and for $f_1={2}f_2/{(3+\epsilon)}$ above $\tau =1200 \,{\rm ps}$. 

Note that the weak  detuning  of $f_1$  from $2f_2/3$ (green curve) to ${2}f_2/({3+\epsilon})$ (orange curve)  reduces $\bar \eta$ by more than $100\,{\rm kW}/{\rm mm}^3$, and pushes the amplification threshold from $300$ to $1200\,{\rm ps}$.
This demonstrates the strong dependence of the net conversion rate  on the commensurability of  $f_1$ and $f_2$ that we discussed in Sec.~\ref{sec:lissajous}.
%\im{didnt we claim that for small enough epsilon we can keep the benefits of Lissajoius? Here it makes it worse than Golden mean case. Perhaps Godlen mean case is also in some sense optimal -- best of all irrationals, most homogenized? and the worst cases are the less-irrational, faster convergent continued fraction cases? Like could try copper mean, silver mean etc. Link in comment below}\fn{I removed this sentence in the previous section, since there are some subtleties, and I don't think it is helpful to get more involved in the details. Ok?}

% https://en.wikipedia.org/wiki/Metallic_mean

%conclusions in Sec.~\ref{sec Sec.~\ref{sec:lissajous_numerics} below [see also Fig.~\ref{fig:dissipation}].

\subsection{Origin of energy conversion}
\label{sec:origin_of_ec}
Next, we  confirm that the amplification of mode 1 (i.e, the positive values of $\bar \eta >0$) we observed is due to topological frequency conversion.
%topological origin of the positive net energy conversion rates we found above.
To this end, we compute  $\bar P(\kv)$ as a function of $\kv$ around the Weyl point.
 %for the parameter sets we probed. 

We first review the  signatures of topological frequency conversion we expect to see. 
For  $\kv$-points where  $H(\kv,t)$ changes adiabatically in time,  electrons should act as  topological frequency converters (as in Ref.~\cite{Martin_2017}) 
that transfer  energy to mode $1$ at  an average rate quantized as   $\pm h f_1f_2W(\kv)$, where the $W(\kv)$   denotes the integer-valued net winding number of the surface $\mathcal B_0$  around $\kv$ [see  Fig.~\ref{fig:weyl_surfaces}(c)].
Here $+$ and $-$ result from  electrons in band $1$ and $2$, respectively. 
We hence expect
\be
\bar P(\kv) = hf_1f_2W(\kv) [\bar \rho_1(\kv)-\bar \rho_2(\kv)] + P_{{\rm dis}}(\kv),
\label{eq:p_hypothesis}
\ee
where  $\bar \rho_\alpha(\kv)$ denotes the time-averaged occupancy of band $\alpha$, and $P_{\rm dis}(\kv)$ denotes the rate of energy loss from mode $1$ due to dissipation.
We expect the latter is always negative, but  only significant around the Fermi surface (due to momentum relaxation), and within the nonadiabatic region (due to nonadiabatic heating).

In Fig.~\ref{fig:conv_power}(a) we plot $W(\kv)$ in the plane $k_y=0$, for  $f_1 = f_2/{\varphi}$ and with all other parameters specified below Eq.~\eqref{eq:eta_numeric_result}. 
We also indicate the Fermi surface (dashed line) and schematically indicate the nonadiabatic region  (shaded region), which surrounds the topological phase boundary (solid line). % the expected nonadiabatic . 
Since $\mu>0$,  band $1$ is fully occupied in equilibrium. 
We therefore expect $\bar \rho_1(\kv)\approx 1$ [see Eq.~\eqref{eq:boltzmann_result}] for all $\kv$ away from the nonadiabatic region  (where  Landau-Zener tunneling can induce holes).
We hence expect topological frequency conversion causes $\bar P(\kv)$ to approximately take value $  hf_1f_2[1-\bar \rho_2(\kv)]$ within the red region of Fig.~\ref{fig:conv_power}(a),  value $ -hf_1f_2[1-\bar \rho_2(\kv)]$ in the blue region, and value $0$ in the white region. 

In Fig.~\ref{fig:conv_power}(b), we plot  $\bar P(\kv)$ in the plane $k_y=0$ in for parameters $f_1= f_2/\varphi$ and $\tau = 51.6 \,{\rm ps}$~\footnote{
We choose this  value of $\tau$  for illustrative purposes, to  make the contributions to $\bar P(\kv)$ from  topological frequency conversion and dissipation visible on the same energy scale.}.
The data  shows clear signatures of topological frequency conversion, in the form of  two ``topological plateaux''  of the Brillouin zone where $\bar P(\kv)$ takes  positive and negative values, respectively.
These plateaus coincide closely with the  regions in Fig.~\ref{fig:conv_power}(a) where $W(\kv)=\pm 1$. 
$\bar P(\kv)$ approximately takes value $ hf_1f_2$ within the red  plateau (away from the Weyl point), and value  between $0$ and $-hf_1f_2$ within the blue plateau (close to the Weyl point). 

We expect $\bar P(\kv)$ differs from $\pm hf_1f_2$ in the topological plateaux due to the finite value of  $\bar \rho_2(\kv)-\bar \rho_1(\kv)$. 
To confirm this,    we  computed the time-averaged number of electrons per $\kv$-point, $\langle \hat n(\kv)\rangle \equiv \lim_{t\to \infty}\frac{1}{t} \int_0^t ds\, \Tr[\hat n \hat \rho(\kv,s)]$, where $\hat n = \hat c^\dagger_1 \hat c_1 + \hat c^\dagger_2\hat c_2$. %for the same parameters as considered in Figs.~\ref{fig:conv_power}(b).
Our expectation that $\bar \rho_1(\kv)\approx 1$ implies that $2-\langle \hat n(\kv)\rangle$ should be  a good proxy for $[\bar \rho_2(\kv)-\bar \rho_1(\kv)]$ away from the nonadiabatic region.
In Fig.~\ref{fig:conv_power}(d) we plot $\langle \hat n(\kv)\rangle$.
Taken in combination with Figs.~\ref{fig:conv_power}(ab), our data are thus  consistent  with $\bar P(\kv)$ taking value $ \pm hf_1f_2[\bar \rho_2(\kv)-\bar \rho_1(\kv)]$ within the topological plateaux. 
%This strongly indicates  that the positive values of $\bar P(\kv)$ (and hence also $\bar \eta$) are due to topological frequency conversion. 

% The data from our numerical simulations confirm additional features of the steady state that we predicted in Sec.~\ref{sec:many_body_tfc}. First, 
In addition to topological frequency conversion, the data in %topological origin of the net amplification of mode $1$, 
Fig.~\ref{fig:conv_power}(b) also  shows clear signatures of the two distinct mechanisms for dissipation that we identified in Sec.~\ref{sec:finite_dissipation}, i.e., momentum relaxation and nonadiabatic heating:
% Specifically, we expect nonadiabatic heating causes  $ P_{\rm dis}(\kv)$ to be significant near the topological phase boundary $\mathcal B_0$, i.e., within  the shaded region in Fig.~\ref{fig:conv_power}(a). 
% We also expect $\bar P_{\rm dis}$ to be nonzero near the Fermi surface [indicated by dashed line in Fig.~\ref{fig:conv_power}(a)] due to momentum relaxation, and to effectively vanish   everywhere else.
 $\bar P(\kv)$ takes  large negative values within the nonadiabatic region, as we expect from nonadiabatic heating, and moderate negative values around the Fermi surface, as we expect  from momentum relaxation. % due to nonadiabatic heating. 

% \subsection{Characteristic features of steady state}
% \label{sec:numerics_steadystate}
% Next, we consider the predicted features of the steady state that exhibited by the data we obtained.

Note also that the data in Fig.~\ref{fig:conv_power}(d) are in good agreement with our prediction that in the regime $\tau \gg 1/\Omega$,  the steady state band populations are effectively   ``smeared'' versions of their  equilibrium counterparts [see Eq.~\eqref{eq:boltzmann_result}]: 
The distribution in Fig.~\ref{fig:conv_power}(d) clearly resembles a ``smeared'' version of the  ellipsoid-profile that occurs in equilibrium [plotted in Fig.~\ref{fig:conv_power}(c)].

Finally, the data in Figs.~\ref{fig:conv_power}(cd)  demonstrate how a nonzero  value of the ``Weyl cone tilt'' $\vec V$ is needed to nonzero net rate of topological frequency conversion,  $\bar \eta_0$, as we discussed in Sec.~\ref{sec:non_dissipative}. 
The nonzero value of $\vec V$, which causes an ellipsoid-profile of $\langle \hat n(\kv)\rangle $ in equilibrium [Fig.~\ref{fig:conv_power}(c)], results in  a ``smeared ellipsoid'' profile of $\langle \hat n(\kv)\rangle$ in the steady state. 
%by the steady state  ellipsoid shape.
As a result of this smeared ellipsoid profile,  the region  characterized by  $W(\kv)=1$  has a larger volume in which $[\bar \rho_2(\kv)-\bar \rho_1(\kv)]>0$ ($\langle \hat n(\kv)\rangle \leq 2$) than  than the volume where $W(\kv)=-1$, allowing for a nonzero value of  $\bar \eta_0$. 
%hf_1f_2\kint W(\kv) [\bar\rho_2(\kv)-\bar \rho_1(\kv)]$.  
%\im{should we comment on the choice of chemical potential? is what we chose optimal in the sense that it allows us to kill the negative contribution from the lower band by the positive contribution from the upper band?}

\subsection{Lissajous Conversion}
\label{sec:lissajous_numerics}
\begin{figure}[t]
\includegraphics[width=0.99\columnwidth]{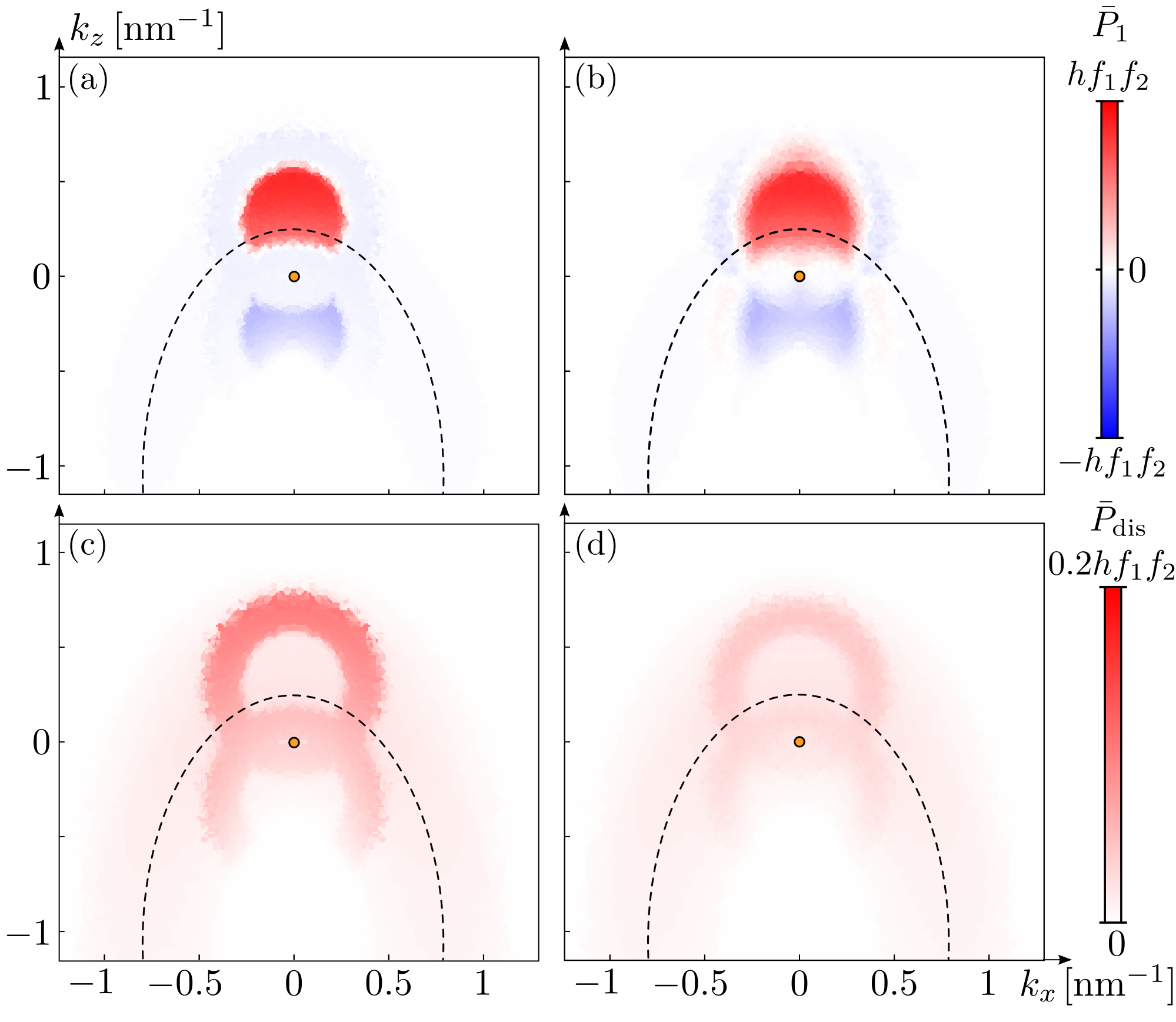}
\caption{ 
Evidence of Lissajous conversion.
(a,b): plot of 
$\bar P(\kv)$ in the  plane $k_y=0$ for parameters $\tau = 516\, {\rm ps}$,  $f_1 = {2}f_2/({3+\epsilon})$(a), and $f_1 = 2f_2/3$  (b). % $\tau = 200\, {\rm ps},\omega_1 = \frac{\sqrt{5}-1}{2}\omega_1$ (c), and $\tau = 200\, {\rm ps},\omega_1 = \frac{2}{3}\omega_2$ (d). 
(c,d): total dissipation rate 
$\bar P_{\rm tot}(\kv)$ for the same  parameters as depicted in panels (a) and (b),respectively. 
Note the different color scales used in panels (a,b) and (c,d). 
}
%\comment{Also show plots with dissipative absorption?}}
%(b) occupation number in the absence of driving for the same parameters as in panel (a). }
\label{fig:dissipation}
\end{figure}

We finally verify that the enhancement of $\bar \eta$ in the Lissajous regime (i.e., at commensurate frequencies) is due to the suppression  of nonadiabatic heating. 
To this end, we plot in Fig.~\ref{fig:dissipation} $\bar P(\kv)$ for the parameters $\tau = 516\,{\rm ps}$ and $f_1=2f_2/3$  (a) and $f_1=2f_2/(3+\epsilon)$ (b). 
The two choices of $f_1$ are very close, but whereas the former choice of $f_1$ is commensurate with $f_2$, the latter choice is not. 
The negative values of $\bar P(\kv)$ within the nonadiabatic region (which we attribute to nonadiabatic heating), are much fainter in panel (a) than in panel (b). 
This is consistent with our expectation that  nonadiabatic heating is indeed significantly suppressed for $f_1= \frac{2}{3}f_2$ compared to $\frac{2}{3+\epsilon}f_2$. 

We also compute the {\it total} dissipated power in the system due to both  driving modes,
$
\bar P_{\rm dis}(\kv)=-\lim_{t\to \infty}\int_0^t\frac{ds}{t} \Tr[\nabla \hat H(\kv,s) \hat \rho(\kv,s)] \cdot (\vec E_1(s)+\vec E_2(s)),
$
 for the same parameters as in panels (a) and (b).
 $\bar P_{\rm dis}(\kv)$ measures  the time-averaged  rate of work done on electrons with wavevector $\kv$  by the two driving modes in combination;  hence it measures the total rate of dissipation, and is guaranteed to be positive due to the second law of thermodynamics. 
In Fig.~\ref{fig:dissipation}(c) and (d) we plot $\bar P_{\rm dis}(\kv)$ for the  parameter sets we considered in panels (a) and (b), respectively. 
While outside the nonadiabatic region,  $\bar P(\kv)$ and $\bar P_{\rm dis}(\kv)$  effectively take the same values for the two frequency ratios, nonadiabatic heating is much weaker in panel (d)  than in panel (c).
% While outside the outside the nonadiabatic region,  $\bar P(\kv)$ and $\bar P_{\rm tot}(\kv)$ effectively take the same values for the two panels.  
%and $\bar P_{\rm dis}(\kv)$ take the same values for the two choices of $f_1$. 
The very different values of $\bar \eta$ at frequencies $f_1= \frac{2}{3}f_2$ and $f_1=\frac{2}{3+\epsilon}f_2$ must therefore be due  to this suppression of nonadiabatic heating in the commensurate case. 
%this confirms that Lissajous conversion is responsible for the enhancment of $\bar \eta$ at frequency ratio  $f_1/f_2= 2/3$.

\subsection{Nonanalytic  amplitude dependence}
\begin{figure}[t!]
    \centering
    \includegraphics[width=0.99\columnwidth]{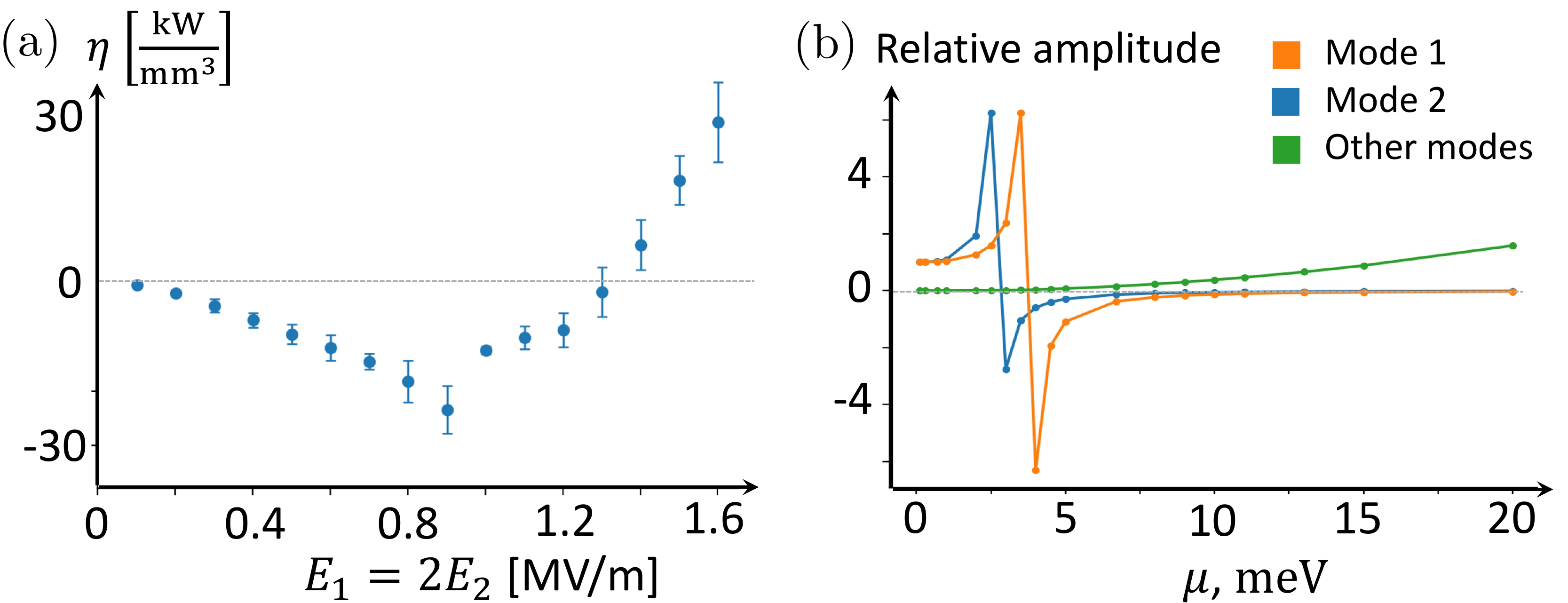}
    \caption{
        \addFN{(a) Net conversion rate as a function of the electric field amplitude inside the material, $E_1$, with $E_2$ fixed to $E_1/2$ throughout, while  $f_1 = \frac{3}{2}f_2 = 3.18\,{\rm THz} $, $\tau =516.3$, and, $\mu=115\,{\rm meV}$ (mode configurations and band structure parameters are given in the main text). 
        (b)
        Relative amplitude of mode $1$ (blue), $2$ (orange), and of all other modes (green) inside the material, due to current-induced plasma oscillations in the grain of Weyl semimetal; see Sec.~\ref{sec:plasmon} for  details of the computation.  }}
    \label{fig:escaling}
\end{figure}

\addFN{We next explore the relationship between the topological frequency conversion rate,  the amplitudes of the incoming modes. %and  and chemical potential of the system. 
% To this end, we computed $\bar \eta$ for various values of the chemical potential, $\mu$, and the electric fields of mode 1 and 2, $E_1$ and $E_2$. 
Fixing $E_2= 2E_1$, Fig.~\ref{fig:escaling}(a) plots $\bar \eta$ as a function of   $E_1$ for the isolated Weyl node studied in the previous subsections, with $f_1 = \frac{3}{2}f_2 = 1.23\,{\rm THz} $,   $\tau= 516.3\,{\rm ps}$, and $\mu=115\,{\rm meV}$. 
The error bars in Fig.~\ref{fig:escaling}(a) indicate the estimated uncertainty  due to the finite number of $k$-points   we sample%at a given parameter set
~\footnote{Specifically, the uncertainty is computed by extrapolating the standard deviation  of $\bar \eta$ that results from restricting the computation to randomly selected subsets of the probed $\kv$-poitns}. 
The conversion rate exhibits a clear cusp when $E_1\approx 1000\,{\rm kV}/{\rm m}$; by inspecting the $k$-dependent frequency conversion rate as in Fig.~\ref{fig:conv_power}, we verified that this is the amplitude where topological frequency conversion sets in due to the surface $\mathcal B_0$ crossing the Fermi surface. 
The cusp of $\bar \eta$ reveals a non-monotonous and nonlinear dependence on driving amplitude, supporting our conclusion that topological frequency conversion  is an effectively nonperturbative response phenomenon.
The amplification threshold is reached at amplitude $E_1\approx 1300\,{\rm kV}/{\rm m}$. }

% \subsection{Role of plasmonic response}
% Next, we consider the dependence of $\bar \eta $ on the chemical potential, $\mu$. 
% Fig.~\ref{fig:escaling}(b) shows $\bar \eta$ as a function of $\mu$ for incoming radiation amplitudes $E_1=1.5\,{\rm MV}/{\rm m}$,  $E_2=0.9\,{\rm MV}/{\rm m}$ (orange), and  $E_1=0.75\,{\rm MV}/{\rm m}$,  $E_2=0.5\,{\rm MV}/{\rm m}$ (green);  all other parameters coincide with the values we considered in panel (a). 
% For both choices of amplitudes, $\bar \eta$ can clearly be positive for a finite, nonzero interval of $\mu$. 
% Interestingly,  smaller chemical potentials   support  frequency conversion at weaker  amplitude.

\section{Conditions for frequency conversion}
\label{sec:conditions}

%In this section 
There are several conditions that a Weyl material must satisfy to realize topological frequency conversion. 
%that the conversion effect would not vanish already on symmetry bases. 
The conditions can be grouped into the conditions that an individual Weyl node must satisfy %to contribute to frequency conversion
(Sec.~\ref{sec:node_condition}),  the conditions on the global band structure and symmetry class of the system (Sec.~\ref{sec:symmetry_condition}), conditions on the driving (Sec.~\ref{sec:driving_conditions}), and conditions on relaxation (Sec.~\ref{sec:relaxation_condition}).

\subsection{Conditions on individual Weyl nodes}
\label{sec:node_condition}

% Except for cases of comensurate frequencies where one has control over the phase difference between the modes, t \im{fragment}
The rate of topological frequency conversion from Weyl node $i$ is given by the $i$th term in the sum over Weyl nodes in Eq.~\eqref{eq:main_result}: 
$
\bar \eta_i =q_i hf_1f_2
 \kint  [\bar\rho_{2}(\kv)-\bar \rho_{1}(\kv)] W(\kv-\kv_i).
%\label{eq:pumping_results_fom}
$
Thus,  Weyl point  $i$ can only contribute to  frequency conversion if  
\be 
\bar \rho_{1}(\kv_i+\delta \kv)\neq \bar \rho_{2}(\kv_i+\delta \kv) \quad\textrm{ for  $\delta \kv$ within $\mathcal B_{0}$.
}%(Condition $1$)}% \quad \textrm{}
\label{eq:condition_1}
\ee  
Therefore only the Weyl nodes sufficiently near the Fermi energy can contribute to topological frequency conversion. If a Weyl node is too far from the Fermi energy, the two touching bands are either both full or empty within a distance $\sim eA/\hbar$ from the Weyl node, implying  $\bar \rho_{1}(\kv_i+\delta \kv) \approx \bar \rho_{2}(\kv_i+\delta \kv) $ for $\delta \kv_i$ within $\mathcal B_0$. 
%then the regions explored by $\mathcal B_0$ in both bands near the Weyl node are either full or empty. 
%This consitutes our first condition.
%%  for $|\kv-\kv_i|\leq e|\vec A(t)|/\hbar$, and hence also for $\kv-\kv_i$ within $\mathcal B_{0}$. % (namely $1$ or $0$ , respectively). 
%Sim
% (namely  $1$ or $0$, respectively). 

For the most natural case where each of the two  modes contains only a single harmonic,
$ 
W(\kv) = -W(-\kv),
\label{eq:w_inversion}
$ 
as is also evident in Figs.~\ref{fig:weyl_surfaces}(c)~and~\ref{fig:conv_power}(a). 
For  Weyl point $i$ to contribute to frequency conversion, $\bar \rho_{1}(\kv)$ or  $\bar \rho_{2}(\kv)$ must hence break inversion symmetry around $\kv_i$. 
Specifically
\be
 \bar \rho_{\alpha}(\kv_i+\delta \kv)\neq  \bar \rho_{\alpha}(\kv_i-\delta \kv )\quad \textrm{ for $\delta\kv$ within $\mathcal B_{0}$.}
\label{eq:rho_inversion}
\ee
This constitutes our second condition.
This symmetry breaking can be achieved with a nonzero value of the Weyl cone tilt, $\vec V$ [see Eq.~\eqref{eq:weyl_hamiltonian}], as we demonstrated in our numerical simulations (Sec.~\ref{sec:numerics}).

\subsection{Condition on symmetry class}
\label{sec:symmetry_condition}
We now identify the  symmetries a Weyl semimetal must break to support topological frequency conversion. 

The two symmetries that are central to the 
Weyl semimetals  % are classified in terms of whether they 
are the inversion and the time-reversal symmetry~\cite{Armitage_2018}; at least one of these symmetries must be broken for the Weyl nodes to exist.
%Weyl semimetals that break both symmetries are in princple also possible 
%A Weyl semimetal must  break time-reversal symmetry to support frequency conversion, while all  crystal symmetries (including inversion) are compatible with the phenomenon. %frequency  conversion.
Both inversion and time-reversal  symmetry results in inversion-symmetric energy bands, $\varepsilon_\alpha(\kv) = \varepsilon_\alpha(-\kv)$ (with $\alpha$ denoting the band index after indexing  them according to their energy).
Thus, for both symmetries,  a  Weyl point at wavevector $\kv$  implies the existence of  a  Weyl point at wavevector $-\kv$.
The conjugate Weyl nodes at $\kv$ and $-\kv$ have equal charges for time-reversal symmetric Weyl semimetals, and opposite charges for inversion-symmetric Weyl semimetals~\cite{Armitage_2018}. 
We expect the   steady-state to approximately inherit the same inversion symmetry, such that 
$
\bar\rho_\alpha(\kv)=\bar \rho_\alpha(-\kv).
$
For the most natural case where  modes 1 and 2 each contain  a single harmonic, $W(\kv)=-W(-\kv)$. 
Hence the contributions to $\bar \eta_0$ from symmetry-conjugate nodes cancel out for Weyl semimetals with time-reversal symmetry, %,in most accessible cases, where the driving itself does not break inversion symmetry (which we argued above is hard to achieve).
but not for Weyl semimetals with inversion symmetry. %, and hence this class of Weyl semimetals can support frequency conversion. 

We conclude that broken time-reversal symmetry is required for topological frequency conversion, while inversion symmetry does not need to be broken. 
%Weyl semimetals without .
Other crystal symmetries, such as reflection and discrete rotation symmetry, do also not preclude  frequency conversion, since the incoming modes (and hence $W(\kv)$) can be configured in  a way that breaks these symmetries. 
Hence magnetic Weyl semimetals, such as ${\rm Co}_3{\rm Sn}_2{\rm S}$ or ${\rm Co}_2{\rm MnGa}$~\cite{Swekis_2021,Xu_2018}, intrinsically support topological frequency conversion, while non-magnetic Weyl semimetals (such as TaAs) require an  externally-provided time-reversal symmetry breaking. %to  support topological frequency conversion. 
\addFN{This external symmetry breaking is already achieved with the circularly-polarized driving; a higher degree of asymmetry can further be achieved with 
e.g.  a current bias or externally applied magnetic field.}
%\fn{Added the last sentences to discuss how TR breaking can be achieved. Does this sound ok?}

\subsection{Condition on driving}
\label{sec:driving_conditions}
Next, we identify the conditions that the driving amplitudes and frequencies must satisfy  to support frequency conversion for a given Weyl semimetal. %   Weyl node. 

Sec.~\ref{sec:nonadiabatic}  concluded that the dynamics of electrons is non-adiabatic within a distance $d_0$ from the boundary $\mathcal B_0$, where $d_0 \sim \sqrt{\frac{eE }{\hbar v_0}}$, and $v_0$ is the smallest singular value of the matrix $R$ in Eq.~\eqref{eq:weyl_hamiltonian} [see Eq.~\eqref{eq:d0_result}].
%I.e., $\v_0$ gives  $\frac{1}{\hbar}$ times the smallest  gradient  of the band gap surrounding  the Weyl point.
For a nonzero number electrons to act as frequency converters, $d_0$ must hence be smaller than the linear dimension of $\mathcal B_0$, which we estimate to be of order $eA/\hbar$. 
% to be located a distance $\sim d_0$ away from   $\mathcal V_{0}$ before the correction in Eq.~\eqref{eq:g_result} is negligible, and they can contribute to topological frequency conversion [indicated by dark red/blue in Fig.~\ref{fig:tfc}(a)]. 
%For there to be a nonzero number of electrons active as  frequency converters, the characteristic  magnitude of the driving-induced vector potential, $A$, must hence be larger than $ d_0$.
These considerations imply that 
\be 
    E\gg \frac{\hbar \Omega^2}{ev_0}
    \label{eq:intensity_condition}
\ee  
is required for frequency conversion.
% Equivalently, we require that the characteristic intensity, $I$, satisfies 
% \be 
% I\gg  \frac{\hbar^2 \Omega^4}{2c \varepsilon_0 v_0^2}.
% \label{eq:intensity_condition}
% \ee 

% Another condition we assume is that the Hamiltonian is well-approximated by the linearized form in Eq.~\eqref{eq:weyl_hamiltonian} throughout the entire region $\mathcal B_0$, i.e., over on reciprocal space length scales of order $eA/\hbar$.
%  We expect this to be justified if this reciprocal space scale is much smaller than the characteristic smallest separation between  Weyl points, $\Delta \vec k$. 
% Using $A\sim E/\Omega$, the considerations above thus suggest the following is a necessary condition for frequency conversion:
% \be
% v_0\gg \frac{\hbar \Omega^2}{eE}\gg \Omega\Delta \vec k%\frac{\hbar \Omega^2}{ e E}
% \ee
Hence, ``steep''   Weyl points (i.e., Weyl points with  large $v_0$) are most useful for frequency conversion, as they support topological frequency conversion at lower intensities. 
%as they require lower intensities to achieve frequency conversion.

Weyl points in known compounds support topological frequency conversion at experimentally accessible parameters:
for example, TaAs has Weyl points for which $v_0~\sim 10^{5}\,{\rm m}/{\rm s}$~\cite{Arnold_2016,Ramshaw_2018}
%https://www.nature.com/articles/s41467-018-04542-9.pdf?origin=ppub; %https://journals.aps.org/prl/pdf/10.1103/PhysRevLett.117.146401}.
At frequency $\Omega \sim 2\pi \, {\rm THz}$, we hence expect these Weyl points  can support topological  frequency conversion at %experimentally accessible 
moderate intensities   of order $100 \,{\rm W}/{\rm mm}^2$  and above.
%(corresponding to $E\gtrsim 0.25 \,{\rm MV}/{\rm m}$). %e expect these parameters are experimentally acessible. 
%{(GR) Typos in the units in above paragraph. could you fix and make sure the numbers are okay?}

Finally, we  require that the bandwidth of the bands containing the Weyl node be larger than the driving frequency; otherwise, driving cannot be considered adiabatic anywhere in the system.
This puts an upper limit for the frequencies that could achieve frequency conversion in a given Weyl semimetal. 
As an example, for TaAs the characteristic band gap between  Weyl points is of order $20 \,{\rm meV}$~\cite{Arnold_2016}, corresponding to a maximum frequency limit of $\sim 5\,{\rm  THz}$. % [Arnold]
%\im{put some number there, some 10s of mV?}

% Hence, we expect the Weyl nodes of real Weyl semimetals support topological frequency conversion for ${\rm THz}$ frequencies. 
% Whether there is a net conversion depends on the filling. 
% However dissipation, and in particular nonadiabatic heating, remains a tricky issue to overcome. 

\subsection{Condition on relaxation}
\label{sec:relaxation_condition}
A final condition for amplification, is that the rate of topological energy conversion, $\bar \eta_0$, must overcome the (negative) rate of dissipation, $\bar\eta_{\rm dis}$. 
%topological frequency conversion is that it must  overcome dissipation in the system.
%the net rate of energy conv
Our analysis and numerical simulations identified two sources of dissipation: momentum relaxation ($\eta_{\rm mr}$) and nonadiabatic heating  ($\eta_{\rm na}$): $\eta_{\rm dis} = \eta_{\rm mr} + \eta_{\rm na}$. Amplification of mode $1$ thus requires % thus be decomposed as
\be 
 \bar \eta_0 + \eta_{\rm mr}+\eta_{\rm na}>0,
\ee 
%where $\bar \eta_0$ denotes the net rate of topological frequency conversion, and  $ \eta_{\rm mr}$, and $\eta_{\rm na}$ denote the (negative) rates of energy dissipation due to momentum relaxation and nonadiabatic heating, respectively. 

In Sec.~\ref{sec:nonadiabatic} we concluded   that the {\it gross} rate of topological frequency conversion, $\eta_{\rm gross}$ (i.e., the rate that results when not taking into account cancellation between electrons that convert energy in opposite directions), can only exceed $\eta_{\rm na}$ if  $\tau \gtrsim 1/\Omega$.
%is required for nonadiabatic heating not to exceed the {\it gross} rate of topological frequency conversion. 
Since the net rate of topological frequency conversion, $\bar \eta_0$, is just a small fraction of $\eta_{\rm gross}$, and since energy is also lost to momentum relaxation, we hence expect net amplification ($\bar \eta >0$) can only be achieved when
\be 
\tau \ll 1/\Omega 
\ee 
%is required to acheive net frequency conversion (i.e., a positive value of $\bar \eta$). 
%Here the left hand side must be many times smaller than the right-hand side. 

The phenomenological discussion in  Sec.~\ref{sec:nonadiabatic}  shows that $\eta _{\rm na}\propto (1-e^{-\Gamma \Delta t})$ where $\Delta t$ denotes the characteristic time between instances where a given wavevector $\kv$ is taken to the Weyl point by the applied drive ($\kv \to \kv+ e\vec A(t)/\hbar$). 
Here $\Delta t$ is significantly smaller for Lissajous conversion (highly rational frequency ratios) than for incommensurate frequency ratios. 
Thus, the threshold for net amplification is significantly lower at commensurate frequencies. 
Indeed, in our  simulations,  a small adjustment of $f_1$ from $2f_2/(3+\epsilon)$ to $2f_2/3$  lowered the amplification threshold from above $1200\,{\rm ps}$ to $\sim 300\,{\rm ps}$.

Our quoted values  in Sec.~\ref{sec:many_body_tfc}  provide an example of how to estimate  the break-even relaxation rate. % breakdown of the net conversion rate: %, which applies to the parameters we used in our simulations: 
For intensity $E \sim 1600\,{\rm kV}/{\rm m}$ and  $\Omega \sim 2\pi {\rm THz}$, we estimated $\eta_{\rm gross}\sim 500\,{\rm kW}/{\rm mm}^3$.
%This value does not take into account %momentum relaxation or 
%partial  cancellation between electrons that convert energy with opposite rates, and hence  t
The net rate $\bar \eta_0$ will be only a fraction of this value. 
For the same parameters, and with isotropic band gap matrix $R$  of order $5\cdot 10^5\,{\rm m}/{\rm s}$ and Fermi surface area $0.06\,{\rm Å}^{-2}$, we found $\eta_{\rm mr}\sim 5\cdot 10^{-8}\,{\rm kJ}/{\rm mm}^3 \tau$, and $\eta_{\rm na}\sim  5\cdot 10^{-8} \,{\rm kJ}/{\rm mm}^3 \tau$ (for incommensurate frequencies) or $3.5\cdot 10^{-9}\,{\rm kJ}/{\rm mm}^3\tau $ (for Lissajous conversion at frequency ratio $2/3$). 
Hence, we expect topological frequency conversion can exceed the rate of dissipation when $\tau$ are several times larger than $100 {\rm ps}$.
We moreover expect the threshold to be significantly lower for  Lissajous  conversion (i.e. commensurate frequencies), than for incommensurate frequencies. 
This is in good agreement with our data in Fig.~\ref{fig:front_page_figure}(b) which indicate that  $\tau$ must exceed $ 300\,{\rm ps}$ in order to achieve  net frequency conversion in the  Lissajous regime for the parameters above, and $1000{\rm ps}$ for incommensurate frequencies.

The different scaling behaviors of dissipation and topological frequency conversion point toward  the parameter regimes  beneficial for amplification. %frequency conversion. 

First, note that (at a fixed area of the Fermi surface), $\eta_{\rm mr}$ scales linearly with electric field $E$, while $\eta_{\rm na}$ and $\eta_{\rm gross}$  scales as $E^{3}$ (specifically, $\eta_{\rm na}\sim I^{3}$ in the Lissajous regime) [see Eqs.~\eqref{eq:momentum_relaxation_estimate},\eqref{eq:eta_na_estimate_general}, and \eqref{eq:eta_gross}]. 
%, and  as $I^{7/4}$ for incommensurate frequencies, though it saturates as $3/2$).
Thus, we expect that the relative contribution of $\eta_{\rm mr}$  decreases at high intensity, while the ratio of $\eta_{\rm na}$ and $\eta_{\rm gross}$ remains fixed, implying that  frequency conversion  becomes more efficient at higher intensities. 

Second, for a given intensity, the topological frequency conversion rate scales as $1/\Omega$, while $\eta_{\rm na}$ scales as $1/\Omega^3$ (for incommensurate frequencies) or $1/\Omega^2$ (for commensurate frequencies). 
Similarly, momentum relaxation scales as $1/\Omega^2$.
Thus, we expect   amplification is most easily reached at the top of the frequency range that supports topological frequency conversion, given the driving intensity and band structure of the system. %range where the system
%maximizing the frequency is helpful. 

The requirements on the relaxation rate pose the biggest current challenge to realizing topological frequency conversion. 
Relaxation times in known Weyl semimetals have been reported to be in the range $0.25-3{\rm ps}$~\cite{Dai_2015,Weber_2015,Ishida_2016,Weber_2017}, although  transient signatures with lifetimes above $100$~\cite{Ishida_2016,Liu_2020} and $1000$~\cite{Jadidi_2020} ps have also been reported in some compounds.
Thus further improvements in the quality of materials are needed to fulfill the requirements of topological frequency conversation in the practically interesting ${\rm THz}$ range.

% \section{Device geometry}
% Here we dicsuss how a Weyl semimetal may be used in practice for frequency conversion. 

% Since a WSM is a semimetal, it may eventually screen electric fields on long length scales, if not for other reasons, then due to TFC. 
% The data suggests full conversion of the beams within a distance $20 \mu {\rm m}$.
% We therefore consider a grain/flake of WSM in a width of of $1{\rm \mu m}$,  much smaller than the $0.3{\rm mm}$ wavelenght of THz radiation and the penetration depth. 

% In this geometry, it is safe to assume that the electric field experienced by the bulk electron is the (effectively uniform) incoming radiation field $\vec E(t)$: the grain is too small to affect the amplitudes of the modes, and the electric field caused by the buildup of charges at the edge of the grain (due to current oscillations) is negligle. 
% For the parameters shown in Fig.~\ref{fig:front_page_figure}, we find a current density 
% %given by $A S_F v_F/h^3 = ES_Fv_F/\Omega h^3$. 
% With $S_F \sim E/\Omega$
% at the driving frequency of $\sim 100{\rm kW}/mm^3/{\rm MV}/{\rm m} \sim 10^8 A/m^2$. 
% Dividing by the angular frequency of $2\pi {\rm THz}$ results in a charge oscillation amplitude of $10^{-7} C/m^2$. 
% This is equivalent to a field of $10 \,{\rm V}/{\rm m}$.
% %10^8 wavelength  

\section{Thinking outside the grain: Global electrodynamics considerations and  implementation using phase arrays}
\label{sec:phased_array}
\addFN{
The full understanding of the frequency conversion effect requires thinking about the global electromagnetic field, and the material response of the Weyl grains to an external drive. Specifically, in this section we incorporate the dielectric response to our analysis, and propose a phase-array geometry of the Weyl grains as a prototype for a Weyl topological amplifier. 
}

\subsection{Renormalization of electric field by plasma oscillations}
\label{sec:plasmon}

% \subsection{Topological frequency conversion from a  metallic grain}
\addFN{Let us begin with considering the macroscopic response of a single grain to external driving. 
%The grain is subject to radiation  by two externally provided plane waves with circular polarizations  and perpendicular directions of propagation.
For $i=1,2$, we let  $\vec E_i(\vec r,t)$ denote the (plane-wave) electric field from mode $i$ as a function of position $\vec r$ and time $t$ and let $\vec E_0(\vec r,t)=\vec E_1(\vec r,t)+\vec E_2(\vec r,t)$ denote the net ``incoming'' field resulting from the driving. 
The current and charge oscillations in the grain induced by  the external driving creates an additional electric field, $\vec E_{\rm ind}(\rv,t)$.
The total electric field inside the sample is thus  given by $\vec E(\vec r,t)\equiv \vec E_0(\rv,t)+\vec E_{\rm ind}(\rv,t)$; this  is the field driving the response of the material, and is the one we considered in the calculation in the previous sections. 
Evidently, the internal field in the  sample  gets renormalized by the  charge and current in  the material.
}

\addFN{
Our first order of business is to find the internal field  $\vec E(\vec r,t)$ (which we used in our analysis above) in terms of the external fields. 
We can find 
 $\vec E(\vec r,t)$   self-consistently by solving Maxwell's equations, taking account   the current and charge dynamics in the grain induced by $\vec E(\vec r,t)$. 
While an exact (geometry-dependent) analysis  is  in principle possible, the small size of the grain allows us to make  some simplifications, such as %effectively take $\vec E_0(\vec r,t)=\vec E_0(\vec 0,t)$ to be uniform. 
%We can also 
ignoring the skin effect. 
% retardation effects are negligible;  here  $\vec 
Thus,  inside the grain  ${\bf E}_{\rm ind}(\vec r,t)$  is approximately given by the electrostatic field  resulting from the instantaneous charge configuration in the system. These in turn arises from the driving-induced oscillations of the grain's bulk plasmonic mode~\footnote{Neglecting these retardation effects corresponds to ignoring the skin effect, as is justified based on the grain's small size}. }

 %and $\Gamma$ the relaxation rate of electrons.% In the TFC regime, these nonlinear effects can have the effect of flipping the sign of $\Gamma$. 
% However, this is a subleading effect; here we are interested in finding the dominant contribution to the field renormalization, which arises from the (energy conserving) current $\vec j_0$; hence we only expect these corrections   to give minor insignificant corrections to the electric field inside the material.
% As a result, we expect  the net electric field inside the grain is effectively  given by 

\addFN{
For a small spherical grain, we can  assume $\vec E_0(\vec r,t)$ uniform within the grain, and moreover ignore retardation effects of the electromagnetic field (this is equivalent to neglecting the skin effect). 
Inside the grain, $\vec E_{\rm ind}(\vec r,t)$  is thus given by the electrostatic field arising from the instantaneous charge distribution at time $t$. 
The charge distribution is nontrivial due to the oscillating currents, which produce surface charges. 
Specifically, $\partial_t \vec \rho(\vec r, t) =\nabla \cdot \vec j(\vec r,t)$, implying $\vec \rho(\vec r,\omega) = \frac{i}{\omega} \nabla \cdot \vec j(\vec r,\omega) $.
% , with $\rho(\vec r,\omega)$ and $\vec j(\rv,\omega)$ denoting the Fourier transforms of the current and charge density within the sample.
We now show that the equations of motion above have a solution in which the current density and $\vec E_{\rm ind}(\vec r,t)$ are  also uniform within the sample. 
To show this, note that a uniform current density in the grain, $\vec j(\vec r,t) = \vec j(t)$, implies that the charge accumulates on the surface. 
The surface charge density at the angle specified by unit vector  $\vec{\hat r}$ on the sphere, is given by $\partial_t \lambda(\hat{\rv},t) = \vec j(t)\cdot \hat {\rv}$. 
Hence $\lambda(\hat{\rv},t) = \hat {\vec r}\cdot \boldsymbol{\lambda}_0(t)$, with $\boldsymbol{\lambda}_0(t)$ denoting the unique zero-mean solution to $\partial_t \boldsymbol{\lambda}_0(t)= \vec j(t)$. 
Inside the sphere, the electrostatic field from a surface charge distribution $\lambda(\hat{\rv}) = \hat {\vec r}\cdot\boldsymbol{\lambda}_0$ is uniform and given by $\vec E_{\rm ind}(t)= \frac{\boldsymbol{\lambda}_0(t)}{3\epsilon_0}$.  }
% Transforming to frequency domain, and using that $\vec j(\omega) = -i\omega \boldsymbol \lambda_0(\omega)$, we arrive at 
% $ 
% \vec E_{\rm ind}(\omega)=\frac{i\vec j(\omega)}{3\epsilon_0 \omega}.
% $ % \ee 
\addFN{
Hence, the electric field is uniform within the sample and given by 
$
\vec E(t) = \vec E_0(t)+  \frac{\boldsymbol{\lambda_0(t)}}{3 \epsilon_0}.
$ 
Thus, a uniform current density $\vec j(t)$ and $\vec E(t)$ solves the dynamics of the grain. }

\addFN{
Transforming to frequency domain, and using that $\vec j(\omega) = -i\omega \boldsymbol \lambda_0(\omega)$, we finally arrive at 
\be 
\vec E(\omega)=\vec E_0(\omega) - \frac{i\vec j(\omega)}{3\epsilon_0 \omega}.
\label{eq:e_renormalization}
\ee  % \ee 
This gives the frequency-dependent renormalization of the electric field inside the grain, and is an exact solution in the limit where the grain size is smaller than the wavelength of the driving modes. 
}

\addFN{
In  linear response theory, the time-derivative of the current response is assumed proportional to the electric field, implying $ \vec j(\omega)  \approx  -i  \kappa \vec E(\omega)/\omega$ for some constant $\kappa$. 
The resulting solution leads to a frequency dependent relative permittivity, $\vec E(\omega) \approx \epsilon(\omega)\vec E_0(\omega)$ with $\epsilon(\omega)= (1-{\omega_p^2}/{\omega^2})^{-1}$  with $\omega_p = \sqrt{\sigma/3\epsilon_0}$ denonting the plasma frequency of the system.  The plasma frequency $\ omega_p$ is estimated for generic Weyl semimetals,  in Ref.~\cite{Zhou_2015}: it is typically given by an $\mathcal O(1)$ constant times the Fermi energy.  }

\addFN{
The linear response analysis above is useful for elucidating the qualitative features of the plasmonic response. 
However, the regime we consider  potentially supports a significant nonlinear response due to the nonquadratic dispersion and large Berry curvature surrounding Weyl nodes  -- indeed, topological frequency conversion is a nonlinear response phenomenon. 
We thus go beyond the linear response regime in our analysis below: 
for  a given internal field configuration, $\vec E(t)$, the current response $\vec j(t)$   can be easily computed in the limit of weak relaxation and adiabatic driving without any linear response approximation,  using Eq.~\eqref{eq:current_result}, $\vec j(t) = \vec j_0(t)+\delta \vec j(t)$  with $\vec j_0(t)$ given in Eq.~\eqref{eq:j0_def} and   the dissipative component $\delta \vec j(t)$ is  negligble in the limit we consider~\cite{NoDissipationForPlasmons}.
%(allowing us to set $\vec j(t) \approx \vec j_0(t)$).
}

\addFN{
The driving frequency controls whether the plasma oscillations amplify or screen the  electric field from the incoming radiation. % inside the material.
This qualitative behavior is evident in the linear response result we quoted above, but also endures after taking into account  the nonlinear response. 
%from the relative permittivity we quoted above.
To see this, consider what external electric field $\vec E_0(\omega)$ is needed to cause a given internal field $\vec E(\omega)$ [which determines the current response $\vec j(\omega)$].
As in the linear response regime,  $\vec j(\omega)$ is controlled by vector potential, and thus scales with $\vec E(\omega)/\omega$ [see Eq.~\eqref{eq:j0_def}].
The plasmon-induced electric field hence is negligible in the limit of large $\omega$ (but grain size still smaller than the wavelength), meaning the grain is effectively transparent to the radiation: $\vec E(\omega)\approx \vec E_0(\omega)$. 
Conversely, for small $\omega$, $\vec j(\omega)/\omega\epsilon_0$ will be considerably larger than $\vec E(\omega)$, implying that $\vec E_0$ in turn has to be much larger in $\vec E(\omega)$ for Eq.~\eqref{eq:e_renormalization} to hold.
Thus, for small frequencies, the plasma oscillations severely screens the electric field inside the sample relative to the external field. 
At some intermediate frequency, $\vec E(\omega) \approx - \frac{i\vec j(\omega)}{3\epsilon_0 \omega}$, and a very weak external  field  thus causes  a large internal field. 
In this case,  driving  resonates with the plasma oscillations, causing significantly enhanced amplitude of the electric field.
As we will see, this mechanism allows for significant enhancement of the topological frequency conversion rate. 
}

% This is a really nice (and exact!) result which we can use to find the dielecric properties directly from our simulations without having to introduce plasmons. 
% Though, to lowest order in $\vec E$, our solution for $\vec j_0(t)$ in the above section gives   $\vec j(\omega) \sim  -i \frac{ \kappa ^2 \vec E(\omega)}{\omega}$
% for some constant $\kappa$ which plays the role of plasma frequency and probably is of order $\mu$. 
% The corrections to this response is responsible for topological frequency conversion, dissipation,  and high-harmonic generation. 

\subsection{Radiation output of grain}
\addFN{Next, we want to compute $\vec E(\vec r,t)$ {\it outside} the grain, to determine the profile of the emitted radiation. 
Here, the grain's small size means that $\vec E_{\rm ind}(\vec r,t)$ to a good approximation takes the form of  dipole radiation generated by some nontrivial  trajectory of the dipole moment.
% Outside the grain, it looks like a dipole moment emmitting radiation (because of its small size).
Using that the surface charge distribution we obtained above, we find  the  dipole moment to simply be given by $   \vec p(\omega) = \frac{4\pi}{3} i r^3\frac{\vec j(\omega)}{ \omega}$. 
}
% The emitted radiation looks like dipole radiation generated by this moment.

\addFN{
The energy converted to mode $1$ leaves the grain as  radiation energy at   frequency $\omega_1$.  
The bulk of the emitted radiation energy comes from  constructive interference between the incoming plane wave $\vec E_0(\vec r,\omega_1)$ and the emitted dipole radiation $\vec E_{\rm ind}(\vec r,\omega_1)$ (i.e., the $\omega_1$-Fourier components of $\vec E_{\rm ind}(\vec r,t)$ and $\vec E_0(\vec r,t)$, respectively). }

\addFN{
To compute the frequency-resolved radiation energy emanating from the grain, we consider the  total energy flux density, given by the  Poynting vector field, $\vec S(\vec r,t) = \frac{1}{\mu_0}\vec E(\vec r,t)\times \vec B(\vec r,t)$. 
% We consider the time-averaged Poynting vector field, $\bar S(\vec r)$. 
By using the Fourier decomposition % of $\vec E$, 
$\vec E(\vec r,t)=  \int d\omega \vec E(\vec r,\omega) e^{-i\omega t}$ along with $\vec E(\vec r,\omega) = \vec E^*(\vec r,-\omega)$ [and likewise for $\vec B(\vec r,t)$],
we find that the time-averaged energy flux density, $\bar{\vec S}(\vec r)$, is thus given by
\be 
\bar {\vec S}(\vec r )=\frac{2}{\mu_0} \int\! d\omega\,   \Re[\vec E(\vec r,\omega)\times \vec B^*(\vec r,\omega)]
\ee 
% Because the emitted dipole radiation is orthogonal to the incoming modes, it follows that the Poynting vector is just given by that of  the emitted dipole radiation, $\vec E_{\rm ind}$. 
}

\addFN{
We   identify $ \vec  S(\vec r, \omega) \equiv \frac{1}{\mu_0}\Re[\vec E(\vec r,\omega)\times \vec B^*(\vec r,\omega)]$ as %the frequency resolved radiation energy flux (,
the   energy flux density from modes with  frequency $\omega$.
The total radiation power from the grain at frequency $\omega$ is  given by 
% integrating $\vec S(\vec r,\omega)$ over an area containing the grain: 
\be 
P(\omega)= \oint_C d\vec A \cdot \vec{  S(\vec r,\omega)}
\ee 
where $C$ is some surface containing the grain.
}

\addFN{
To compute $P(\omega)$, we use the divergence theorem to find
$ 
% \oint d\vec A \cdot \vec{  S(\omega)} =
P(\omega) = \frac{2}{\mu_0}\int_C   d V \, \Re[\nabla \cdot( \vec E(\omega) \times \vec B^*(\omega) ], 
$ 
with $\int_C dV$ denoting the  integral over the volume contained in $C$. 
% with the volume integral taking over some volume that contains the grain. 
Next, we apply the cross product identity $\nabla \cdot( \vec E \times \vec B^*)  =- \vec E \cdot (\nabla  \times \vec B^*)$. 
Using Ampere's law $\nabla \times \vec B(\vec r,\omega) = -i\mu_0\epsilon_0 \omega\vec  E(\vec r,\omega)-\mu_0\vec j(\vec r,\omega)$, where $\vec j(\vec r,\omega)$ is the Fourier transform of the current density, yields 
$
P(\omega)=   \frac{2}{\mu_0} \int\! dV \, \Re[i\omega\epsilon_0  |\vec E^2(\omega)|+  \vec E(\omega)\cdot \vec j^*(\omega)].
%\\cdot \nabla \times B(\omega).
$ 
The first term in the parenthesis  evidently is fully imaginary, and thus gives a vanishing contribution to the integral. 
This leave us with 
$   
P(\omega)=  2  \int dV  \Re[ \vec E(\vec r,\omega)\cdot \vec j^*(\vec r,\omega)].
%\\cdot \nabla \times B(\omega).
$ 
Since $\vec E(\vec r,\omega) $ and $\vec j(\vec r,\omega)$ are uniform within the grain, we find 
\be 
P(\omega)=  2  V  \Re[ \vec E(\omega)\cdot \vec j^*(\omega)],
\ee 
which is exactly the quantity we calculated in Sec. III. 
Interestingly, the plasma-induced electric field does not directly contribute to the power output, since it is proportional to $-i\vec j(\omega)$; rather it indirectly modifies the power output through its effect on the current response.
%\im{so, point is that current has become much bigger near the resonance, but we still have to multiply it bu the external field to compute the power, right?}
We thus arrive at 
% \vec E(\vec r,\omega_1) =\frac{1}{2\pi} \int dt \vec E_1(\vec r,t) e^{i\omega t}$, we identify 
\be 
P(\omega_1) = \bar \eta V .
%\\cdot \nabla \times B(\omega).
\ee 
with $\bar \eta$ denoting the frequency conversion rate within the grain. This gives the output intensity of the dipole radiation emitted with frequency $\omega_1$.}

% The phase of the output radiation interferes constructively with the incoming electric field. % the topological frequency conversion density, taking into account the local variations over electric field and current density (which we expect to be much longer than the Fermi wavelength). 

\subsection{Implementation using a phase array}
\begin{figure}[t]
\includegraphics[width=0.99\columnwidth]{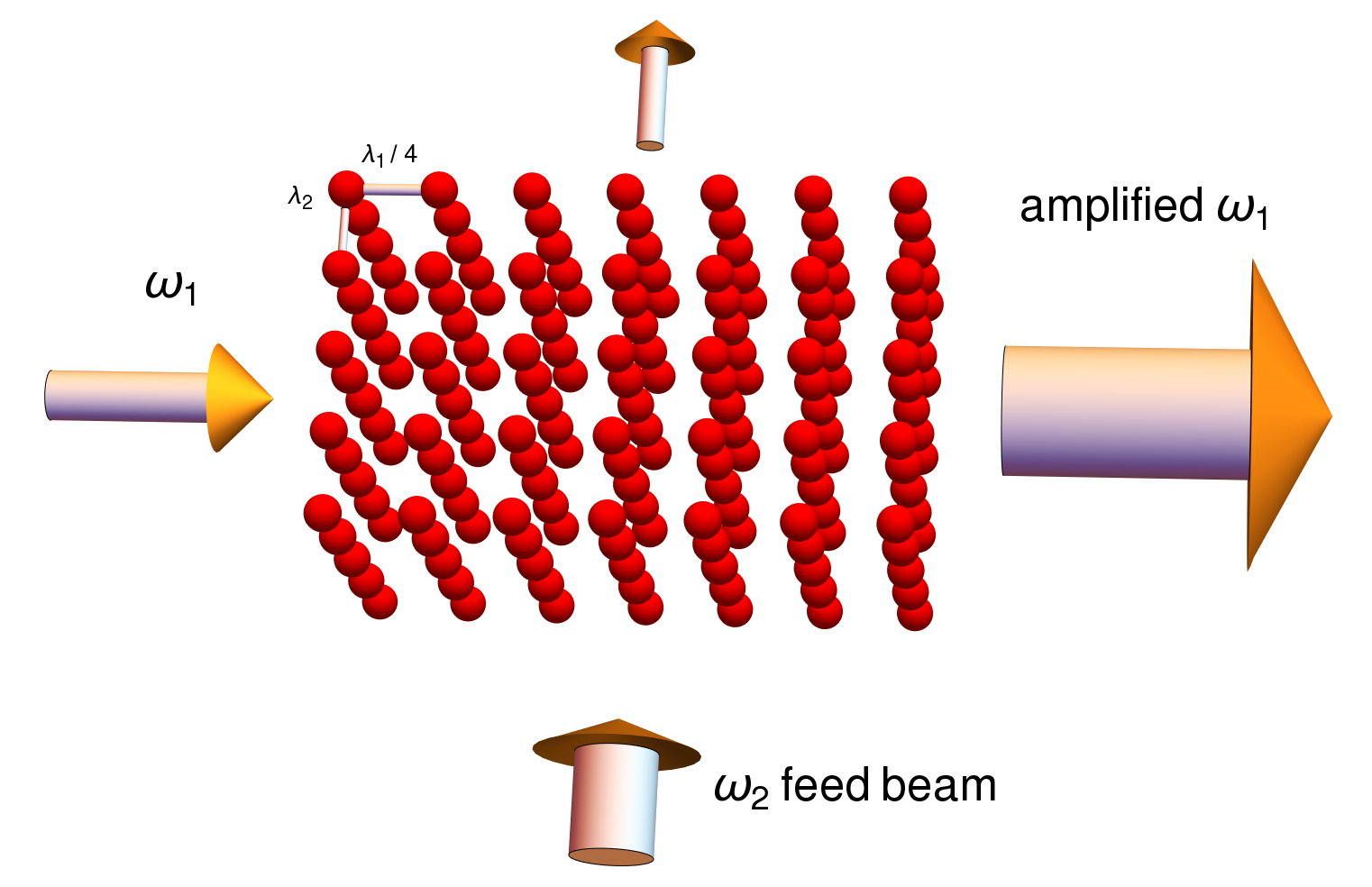}
\caption{Phase-array proposal for combining the gain from Weyl grains into a topological amplifier. The spacing of the grains in the direction of propagation should match a quarter of the wavelength of the amplified wave. This will suppress reflection, and will concentrate the contribution of each grain into a single forward-propagating beam. The pump beam is expected to be normal to the source beam, and the spacing along its direction should be its wavelength.\label{array-fig} }
\end{figure}

\addFN{
In order to produce an amplifier out of the frequency-conversion effect, we must consider combining many grains together to create a phase array. While we do not intend to analyze such a device in detail in this manuscript, we will here outline its design. The phase-array geometry we envision has Weyl  grains arranged in a 3d cubic-lattice, with one axis along the propagation direction of an external plane wave with circularly polarization (mode $1$), which we intend to amplify. In the direction of the propogation of mode 1 the structure will have a lattice constant of a quarter wavelength, such that the backscattering element of the amplified mode 1 is be eliminated by destructive interference.  

The array will also be subject to a normal-incident radiation from mode $2$, which is the amplification source beam to be converted. In order to maximize the amplification effect, we anticipate that in-phase arrangement of all layers with respect to mode 2 would be beneficial, hence the lattice constant along the mode 2 direction should be mode-2 wavelength,$\lambda_2$ (see Fig. \ref{array-fig}).

Indeed, obtaining amplifiers from single gain elements is a common practice.  Josephson traveling wave amplifiers (JTWA; see, e.g., Ref. \onlinecite{Siddiqi2015}) , for instance,  are essentially a chain of individual Josephson parametric amplifiers. Other traveling wave optical amplifiers rely on nonlinear crystals such as LiNb or $\beta{\rm -BaB}_2{\rm O}_4$ (BBO). It is only in the macroscopic constructs of parametric JTWA, and non-linear crystals which mix light modes, that issues related to phase-matching arise. Similarly, a single Weyl grain in a topological-frequency-conversion regime requires no mode-matching beyond the need to have a rational frequency ratio to be in the Lissajous regime. In fact, the grains are expected to be smaller than the wavelengths involved. It is only when these grains are combined into an array that we need to consider how the wavelengths of the amplified radiation correspond to the spatial structure of the amplifier. }

% So we need to make the chemical potential smaller. 
% If we can make it a lot smaller it is very helpful, if it allows us to get near the plasma resonance. {\bf Let's see what the simulation says. Maybe it will work???}
% Then very little external field can have a big impact. 
% But that's all the way down to $4{\rm THz}$, so perhaps it's not so realistic, because we start to run into trouble with nonadiabatic heating. 
% At least we need a lot weaker fields then. 
% \addFN{\bf Maybe tiny doping an an asymmetric fermi surface is actually the way to go.}

% If we have a higher Fermi velocity, the plasma frequency is smaller relative to the Fermi energy. 
% So higher Fermi velocities are clearly beneficial. 
% With $10^6 {\rm m}/{\rm s}$, we have $\omega_p \approx \mu$. 

\subsection{Numerical simulations of plasmon-enhanced amplification}
\label{sec:plasmonnumerics}
\addFN{We now demonstrate that topological frequency conversion remains possible even after including screening effect of plasma oscillations. Furthermore, we show that  tuning frequencies  near the plasma resonance  dramatically {\it enhances} the frequency conversion effect. 
}

\addFN{
We consider  an inversion-symmetric Weyl semimetal whose   Fermi surface  consists of two Weyl nodes that are related by inversion symmetry. The Hamiltonian of one Weyl node is given by $H(\kv) = \hbar  v \vec k\cdot {\bs \sigma} +   \hbar \vec k \cdot {\vec V}$ with 
 $v=3.87\cdot 10^5\, {\rm m}/{\rm s}$ and $\vec V = (0,0,3.1\cdot 10^5\, {\rm m}/{\rm s})$  (i.e., the same dispersion as considered in Sec.~\ref{sec:numerics}); the Hamiltonian of the other Weyl node is given by $H(-\kv)$. 
We consider the case where the  electric field inside the grain, $\vec E(t)$, is  fixed and given by two circularly polarized modes as in  Sec.~\ref{sec:numerics} [see  Eqs.~\eqref{eq:e1}-\eqref{eq:e2}] 
%$\vec E(t) = \vec E_1(t)+\vec E_2(t)$ 
with amplitudes $\mathcal E_1 = 100\,{\rm kV}/{\rm m} $ and $\mathcal E_2 = 50 \,{\rm kV}/{\rm m}$; we allow the incoming radiation field $\vec E_0(\omega)$ to vary.
}

\addFN{
We first compute the external electric field $\vec E_0(t)$ which causes the  internal electric field to be given by $\vec E(t)$ as specified above.   
We compute $\vec E_0(\omega)$ as described in Sec.~\ref{sec:plasmon}. I.e, we use $\vec E(\omega)=\vec E_0(\omega) - {i\vec j(\omega)}/{3\epsilon_0 \omega}$, 
 along with $\vec j(t) \approx -e \kint \sum_\alpha \bar \rho_\alpha(\kv) \dot{\vec r}_\alpha(\kv,t)$, where $\bar \rho_\alpha(\kv,t)$ denotes the time-averaged value of the equilibrium state occupation of band $\alpha$, $(1+\exp[-(\varepsilon_\alpha(\kv+ e\vec A(t)))-\mu)/k_{\rm B}T])^{-1}$, with $\varepsilon_\alpha(\kv)$ denoting the $\alpha$th band of $H(\kv)$.
We use temperature $T=20\, {\rm K}$ in our simulation and consider different values of $\mu$. 
}

\addFN{
In Fig.~\ref{fig:escaling}(b) we plot the resulting relative increase of mode $1$ inside the material, $G = |\vec E(\omega_1)\cdot \vec E_0^* (\omega_1)|/|\vec E_0(\omega_1)|^2$, as a function of chemical potential, $\mu$ (blue). 
We also plot the corresponding relative increase of mode $2$ (orange), as well as the net relative gain of all remaining modes (induced by the nonlinear oscillations of the plasmons), $G_3 = (\int d\omega |\delta \vec E_0(\omega)|^2)^{1/2}$, with $\delta \vec E_0(\omega)$ denoting the component of external field $\vec E_0(\omega)$ which is orthogonal to the internal field $\vec E(\omega)$ (for $\omega \notin \{ \omega_1,\omega_2\}$, $\delta \vec E(\omega) = \vec E_0(\omega)$). 
The data in Fig.~\ref{fig:escaling}(b) shows that the plasma oscillations do not affect  the modes when the chemical potential is smaller than $2.5\,{\rm meV}$. 
Moreover, for $\mu \lesssim 5\,{\rm meV}$, $G_3\ll 1$, implying that the external radiation field does not need lead to appreciable  amplitude of  any   higher-harmonic or orthogonal modes to provide the  bichromatic electric field inside the grain which we require.  In other words,  the plasma oscillations do not significantly excite any  modes other than the pump and signal mode when $\mu \lesssim 5\,{\rm meV}$. 
For values of $\mu$ above $5\,{\rm meV}$, the internal electric field gets severely suppressed by the plasmon screening, while the plasma oscillations begin to significantly excite modes  other than the pump and signal modes. 
Here, the frequency conversion rate is significantly reduced. Furthermore, a more careful analysis is needed in this regime to account for the higher harmonics of $\vec E(t)$ induced by the nonlinear plasma oscillations. 
}

\addFN{
In the range $2\,{\rm meV}\lesssim \mu \lesssim 4\,{\rm meV}$,  the internal field is significantly enhanced by the plasma oscillations, without nonlinear corrections playing a role. 
This plasma resonance    dramatically {\it enhances} topological frequency conversion:
}
\addFN{
we first compute the frequency conversion rate for the same parameters considered for Fig.~\ref{fig:escaling}(b),  using the  approach of Sec.~\ref{sec:numerics}. 
%for various values of the relaxation time $\tau$,.
From our obtained  frequency conversion rate, $\bar \eta$, we compute the gain coefficient of the material, $g = \bar \eta/I_1$ with $I_1 = c\varepsilon_0 |\vec E_0(\omega_1)|^2$ denoting the intensity of mode $1$ outside the material; we use the data from Fig.~\ref{fig:escaling}(b) to compute $\vec E_0(\omega_1)$ (recall we consider a fixed value of the internal field, $\vec E(\omega_1)$, but allow $\vec E_0$ to vary). % to be fixed. 
The gain coefficient has dimension of inverse length, and gives the characteristic rate at which mode $1$ gets amplified inside the material. 
In Fig.~\ref{fig:front_page_figure}(b), we plot the gain coefficient as a function of chemical potential, using $\tau = 200\,{\rm ps}$ (blue), $400\,{\rm ps}$ (orange) and $600\,{\rm ps}$ (green). 
 When the plasmon resonance is reached at $\mu \approx 4\,{\rm meV}$, the gain coefficient increases dramatically, reaching values of order $100\,{\rm cm}^{-1}$, exceeding, e.g., the  THz gain coefficients of $20$-$50$ ${\rm cm}^{-1}$ reported in Refs.~\cite{Jukam2009,Kao2017}. }

\section{Discussion}
\label{sec:discussion}
In this manuscript, we showed that Weyl semimetals can efficiently convert energy between two driving modes, through the mechanism of topological frequency conversion~\cite{Martin_2017}. %~[martin2017] realize topological frequency conversion. 
This effect makes Weyl semimetals promising media for ${\rm THz}$ and possibly even infrared amplification. Our analysis shows that Weyl semimetal with feasible band dispersions support topological frequency conversion in the ``THz gap'' at  experimentally accessible intensities of order $\sim 50\,{\rm W}/{\rm mm}^2$, or even less ($\sim 1 \,{\rm W}/{\rm mm}^2$) if one drives near the plasma frequency. 
Topological frequency conversion is supported both for incommensurate frequencies and commensurate frequencies, but is most efficient in the  latter case, due to the mechanism of Lissajous conversion. 
Our numerics and estimates focused on topological frequency conversion in the THz regime, where there is the biggest need for new photonic control elements, but the effect may also be supported at other frequency ranges. 

%We characterized the topological frequency conversion effect in Weyl semimetals and discussed its limitations.  

The primary obstacle to Weyl semimetals operating as topological frequency converters is drive-induced heating. Heating both wastes energy from the beams we would like to amplify, and may even damage the material. Phonons, interactions and impurities all lead to electron relaxation processes which cause this heating.

Through phenomenological arguments and numerical simulations, we  identified two important mechanisms for dissipation: momentum relaxation and nonadiabatic heating. 
Momentum relaxation occurs when electrons near the Fermi surface relax their energy by changing their momentum, and which is common to all irradiated materials. 
Nonadiabatic heating emerges when electrons undergo  Landau-Zener transitions between the valence and the conduction band.
This mechanism is particularly relevant in topological semimetals, due to the existence of gap-closing points in these materials. 
Even so, non-adiabatic heating is strongly suppressed in the Lissajous regime, which makes it much preferred for amplification. 
%ltaneously with a moderate average energy influx.

In our simulations and phenomenological discussion, relaxation was parameterized through a single relaxation time, $\tau$. 
In particular, we took electron-hole recombination, and intra-band momentum relaxation (which is supported by phonons) to have the same characteristic rates.  Needless to say this treatment could be made more realistic by considering separate relaxation rates for these processes, as suggested by experiments ~\cite{Ishida_2016,Weber_2017,Liu_2020}. Nonetheless, we believe our simple dynamical model  captures the conditions for amplification. 

To achieve amplification, where energy gain due to topological frequency conversion exceeds the loss due to dissipation, the characteristic relaxation time $\tau$ must be sufficiently long. To limit non-adiabatic heating as well as momentum relaxation we need $\tau f \gg 1$. This condition was clearly evident in our simulations: even for optimal parameters, and in the Lissajous regime, break-even was only reached when $1/\tau \gtrsim 300 f$,% \im{should we replace f by some actual frequency? the cofficients 300 and 1000 are anyway probably obtained for specific case, so there is not much gained by this generality?} eeven in the optimal Lissajous conversion regime. 
(for incommensurate frequencies, amplification required $\tau f>1000$). 
So far, $\tau$'s were reported in the range $0.25-5\,{\rm ps}$~\cite{Dai_2015,Weber_2015,Weber_2017} in pump-probe experiments.
This suggests  net amplification of continuous-wave ${\rm THz}$ frequencies %from topological frequency conversion
is currently beyond reach. % with current material properties. 
That said, transient experimental signatures  with  $\tau>100\,{\rm ps}$ have been seen in Weyl semimetals~\cite{Ishida_2016,Liu_2020}, emphasizing that a more discriminating analysis may reveal a broader amplification regime. 

Notwithstanding, signatures of topological frequency conversion effect could be observed even if the relaxation time is too short to allow for amplification. 
That is because the direction of energy conversion  of  a Weyl semimetal driven bichromatically by two circularly-polarized modes is determined by the product of the two mode's polarizations. Hence a reversal of the circular polarization of either mode should will lead to an increase in the output intensity of one mode, and a decrease for the other mode.
This topological effect  could be accessed experimentally.
\addFN{Another group recently proposed to utilize this chirality-sensitive intensity shift  to extract enantioselective information from a gas of chiral molecules~\cite{Schwnnicke_2022}. }

Furthermore, we can suggest several strategies to approach the amplification regime. Commensurate frequency conversion, i.e.  Lissajous conversion, already provides a dramatic improvement by suppressing non-adiabatic effects by an order of magnitude. Momentum relaxation is harder to control. Note, however, that momentum relaxation energy loss  scales linearly with radiation intensity, $I$. In contrast, topological energy conversion (and nonadiabatic heating) scale as $I^{3/2}$.
Therefore the relative significance of momentum relaxation should decrease at larger intensities. 
Moreover, at a given intensity, topological frequency conversion scales inversely with the driving frequency, $f$, while dissipative energy absorption decreases as $f^{-2}$ (specifically, $\eta_{\rm na}\sim f^{-2}$ in the Lissajous conversion regime). %{\bf GR: is this for both MR and NA?} or faster. 
The amplification threshold of $\tau f$ will therefore be lower at higher frequencies. 

%The regime of higher intensities and frequencies puts different constraints on the materials needed for topological frequency conversion. 
%Larger intensities induce heating which may cause material damage. 
%This problem, however, 
If an issue, driving-induced heating  can possibly be circumvented  by using pulsed lasers instead of  continuous wave beams: by allowing the system to dissipate heat between pulses, such a scheme would allow us to reach the high-intensity regime without causing material damage; \addFN{while a detailed investigation would be an interesting topic for future studies, we expect pulses with durations more than a few periods, or  randomly-timed pulses, will  yield conversion rates consistent with topological frequency conversion at continuous-wave operation.}
This way the large-amplitude regime required for frequency conversion could be realized while allowing time for the system to dissipate absorbed heat between pulses even if relaxation times are short. In addition to these considerations, materials with a steeper velocity makes realizing the large frequency regime easier, as the velocity at the Weyl point  is the 
``coupling constant'' that converts the electric-field amplitude into an energy scale. 

Weyl nodes need to be located near the Fermi surface to support topological frequency conversion, and moreover need to be surrounded by an asymmetric electron distribution, in order to ensure an imbalance in the numbers of electrons that convert energy at opposite rates. 
Optimal imbalance can be reached in the presence of a ``Weyl cone tilt'', and through appropriate tuning of the chemical potential.
Additionally, our analysis indicates that time-reversal symmetry needs to be broken to acheive frequency conversion. 
Hence, we expect magnetic Weyl semimetals, such as ${\rm Co}_3{\rm Sn}_2{\rm S}$ or ${\rm Co}_2{\rm MnGa}$~\cite{Xu_2018,Swekis_2021} are best-suited for topological frequency conversion. 

%\im{say smth about doping/chemical potential tuning?}

%Weyl nodes with steeper band gaps are better suited for frequency conversion, as they support topological energy conversion at higher frequency ranges. 

%Topological frequency conversion is best realized with Weyl nodes with steeper band gaps. 
%Here the gap steepness is parameterized by the smallest singular value of  $R$ in Eq.~\eqref{eq:weyl_hamiltonian}, $v_0$.
%feature  a rapid increase of the  band gap in all direction away from the Weyl point. 
% $v_0$ controls the  frequency range that supports frequency conversion at a given intensity through Eq.~\eqref{eq:intensity_condition}. %; the  $\Omega$ is of order $I_0\sim \frac{\hbar^2 \Omega^4}{2\varepsilon_0 v_0}$.
%Known compounds have Weyl points in which $v_0\sim 10^5\,{\rm m}/s$, which allows for ${\rm THz}$ conversion at intensities of order $100{\rm W}/{\rm mm}$ (see above). %and optical (${\rm PHz}$
%We expect identifying compounds with  higher values of $v_0$ will  be an important future avenue of research, as this would extend the feasible frequency range for topological frequency conversion, and will lead to lower amplification threshold. 
%closer to realization. 

Topological frequency conversion could also be achieved in non-magnetic Weyl semimetals, or even be  enhanced in magnetic Weyl semimetals, by ``priming'' the particle distribution into an  out-of-equilibrium state.
Such priming could e.g. be achived by driving the system with ultrashort laser pulses or with a DC current, and would create a transient state  more suited for frequency conversion than the steady states we  have considered in this work.  Similarly, purification of the material, alongside bath or substrate engineering are other potentially important directions for realizing amplification by suppressing dissipation. Indeed, these research directions are also important for the general nonlinear response of Weyl semimetals (e.g., chiral photogalvanic effect)~\cite{de_Juan_2017}.

{\it Acknowledgements ---}
We thank N. Peter Armitage, Chris Ciccarino, Cyprian Lewandowski, Prineha Narang, and Mark Rudner for valuable discussions. 
FN gratefully acknowledges the support of the European Research Council (ERC) under the European Union Horizon 2020 Research and Innovation Programme (Grant Agreement No. 678862) and  the Villum Foundation. IM was supported by the Materials Sciences and Engineering Division, Basic Energy Sciences, Office of Science, U.S. Department of Energy.
GR is grateful for support from the Simons Foundation as well as support from the NSF DMR grant number 1839271, and This work is supported
by ARO MURI Grant No. W911NF-16-1-0361. This work was performed in part at Aspen Center for Physics, which is
supported by National Science Foundation grant PHY-1607611.

\bibliography{Weyl_bibliography}

\begin{thebibliography}{114}
\bibitem [{NoD()}]{NoDissipationForPlasmons}%
  \BibitemOpen
  \href@noop {} {}\bibinfo {note} {While the dissipative component of the
  current $\delta \vec j(t)$ is important for the energy conversion properties
  of the system and thus included in the simulations of
  Sec.~\ref{sec:numerics}, we expect the plasmon-induced electric field to be
  dominated by the non-dissipative component, $\vec j_0(t)$.}\BibitemShut
  {Stop}%
\end{thebibliography}
\appendix
\section{Derivation of Eq.~(\ref{eq:pn_def})}
\label{app:p_alpha_derivation}

In this Appendix, we derive the expression %in Eq.~\eqref{eq:pn_def} 
for the time-averaged energy conversion rate from a single electron in band $\alpha$, $\bar P_\alpha(\kv)$, that we quote in Eq.~\eqref{eq:pn_def} in  the main text. %transfer to mode $1$ .

To recapitulate, the equation we aim to derive [Eq.~\eqref{eq:pn_def}] reads
\be
\bar P_\alpha(\kv)
    = f_1f_2\frac{e^2 }{4\pi^2\hbar}
        \int_0^{2\pi}\!\!\! {\rm d}\phi_1 {\rm d}\phi_2\, 
        (\partial_{\phi_1}{\bs \alpha}\times \partial_{\phi_2}{\bs \alpha} )
    \cdot 
        \vec \Omega_\alpha(\kv+e{\bs \alpha}/\hbar).
\label{eqa:p_alpha_result}
\ee
where  we suppressed the $(\phi_1,\phi_2)$-dependence of the integrand.
The quantities above are defined in the main text.
For brevity,  we will use the shorthand notation $\phi=(\phi_1,\phi_2)$ and  $\vec \Omega_\alpha(\kv,\phi)\equiv \vec \Omega_\alpha(\kv+e{\bs \alpha}/\hbar)$ in the following.

%To arrive at 
We derive Eq.~\eqref{eqa:p_alpha_result} starting from Eq.\eqref{eq:phase_integral} in the main text: % which states that 
\be 
\bar P_\alpha(\kv)=\frac{-e}{4\pi^2}\int_0^{2\pi}\!\!\!{\rm d}^2\phi\,  {\bs \epsilon}_1 (\phi) \cdot \vec v_\alpha(\kv;\phi).
\ee
%(See also main text for definitions of the quantities above). 
Here $\bs \epsilon_i(\phi)$ denotes the electric field of mode $i$ as a function of $\phi$~\footnote
{
    Since he electric field from mode $i$ only depends on $\phi_i$,  $\bs \epsilon_i(\phi)$ only depends on the $i$th component of $\phi=(\phi_1,\phi_2)$.
    We choose the notation $\bs \epsilon(\phi)$ for convenience.
    },
while  $\vec v_\alpha(\kv;\phi)$ denotes the wavepacket velocity in band $\alpha$ as a function of  $\phi$, and is given by 
\be 
\vec v_\alpha(\kv;\phi) 
    = \frac{1}{\hbar}\nabla \varepsilon_\alpha(\kv,\phi)
    - \frac{e}{\hbar}{\bs \epsilon}(\phi)\times \vec \Omega _\alpha(\kv,\phi)
\label{eqa:v_alpha_def}
\ee 
with ${\bs \epsilon}(\phi) = {\bs \epsilon}_1(\phi)+{\bs \epsilon}_2(\phi)$,  $\varepsilon_\alpha(\kv,\phi)\equiv \varepsilon_\alpha(\kv+e{\bs \alpha}(\phi)/\hbar)$ and $\varepsilon_\alpha(\kv)$ denoting the energy of band $\alpha$. 
 %and $\nabla$ denotes the gradient with respect to $\kv$. 

%and we again suppressed  quantities' dependence on $\phi_1$ and $\phi_2$ for brevity. \comment{Use $\phi$ notation}

First, we consider the contribution to $\bar P_\alpha(\kv)$ from the group velocity component of $\vec v_\alpha$:
\be 
\bar P_{\alpha;{\rm gv}}(\kv)\equiv \frac{-e}{4\pi^2\hbar }\int_0^{2\pi}\!\!\!{\rm d}^2\phi\,  {\bs \epsilon}_1 (\phi) \cdot \nabla \varepsilon_\alpha(\kv,\phi) %\vec v_\alpha(\kv;\phi_1,\phi_2).
\label{eqa:p1alpha_expr}
\ee
Using  ${\bs \epsilon}_i(\phi) = \omega_i \partial_{\phi_i}\bs \alpha(\phi)$ along with the chain rule, one can   verify that
\be 
-e {\bs \epsilon}_1 (\phi) \cdot \nabla \varepsilon_\alpha(\kv,\phi) = \hbar \omega_1\partial_{\phi_1}\varepsilon_\alpha(\kv,\phi).
\ee
Since $\varepsilon_\alpha(\kv,\phi)$ is $2\pi$-periodic in $\phi_1$, we conclude
$
\bar P_{\alpha;{\rm gv}}(\kv)=0
\label{eqa:p1alpha=0}
$,
implying that
\be 
\bar P_\alpha(\kv)=\frac{-e^2}{4\pi^2\hbar }\int_0^{2\pi}\!\!\!{\rm d}^2\phi\,  {\bs \epsilon}_1 (\phi) \cdot 
\left[{\bs \epsilon}(\phi)\times \vec \Omega _\alpha(\kv,\phi) \right]
\label{eqa:p_anonm_v}
\ee

To evaluate Eq.~\eqref{eqa:p_anonm_v}, we use that ${\bs \epsilon}(\phi)= {\bs \epsilon}_1(\phi)+ {\bs \epsilon}_2(\phi)$, along with $\vec a \cdot(\vec b\times \vec c) = \vec c \cdot(\vec a\times \vec b)$, we obtain
\be 
\bar P_\alpha(\kv)=\frac{-e^2}{4\pi^2\hbar }\int_0^{2\pi}\!\!\!{\rm d}^2\phi\,  [ {\bs \epsilon}_1 (\phi)\times  {\bs \epsilon}_2(\phi)] \cdot \vec \Omega _\alpha(\kv,\phi)\cdot 
\ee
Using ${\bs \epsilon}_i = \omega_i \partial_{\phi_i}{\bs \alpha}$ and $\omega_i=2\pi f_i$, we identify 
\be 
{\bs \epsilon}_1 (\phi)\times  {\bs \epsilon}_2(\phi) = \omega_1\omega_2 \partial_{\phi_1}{\bs \alpha}\times  \partial_{\phi_2}{\bs \alpha}.
\ee
Inserting this in the above establishes Eq.~\eqref{eqa:p_alpha_result}.

\section{Solution of master equation}
\label{app:master_equation_solution}
Here solve the master equation in Eq.~\eqref{eq:vn_modified}, and use the solution to obtain the expression for the current density in Eq.~\eqref{eq:current_result}.

The Appendix is structured as follows: 
We provide a summary of the results in Sec.~\ref{seca:summary}.
In Sec.~\ref{seca:steady_state} we derive the steady state solution to the master equation. 
We demonstrate our solution for the Boltzmann-form dissipator in Sec.~\ref{seca:boltzmann}.
Using  our steady state solution, in Sec.~\ref{seca:current_result} we obtain the current density, while Sec.~\ref{seca:eqs_derivations} contains derivations of auxiliary results which enter in our calculation.

\subsection{Summary of solution}
\label{seca:summary}
Our goal is to obtain the steady-state  of the master equation  
\be 
\partial_t \hat \rho(\kv,t) = -\frac{i}{\hbar}[\hat H(\kv,t),\hat \rho(\kv,t)] + \mathcal D(\kv,t)\circ \hat \rho (\kv,t).
\label{eqa:master_equation}
\ee
Here  $\hat \rho(\kv,t)$ and $\hat H(\kv,t)$ %=\sum_{ij}\langle i|H(\kv,t)|j\rangle \hat c^\dagger_i \hat c_j$ d
denote the momentum-resolved density matrix and Hamiltonian in the second-quantized Bloch space of the system, $\mathcal H_2$, %(see Sec.~\ref{sec:many_body_tfc}), 
while   $\mathcal D(\kv,t)$   is  Lindblad-form superoperator.
$\hat H(\kv,t)$ is given by $\sum_{ij}\langle i|H(\kv,t)|j\rangle \hat c^\dagger_i \hat c_j$, where $H(\kv,t)$ denotes the ordinary (first-quantized) Bloch Hamiltonian of the system, and $\hat c_i$ annihilates a fermion in orbital $\alpha$; see Sec.~\ref{sec:many_body_tfc} for further details of the notation.

We solve Eq.~\eqref{eqa:master_equation} in the  limit where dynamics are adiabatic, and the characteristic relaxation rate $\Gamma  = \norm{\mathcal{D}(\kv,t)}$ is slower than the characteristic angular driving frequency, $\Omega$.
This limit is summarized through the following conditions:
\be
\Gamma 
    \ll
    \Omega,\quad \hbar \Omega \ll \delta \varepsilon(\kv,t), 
    \quad \hbar \partial_t H(\kv,t)
    \ll 
    \delta \varepsilon^2(\kv,t),
\label{eqa:coh_adiabatic_limit}
\ee
where $\delta \varepsilon(\kv,t)$ denotes the (smallest) spectral gap of  $H(\kv,t)$. 
The second and third inequality  are independent conditions that are both needed to ensure adiabatic dynamics.

To quantify the extent to which the system satisfies the conditions in Eq.~\eqref{eqa:coh_adiabatic_limit}, we use the dimensionless parameter
\be 
\lambda(\kv) \equiv \max_t \left(\frac{\Gamma}{\Omega},\frac{\hbar \Omega}{\delta \varepsilon(\kv,t) },\frac{\hbar \partial_t H(\kv,t)}{ \delta \varepsilon^2(\kv,t)}\right).
\label{eq:lambda_def}
\ee
The system satisfies the  conditions in Eq.~\eqref{eqa:coh_adiabatic_limit} for wavevectors $\kv$ where $\lambda(\kv) \ll 1$.
In Sec.~\ref{seca:steady_state}, we derive the steady-state solution of Eq.~\eqref{eqa:master_equation} up to a correction of order  $\lambda^2(\kv)$. %and use this solution 

From our steady-state solution we obtain the current density using %q.~\eqref{eqa:current_density}.
\be 
\vec j(t) 
    = 
    -\frac{e}{\hbar} \kint \Tr[\nabla \hat H(\kv,t)\hat \rho(\kv,t)].
\label{eqa:current_density}
\ee 
The relevant property of the steady state in  this computation  are  the band occupancies of the instantaneous Hamiltonian:
\be 
\rho_\alpha(\kv,t) \equiv  \Tr[\hat \rho(\kv,t)\hat \psi_\alpha^{\dagger}(\kv,t)\hat \psi_\alpha(\kv,t)],
\label{eqa:rho_alpha_def}
\ee 
where 
$ 
\hat \psi_\alpha (\kv,t) = \sum_i \langle i |\psi_\alpha(\kv,t)\rangle \hat c_i,
$ 
denotes the $\alpha$th eigenmode of $\hat H(\kv,t)$, with  $|\psi_\alpha(\kv,t)\rangle$ denoting the $\alpha$th eigenstate of $H(\kv,t)$ with associated energy $\varepsilon_\alpha(\kv,t)$.
%i.e.,  $H(\kv,t)|\psi_\alpha(\kv,t)=\varepsilon_\alpha(\kv,t)|\psi_\alpha(\kv,t)\rangle$.
In Sec.~\ref{seca:current_result}, we show that the integrand in Eq.~\eqref{eqa:current_density}  can be expressed in terms of $\rho_\alpha$ as follows: 
\be 
\frac{1}{\hbar}
\Tr[\nabla \hat H(\kv,t) \hat \rho(\kv,t)] = \sum_\alpha \rho_\alpha(\kv,t)\dot{\vec r}_\alpha(\kv,t)
+\mathcal O\left(\lambda^2 (\kv) v_{\rm F}\right),
\label{eqa:current_result}\ee 
where  $\dot{\vec r}_\alpha(\kv,t)\equiv \nabla\varepsilon_\alpha(\kv,t)-\frac{e}{\hbar} \vec E(t)\times \vec\Omega_\alpha(\kv,t)$ 
denotes the  group velocity in band $\alpha$, and 
and $v_{\rm F}$ denotes the characteristic magnitude of $\norm{\nabla H(\kv,t)}/\hbar$.
This constitutes the main result of this appendix.

We provide a prescription for computing $\rho_\alpha(\kv,t)$ in Sec.~\ref{seca:rho_alpha_prescription}, and demonstrate the computation for the case of a Boltzmann-type dissipator in Sec.~\ref{seca:boltzmann}.

\subsubsection{Decomposition of current density}
We now show how Eq.~\eqref{eqa:current_result}  allows us to decompose  the current density as  
\be 
\vec j(t) = \vec j_0(t) +\delta \vec j(t),
\label{eqa:current_decomp}
\ee 
where 
\begin{eqnarray}
\vec j_0(t) &=& 
    -e
    \kint \sum_\alpha \bar \rho_\alpha(\kv)\dot{\vec r}_\alpha(\kv,t),
    %\vec v_0(\kv,t),
\end{eqnarray}
with $\bar \rho_\alpha(\kv)$ denoting the time-average of $\bar \rho_\alpha(\kv,t)$, and $\delta \vec j(t)$ denotes a dissipative component of the current density, which we define below, and which is small in the limit $\lambda(\kv)\ll 1$.
This result was  quoted in Eq.~\eqref{eq:current_result} of the main text.
%Our discussion below moreover  provides a definition of $\delta \vec j(t)$.

As our first step,  we find  that $\rho_\alpha(\kv,t)$ is nearly stationary in the limit $\Gamma\ll \Omega$:
\be 
\rho_\alpha(\kv,t)= \bar \rho_\alpha(\kv) + \mathcal O (\Gamma/\Omega)+\mathcal O(\lambda^2(\kv)). 
\label{eqa:rho_nearly_stationary}
\ee 
% where 
% $
% \bar \rho_\alpha(\kv)\equiv\lim_{t\to \infty}\frac{1}{t}\int_0^t{\rm  d}t' \rho_{\alpha}(\kv,t')
% $ 
% denotes the time-average of $\rho_\alpha(\kv,t)$.
This result is established in Sec.~\ref{seca:near_stationary}.  
Note that $\Gamma/\Omega \leq \lambda(\kv)$, such that $v_{\rm mr}(\kv,t)\lesssim \lambda (\kv) v_{\rm F}$. 
However, we expressed this $\mathcal O(\lambda)$ correction as above to make it explicitly clear that it  is  controlled by  $\Gamma/\Omega$. 
%since
%we use this fact in Sec.~\ref{sec:finite_dissipation} of the main text. \comment{Double check that we do}

Next, we use that the two components of the group velocity  satisfy
\be
    \frac{1}{\hbar}\nabla\varepsilon_\alpha(\kv,t)
    \lesssim 
    v_{\rm F}, 
\quad  
    \frac{e}{\hbar}|\vec E(t)\times \vec\Omega_\alpha(\kv,t)| 
    \lesssim \lambda(\kv)v_{\rm F}.
\label{eqa:group_velocity_result}
\ee 
These  results are  established in Sec.~\ref{seca:group_velocity_bounds}.

The above two results motivate us to decompose  Eq.~\eqref{eqa:current_result} as % follows: % in terms of $\rho_\alpha(\kv,t)$:
\be 
\frac{1}{\hbar}\Tr[\nabla \hat H(\kv,t) \hat \rho(\kv,t)]   
    = 
    \vec v_0(\kv,t) 
    +
    \vec v_{\rm mr}(\kv,t)
    +
    \vec v_{\rm na}(\kv,t).
\label{eqa:velocity_decomposition}
\ee
where
\begin{eqnarray}
\vec v_0(\kv,t) 
    &\equiv & \sum_\alpha \bar \rho_\alpha(\kv) \dot{\vec r}_\alpha(\kv,t)
    \label{eqa:velocity_components}\\
    \vec v_{\rm mr}(\kv,t)
    &\equiv &
    \frac{1}{\hbar}\sum_\alpha (\rho_\alpha(\kv,t)-\bar \rho_\alpha(\kv))\nabla\varepsilon_\alpha(\kv,t) %\dot{\vec r}_\alpha(\kv,t), % \nabla \varepsilon_\alpha(\kv,t), %\dot{\vec r}_\alpha(\kv,t)
    \notag\\
\vec v_{\rm na}(\kv,t)
    &\equiv &
    \frac{1}{\hbar}\Tr[\nabla \hat H(\kv,t) \hat \rho(\kv,t)]-\vec v_0(\kv,t)-\vec v_{\rm mr}(\kv,t) %\frac{1}{\hbar}\sum_\alpha \rho_\alpha(\kv,t)\dot{\vec r}_\alpha(\kv,t), %-\bar \rho_\alpha(\kv))\dot{\vec r}_\alpha(\kv,t), 
    \notag 
   % \sum_\alpha \rho_\alpha(\kv,t)\dot{\vec r}_\alpha(\kv,t)\notag 
\end{eqnarray}
Due to Eqs.~\eqref{eqa:rho_nearly_stationary} and \eqref{eqa:group_velocity_result}, the latter two components in  particular satisfy
\begin{eqnarray}
\vec v_{\rm mr}(\kv,t)& \lesssim& \mathcal O ( v_{\rm F}\Gamma /\Omega ) + \mathcal O(\lambda^2(\kv)v_{\rm F}),\label{eqa:v_mr_result}
\\ 
\vec v_{\rm na}(\kv,t) &\lesssim& \mathcal O (\lambda^2(\kv)  v_{\rm F}).
\end{eqnarray}

The decomposition  above  allows us to express the current density as 
\be 
\vec j(t) 
    = \vec j_0(t) +  \vec j_{\rm mr}(t)+  \vec j_{\rm na}(t)
\ee 
where 
\begin{eqnarray}
\vec j_{\rm mr}(t) &=& 
    -\frac{e}{\hbar}
    \kint \vec v_{\rm mr}(\kv,t),
\\
\vec j_{\rm na}(t) &=& 
    -\frac{e}{\hbar}
    \kint \vec v_{\rm na}(\kv,t).
\end{eqnarray}
%while $\vec j_{\rm mr}(t)$ and $\vec j_{\rm na}(t)$ are defined likewise from $\vec v_{\rm mr}(\kv,t)$ and $\vec v_{\rm na}(\kv,t)$. 
We identify $\delta \vec j(t) =\vec j_{\rm na}(t) +\vec j_{\rm mr}(t)$. 
Evidently, $\vec j_0$ dominates in the limit of adiabatic driving and coherent dynamics, where $\lambda(\kv) \ll 1$.

 $\vec j_{\rm mr}$ is the current density correction due to relaxation-induced fluctuations in the band-occupancy, while $\vec j_{\rm na}$ as the   correction due to the finite driving frequency and relaxation rate (relative to the band gap). % driving. 
Note that  $\vec j_{\rm na}(t)$ is only significant for $\kv$-points where dynamics are non-adiabatic, while $\vec j_{\rm mr}(t)$ can be nonzero for all $\kv$-points where the electron density fluctuates.
For this reason, we heuristically identify $\vec j_{\rm mr}$ and $\vec j_{\rm na}$ as the components of $\delta \vec j(t)$ that arise due to momentum relaxation and nonadiabatic heating, respectively.

% We now derive the steady state and  provide a prescription for computing $\rho_\alpha(\kv,t)$. 
% Subsequently, in Sec.~\ref{sec:..} we use these results to establish Eqs.~\eqref{eqa:current_result}, \eqref{eqa:rho_nearly_stationary} and \eqref{eqa:group_velocity_result}, which we used above. 

% \subsection{Derivation of steady state}

% % This part of the Appendix proceeds as follows: in Sec.~\ref{seca:adiabatic_transf} we  solve the master equation in Eq.~\eqref{eqa:master_equation} up to a correction of order $\lambda^2$, and provide a prescription for computing  $\rho_\alpha(\kv,t)$  [Eq.~\eqref{eqa:rho_alpha_def}]. 
% % From this solution, we establish Eq.~\eqref{eqa:v_mr_result} (Sec.~\ref{seca:v_mr_result}) and  Eq.~\eqref{eqa:v_na_result} (Sec.~\ref{seca:correlation_matrix}).
% % Finally,   in Sec.~\ref{seca:boltzmann} we demonstrate how to compute $\rho_\alpha(\kv,t)$ for the Boltzmann-form dissipator in Eq.~\eqref{eq:boltzmann_dissipator}.

\subsection{Derivation of steady state}
\label{seca:steady_state}
In this subsection we derive the  steady state solution of Eq.~\eqref{eqa:master_equation}.

We first show that such a steady state exists.
Given an initial condition  specified at some time $t_0$, the solution of  Eq.~\eqref{eqa:master_equation} can formally be written  as 
\be 
\hat \rho(\kv,t) = \mathcal T e^{\int_{t_0}^t ds {\mathcal L}(s)}\circ \hat \rho(\kv,t_0).
\ee 
where $\mathcal T$ denotes time-ordering, and  $\mathcal L(\kv,t)$ denotes the Liouvillian generating the time-evolution: $\mathcal L(\kv,t) \circ \hat{\mathcal O } = - (i/\hbar)[\hat H(\kv,t),\hat{\mathcal O}]+\mathcal D(\kv,t)\circ \hat{\mathcal O}$.

Due to its Lindblad form, ${\mathcal L}(\kv,t)$  is negative semidefinite. 
Except in cases of fine-tuning or in the presence of conserved integrals of motion (which we do not consider here), all eigenvalues of  ${\mathcal L}(\kv,t)$ except for one  are negative; the last eigenvalue takes value $0$.
The  left eigenvector corresponding to this unique zero-eigenvalue is the identity operator, $\hat I$~\footnote
{
    Here we take the inner product in operator space to be the Hilbert-Schmidt product, $(\hat A,\hat B)=\Tr[\hat A^\dagger \hat B]$.
}
It follows that $\lim_{t_0\to -\infty}\mathcal T e^{\int_{t_0}^t ds {\mathcal L}(\kv,s)}$ has a single left eigenvector   with eigenvalue $1$ (namely $\hat I$), while all other eigenvalues vanish.   
Letting $\hat \rho(\kv,t;t_0)$ denote the corresponding right eigenvector, we hence have
\be
\lim_{t_0\to -\infty}\mathcal T e^{\int_{t_0}^t ds {\mathcal L}(\kv,s)}\circ \hat M = \hat \rho_{0}(\kv,t;t_0) \Tr[\hat M]
\label{eqa:steady_state_def}
\ee %in the limit $t_0\to -\infty$. 
$\mathcal L(\kv,t)$ preserves the trace and positivity of any operator and we may choose $\hat M$ positive-definite. 
Hence $\hat \rho_0(\kv,t;t_0)$ must be positive-definite and have unit trace. 
In other words, $\hat \rho_0(\kv,t;t_0)$ corresponds to a physical density matrix.

The semigroup property, 
$
    \mathcal T e^{\int_{t_0}^t ds {\mathcal L}(\kv,s)}\circ M =
    \mathcal T e^{\int_{t_1}^t ds {\mathcal L}(\kv,s)} 
    \circ 
    (\mathcal T e^{\int_{t_0}^{t_1} ds {\mathcal L}(\kv,s)}\circ M)
$ 
implies that $\hat \rho_{0}(\kv,t;t_0)$ must be independent of  $t_0$ in the limit $t_0\to -\infty$.
We  thus simply refer to this operator as $\hat \rho_{0}(\kv,t)$. 
This operator defines the time-dependent steady state of the system. 
Our goal is to obtain this steady state. 

Eq.~\eqref{eqa:steady_state_def} implies we can obtain the steady state by evolving Eq.~\eqref{eqa:master_equation} from any initial state with unit trace; for our purpose it is convenient to choose the initial state $\hat \rho(\kv,t_0) = \hat I/2^d$, where $\hat I$ denotes the identity operator. % from 
Our derivation proceeds as follows: we first identify a time-dependent unitary transformation (or ``comoving frame transformation'')  that maps Eq.~\eqref{eqa:master_equation} into a new  master equation of the same form  in which the eigenbasis of the Hamiltonian is constant up to a correction of order $\lambda^2$; this approach  was e.g. also used in Ref.~\cite{Berry_1987}.
We then solve the  master equation in this new frame using a rotating wave approximation, by exploiting that the spectral gap of the Hamiltonian is the largest energy scale of the system in the limit $\lambda(\kv) \ll 1$~\cite{Breuer,GardinerZoller}.

In the following,  we consider the dynamics of electrons with a fixed given  wavevector, $\kv$.
For brevity, we  suppress all quantities' dependence on $\kv$, unless otherwise noted.

\subsubsection{Rotating frame transformation}
\label{seca:adiabatic_transf}
Here we  map the master equation in Eq.~\eqref{eqa:master_equation} into one where the Hamiltonian has an effectively time-independent eigenbasis. 
To this end, we sequentially apply two  comoving frame transformations that each reduce the time-dependence of the Hamiltonian's eigenstates by a factor $\lambda$~\cite{Berry_1987}. 
The first  transformation, $\hat Q_1(t)$, maps $\hat \psi_\alpha(t)$ into  the orbital annihilation operator $\hat c_\alpha$, for all  $\alpha$:
\be 
\hat Q_1^\dagger(t) \hat \psi_\alpha(t) \hat Q_1(t) 
    =  
    \hat c_\alpha
\ee
As we show in Sec.~\ref{app:unitary_proof}, the above is realized when
\be 
\hat Q_1(t) 
    =
    \mathcal T e^{
    -i\int_0^t ds \sum_{\alpha\beta} \mathcal M_{\alpha\beta}(s)\hat \psi^\dagger_\alpha(s)\hat \psi_\beta(s)
    }
    \hat V_1,
\label{eqa:q1_exp}
\ee
where %$\mathcal T$ denotes time-ordering, 
\be 
\mathcal M_{\alpha\beta}(t) 
    =
    i   \langle  \psi_\alpha(t)|\partial_t \psi_\beta(t)\rangle
\label{eqa:m1_expr}
\ee
and 
$
\hat V_1 
    =
    \exp( 
    \sum_{ij}\hat c^\dagger_i \hat c_j \log(M)_{ij}
    )
$, 
with $\log(M)$ denoting the  logarithm of the matrix with entries  $M_{ij}=  \langle \psi_i(0)|j\rangle $.
Since % $\langle \psi_\alpha(t)|\psi_\beta(t)\rangle = \delta_{\alpha\beta}$, implying   
$\langle  \psi_\alpha(t)|\partial_t \psi_\beta(t)\rangle = - \langle \partial_t \psi_\alpha(t)| \psi_\beta(t)\rangle$,   $\mathcal M_{\alpha\beta}(t)$ is Hermitian.
Without loss of generality, we choose to work in a gauge where $\langle \psi_\alpha|\partial_t \psi_\alpha\rangle = 0$, implying $\mathcal M_{\alpha\alpha}(t)=0$.
%due to our gauge choice for $|\psi_\alpha(t)\rangle$ (see Footnote~\cite{fn_gauge}). % below Eq.~\eqref{eqa:h_spectral_decomp}]. 

We consider the evolution of the system in the rotating frame that results after applying  $\hat Q_1(t)$. 
I.e., we consider the evolution of 
\be 
\hat \rho_1(t) \equiv \hat Q_1^\dagger (t) \hat \rho(t) \hat Q_1(t).
\ee
By  taking the time-derivative of $\hat \rho_1(t)$ and exploiting  
$
\partial_t \hat Q_1(t) 
    = 
    -i\sum_{\alpha\beta}
    \mathcal M_{\alpha\beta}(t) \hat \psi^\dagger_\alpha(t)\hat\psi_\beta(t)\hat Q_1(t)
$, 
we find that 
%$ \hat \rho_1(t) $ evolves according to the % a 
%master equation % of the same form as Eq.~\eqref{eqa:master_equation}, namely
\be 
\partial_t \hat \rho_1(t) 
    = 
    -\frac{i}{\hbar }[\hat H_1(t),\hat \rho_1(t)] 
    + \mathcal D_1(t)\circ \hat \rho_1(t).
    \label{eqa:me_1}
\ee
where %, $ \mathcal D_1(t)$ is a Lindblad-form  superoperator defined by 
\be 
\hat H_1(t) 
    = 
      \sum_\alpha           \varepsilon_\alpha(t)\hat c^\dagger_\alpha \hat c_\alpha
    - \sum_{\alpha\beta}    \mathcal M_{\alpha\beta}(t)\hat c^\dagger_\alpha \hat c_\beta.
\label{eqa:h1_def}
\ee
and 
\be 
 \mathcal D_1\circ \hat { \mathcal O} = Q^\dagger_1\Big[ \mathcal D\circ (\hat Q_1 \hat {\mathcal O}\hat Q_1^\dagger) \Big]\hat Q_1,
 \ee
with time-dependence suppressed for brevity.
Note that  $ \mathcal D_1(t)$ is in the Lindblad-form. 

%Importantly, 
Eq.~\eqref{eqa:me_1} is of the same form as the original master equation we considered, Eq.~\eqref{eqa:master_equation}. 
However, the eigenmodes of the new Hamiltonian $\hat H_1(t)$, $\hat \psi^{1}_\alpha(t)$ are nearly stationary. 
To see this, note that, for $\alpha\neq \beta$,
\be 
\mathcal M_{\alpha\beta}(t)
=-i\frac{\langle \psi_\alpha(t)|\partial_t H(t)|\psi_\beta(t)\rangle}{2(\varepsilon_\alpha(t)-\varepsilon_\beta(t))}
\label{eqa:m_expr}
\ee 
implying
\be 
|\mathcal M_{\alpha\beta}(t)|\lesssim  \lambda \delta \varepsilon(t). 
\label{eqa:m_bound}
\ee 
%\sim 
%    \mathcal O(\Omega)$.
Thus, in the adiabatic limit, $\lambda \ll 1$,
%in the adiabatic limit ($\hbar \Omega \ll \delta \varepsilon$),  
$\hat \psi^{1}_\alpha(t)$ can be computed using canonical perturbation theory with respect to the term 
$
\sum_{\alpha\beta}\mathcal M_{\alpha\beta}(t)\hat c^\dagger_\alpha \hat c_\beta
$
in Eq.~\eqref{eqa:h1_def}.
The $n$th term in this expansion will be of order $\lambda^n$, and first order expansion  thus yields
\be 
\hat \psi^{1}_\alpha(t)
    =
    \hat c_\alpha 
    -\sum_{\beta \neq \alpha} 
        \frac{\mathcal  M_{\alpha\beta}(t)}{\varepsilon_\alpha(t)-\varepsilon_\beta(t)}
        \hat c_\beta 
    + \mathcal O\left(\lambda^2\right),
\label{eqa:f1_expr}
\ee
The expression above  gives $\hat \psi^{1}_\alpha(t)$ up to an  overall (time-dependent) phase factor which we are free to choose due to gauge symmetry. 
Similar perturbative arguments show that the associated energies of $\hat H_1(t)$ are given by  $\varepsilon^{1}_\alpha(t) = \varepsilon_\alpha(t) + \mathcal O (\lambda^2\delta \varepsilon)$, since we chose a gauge for $|\psi_\alpha(t)\rangle$ where $\mathcal M_{\alpha\alpha}(t) = 0$.
Evidently, the $\alpha$th eigenmode of the transformed Hamiltonian, $\hat \psi^{1}_\alpha(t)$, is   given by $\hat c_\alpha$, up  to a time-dependent correction of order $\lambda$.
Hence the eigenmodes of $\hat H_1(t)$ are nearly stationary in the limit $\lambda\ll 1$.

%By recursively applying the above procedure,  we can  reduce the time-dependence of the eigenmodes of the Hamiltonian even further: 
We now apply the above procedure one more time. 
We first apply a comoving frame transformation  $\hat Q_2(t)$  
to  $\hat H_1(t)$ which transforms each eigenmodes $\hat \psi^1_\alpha(t)$ into the orbital  annihilation operator $\hat c_\alpha$:  
\be 
\hat Q_2^\dagger (t) \hat \psi_\alpha^{1}(t) \hat Q_2(t) 
    = 
    \hat c_\alpha.
\label{eqa:q2_def}
\ee
Since $\hat \psi_\alpha^1(t) = \hat c_\alpha+ \mathcal O(\lambda)$,  $\hat Q_2(t)=1+\mathcal O(\lambda)$. % is nearly identical to the identity. 
We can find $\hat Q_2(t)$ exactly using the same procedure we used to obtain $\hat Q_1(t)$. 
Following this procedure, we find the density matrix in this frame, $\hat Q_2(t) \hat \rho_1(t)\hat Q_2^\dagger(t)$ evolves according to the master equation 
\be 
\partial_t \hat \rho_{ 2}(t) 
    = 
    -\frac{i}{\hbar } [\hat H_2(t),\hat \rho(t)] 
    - \mathcal D_{2}(t)\circ\hat \rho_{0}(t). %+ \mathcal O (\lambda^2 \delta \varepsilon)
\label{eqa:rf_me}
\ee
where 
$
\mathcal D_{2}\circ \hat { \mathcal O} 
    = \hat Q_2 \big[ \mathcal D_1 \circ (\hat Q_2\hat {\mathcal O}\hat Q_2^\dagger\big]
    \hat Q_2$ 
(with time-dependence suppressed for brevity), and 
\be
\hat H_2(t) 
    = 
    \sum_\alpha \varepsilon^{1}_\alpha(t) \hat c^\dagger_\alpha \hat c_\alpha 
    + \sum_{\alpha\beta}\mathcal M^{1}_{\alpha\beta}(t)\hat c^\dagger_\alpha \hat c_\beta.
\label{eqa:h2_def}
\ee
Here 
$
\mathcal M^{1}_{\alpha\beta}(t) 
    \equiv 
    i\langle \psi^1_\alpha(t)|\partial_t \psi_\beta^1(t)\rangle
$, with 
$
|\psi_\alpha^1(t)\rangle  
    =
    \hat \psi^{1\dagger}_\alpha(t)|0\rangle
$ 
denoting the single-particle eigenstate of $\hat H_1(t)$ with associated energy $\varepsilon_\alpha^1(t)$. 

We now seek to 
bound $\mathcal M^1_{\alpha\beta}$. % is of order $\lambda^2 \delta \varepsilon$. 
To this end, we use that $\partial_t ({\varepsilon_\alpha(t)-\varepsilon_\beta(t)})^{-1}\lesssim \lambda$~
\footnote{
    This follows using the chain rule along with $|\partial_t \varepsilon_\alpha|\leq |\partial_t H |\leq \lambda \delta \varepsilon^2$
    },  
$\partial_t|\psi_\alpha(t)\rangle  \lesssim \lambda\delta \varepsilon(t)$~
\footnote{
    This follows since $\partial_t|\psi_\alpha(t)\rangle =-i\sum_\beta \mathcal M_{\beta \alpha}(t) |\psi_\beta(t)\rangle$, and $|\mathcal M_{\alpha\beta}(t)|\lesssim \lambda \delta \varepsilon(t)$
    }, 
and $\partial_t^2 H(t) \sim \Omega \partial_t H(t)\leq \lambda^2 \delta \varepsilon^2(t)$. 
Combining these results with Eq.~\eqref{eqa:m_expr}, we  conclude 
$ 
\partial_t \mathcal M_{\alpha\beta}(t)\sim \lambda^2\delta \varepsilon(t).
$
Using Eq.~\eqref{eqa:f1_expr}, $\partial_t (\varepsilon_\alpha(t)-\varepsilon_\beta(t))^{-1}\lesssim \lambda$, and the definition of $\mathcal M_{\alpha\beta}^1(t)$, we hence obtain 
\be 
\mathcal M^{1}_{\alpha\beta}(t) 
    \sim 
    \lambda^2\delta \varepsilon(t)
\ee

In principle we could iterate the comoving frame transformation procedure further to obtain increasingly accurate master equations for $\hat \rho(\kv,t)$~\cite{Berry_1987}. 
However, since we are just interested in obtaining $\hat \rho(\kv,t)$ to corrections of order $\lambda^2$, this  second step is enough for our purpose.

\subsubsection{Rotating wave approximation}
\label{seca:rwa}
We now solve %the master equation in
Eq.~\eqref{eqa:rf_me} with a rotating wave approximation. 
%by exploiting that the term that couple off-diagonal elements  dissipator $ \mathcal{\mathcal D}_2$ is much smaller than the spectral gap of $\hat H_2(t)$ in the limit $\Gamma \ll \delta \varepsilon$~\cite{GardinerZoller,Breuer}. 
To this end, we  first apply a {\it final}   unitary transformation to Eq.~\eqref{eqa:rf_me} which is generated by diagonal part of $\hat H_2(t)$:
\be 
\hat V(t)
    = 
    \exp\left[
    -i\int_0^t{\rm d}t
    \sum_\alpha \varepsilon^{1}_\alpha(t) \hat c^\dagger_\alpha \hat c_\alpha 
    \right].
\ee 
%It is straightforward to verify from Eq.~\eqref{eqa:rf_me} that t
The density matrix in this frame, 
$
\tilde \rho(t) = \hat V^\dagger(t) \hat \rho_{2}(t)\hat V(t),
$
evolves according to the master equation
\be 
\partial_t \tilde \rho = \hat V ^\dagger\Big( \mathcal L_2 \circ [\hat V\tilde  {\rho}\hat V^\dagger] \Big)\hat V.%\tilde {\mathcal L}(t) \circ \tilde \rho(t),
\label{eqa:final_step_before_rwa}
\ee
where %the dissipator $\tilde{\mathcal D}$ is defined  as follows (through its action on an arbitrary operator $\hat O$): 
\begin{align}
{\mathcal L}_2\circ \hat { \mathcal O} =&-\frac{i}{\hbar} [\sum_{\alpha\beta}\mathcal M^{1}_{\alpha\beta}\hat c^\dagger_\alpha \hat c_\beta %e^{-i\int_0^t ds (\varepsilon_\alpha(s)-\varepsilon_\beta(s))}
,\hat{\mathcal O}] + % \hat V ^\dagger(t)\Big( 
\mathcal D_{2} \circ\hat{\mathcal O}. %[\hat V(t)\hat {\mathcal O}\hat V^\dagger(t)] \Big)\hat V(t).
\end{align}
Here we suppressed time-dependence for brevity. 

%and we expect it oscillates with characteristic frequency $\Omega$. 

We consider the matrix elements of  $\tilde \rho(t)$ in the basis of states corresponding to the $2^d$ unique  configurations of electrons in  the orbitals in the system, $\{|\bs n\rangle\}$, 
\be 
p_{\bs m\bs n}(t)\equiv \langle \bs m|\hat \rho(t)|\bs n\rangle .
\ee 
Here,
${\bs n}=(n_1,\ldots n_d)$  with  $n_i=0,1$ for each $i$ and $ 
\hat c^\dagger_i \hat c_i |\bs n\rangle = n_i|\bs n\rangle
$.
I.e., $|{\bs n}\rangle$ denotes the state in $\mathcal H_2$ with $n_i$ Fermions in orbital $i$.
In this basis t orbital basis of $\mathcal H_2$.
% \be 
% \tilde \rho(t) = \sum_{\bs n,\bs m} p_{\bs n\bs m}(t)|\bs n\rangle\langle \bs m |.
% \label{eqa:rho_coefficient_decomposition}
% \ee
% % where ${\bs n}=(n_1,\ldots n_d)$  (with  $n_i$ taking value $0$ or $1$ for each $i$) and likewise for $\bs m$, while $|{\bs n}\rangle$ denotes the state in $\mathcal H_2$ with $n_i$ Fermions in orbital $i$: 
% % $ 
% % \hat c^\dagger_i \hat c_i |\bs n\rangle = n_i|\bs n\rangle
% % $
Here and below, we use bold italic symbols to indicate configurations of orbital occupancies, as above. % ${\bs n} =(n_1,\ldots n_d)$. 

According to  Eq.~\eqref{eqa:final_step_before_rwa},  $p_{\bs m\bs n}(t)$ evolves according to  the coupled differential equation 
\be 
\partial_t p_{\bs n\bs m}(t) 
    =\sum_{\bs k\bs l}^N
     R_{\bs n\bs m}^{\bs k\bs l}(t) p_{\bs k\bs l}(t)
\label{eqa:rwa_0}
\ee
where
\be
R_{\bs n\bs m}^{\bs k\bs l}(t)
    \equiv 
    \langle \bs n|  {\mathcal L}_2(t)\circ  (|\bs k\rangle\langle \bs l|)|\bs m\rangle  e^{i\!\int_0^t \!ds \sum_{i}\! \varepsilon^2_i(t) (n_{i}-m_{i}+k_{i}- l_{i})}.
\label{eqa:l_matrix} 
\ee

Note that ${\mathcal L}_2(t)$ is of order $\lambda^2 \delta \varepsilon(t)$.
This follows since $\mathcal M^1_{\alpha\beta}\sim \lambda^2 \delta \varepsilon(t)$ and $\norm{\mathcal D_2}=\norm{\mathcal D}\leq\Gamma$, while $\Gamma \leq \lambda^2 \delta \varepsilon(t)$. 
At the same time, we expect $\langle \bs n|\tilde  {\mathcal L}_{ 2}(t)\circ  (|\bs k\rangle\langle \bs l|)|\bs m\rangle$
oscillates with characteristic frequency $\Omega$. 
%and has magnitude of order $\Gamma$. 
%Note that $\Gamma \leq \lambda^2 \delta \varepsilon/\hbar$.
In the limit  $\lambda \ll 1$, $R_{\bs n \bs m}^{\bs k\bs l}(t)$ hence   oscillates rapidly  relative to its characteristic  magnitude for choices of $\bs n,\bs m,\bs k,\bs l$ where $n_i- m_i \neq k_i-l_i$ for one or more choices  $i$.
This allows us to make a rotating wave approximation, where we  only keep the terms in Eq.~\eqref{eqa:l_matrix}   where $n_i- m_i = k_i-l_i$  for all $i$. 
We expect this approximation  yields the correct steady state  up to a correction of order $\norm{\mathcal L_2}/\delta \varepsilon\sim \lambda^2$~\cite{Breuer}. 

After the rotating wave approximation above, Eq.~\eqref{eqa:l_matrix} in particular only couples diagonal matrix elements of $\tilde \rho$ with other diagonal elements:
\be %\begin{eqnarray}
\partial_t p_{\bs n\bs n}(t) = -  \sum_{\bs m}  R_{\bs n\bs n}^{\bs m \bs m}(t)p_{\bs m\bs m}(t).
\label{eqa:diag_me}
\ee 
Since we chose the initial condition $\hat \rho(t_0)=\hat I/2^d$, implying $ p_{\bs n \bs m}(t_0)=1/2^d \delta_{\bs n\bs m}$, we hence conclude 
\be
\tilde \rho(t)= \sum_{\bs n} f_{\bs n}(t)|\bs n\rangle \langle \bs n| +  \mathcal O(\lambda^2). % \quad {\rm for} \quad \neq j.
\label{eqa:steady_state}
\ee
where   $f_{\bs n}(t)$ denotes the steady state of 
\be 
\partial_t f_{\bs n}(t) = -  \sum_k  R_{\bs n\bs n}^{\bs m\bs m}(t)f_{\bs m}(t)
\label{eqa:f_expr}
\ee 
and is normalized such that $\sum_{\bs n} f_{\bs n}(t) =1 $.
Evidently, $\tilde \rho(t)$ is diagonal in the orbital eigenbasis up to a correction of order $\lambda^2$, while the off-diagonal elements decohere (this is a general feature for open quantum systems where relaxation is slow compared to the level spacing of the system~\cite{Breuer}).

We obtain the  steady state in the lab frame, $\hat \rho(t)$  by reverting the net unitary transformation that we applied to obtain $\tilde \rho$,   
\be 
\hat U(t) = \hat V(t)\hat Q_2(t)\hat Q_1(t).
\ee 
Thus we conclude %\hat U(t)$:
\be 
\hat \rho(t) = \sum_{\bs n} f_{\bs n}(t)\hat U^\dagger(t)|\bs n\rangle\langle \bs n|\hat U(t)+\mathcal O (\lambda^2)
\label{eqa:steady_state_solution}
\ee 
Here   $f_{\bs n}(t)$ is computed from Eq.~\eqref{eqa:f_expr}.

%However, we provide a more direct method for computing $f_{\bs n}(t)$ in Sec.~\ref{seca:di} below. 

\subsubsection{Direct method for computing $R_{\bs n\bs n}^{\bs m\bs m}$}
The matrix elements   $R_{\bs n\bs n}^{\bs m\bs m}$, which determine the steady state through the coefficients $f_{\vec n}(t)$, can in principle be obtained from  the definition in Eq.~\eqref{eqa:l_matrix}.
However, we can  obtain them  directly from the eigenstates of the instantaneous Hamiltonian $\hat H(t)$ and the lab frame dissipator $\mathcal D(t)$ without having to go through the   procedure we described in Sec.~\ref{seca:adiabatic_transf}.

First note  that $\mathcal M^1_{\alpha\alpha}=0$, implying 
\be 
%\langle \bs n|\mathcal L_2\circ(|\bs m\rangle\langle m|)|\bs n\rangle
R_{\bs n \bs n}^{\bs m \bs m}(t)= \Tr\big[
\mathcal D_2 (t)\circ (| {\bs m}\rangle\langle \bs m|) |{\bs n}\rangle \langle {\bs n}|\big]
\ee 
Next, we recall that 
\be 
\mathcal D_2 \circ \hat{ \mathcal O} = \hat Q_1^\dagger \hat Q_2^\dagger \big[ \mathcal D \circ (\hat Q_2 \hat Q_1\hat {\mathcal O} \hat Q_1^\dagger\hat Q_2^\dagger)\big]\hat Q_2\hat Q_1,
\ee 
where we suppressed $t$. 
%and $\mathcal D(t)$ denotes the  original lab frame dissipator that enters in Eq.~\eqref{eqa:master_equation}. 
We now use that $\hat Q_2 = 1+\mathcal O(\lambda)$ and 
\be 
\hat Q_1(t) |{\bs n}\rangle = |\Psi_{\bs n}(t)\rangle  
\ee 
where $|\Psi_{\bs n}(t)\rangle$ is the eigenstate of $\hat H(t)$ satisfying 
\be 
\hat \psi^\dagger_\alpha(t)\hat \psi_\alpha(t)|\Psi_{\bs n}(t)\rangle = n_\alpha|\Psi_{\bs n}(t)\rangle.
\ee 
Combining the above results and using $\mathcal D(t) \sim \Gamma$,  we thus obtain 
\be
{ R}_{\bs n\bs n}^{\bs m\bs m}
    = 
    %\lim_{t \to \infty} \frac{1}{t} \int_0^t \!\! {\rm d}t' 
    \Tr\big[{\mathcal D}\circ  (|\Psi_{\bs m}\rangle\langle \Psi_{\bs m}|)|\Psi_{\bs n}\rangle \langle \Psi_{\bs n}|\big]
    +
    \mathcal O (\lambda \Gamma). %\frac{\Gamma\Omega}{\delta \varepsilon}\right)
    \label{eqa:r_matrix_simplified}
\ee
where we suppressed the time-dependence of $|\Psi_{\bs n}(t)\rangle$, $\mathcal D(t)$ and ${ R}_{\bs n\bs n}^{\bs m\bs m}(t)$.
Since ${ R}_{\bs n\bs n}^{\bs m\bs m}(t)\sim \Gamma$, we expect neglecting the correction above yields the correct value of $f_{\bs n}(t)$ state up to a correction of order $\lambda$.

\subsubsection{Calculation of band occupancies}
\label{seca:rho_alpha_prescription}
Now, we compute the instantaneous occupancy in band $\alpha$ in the system, $\rho_\alpha(\kv,t)$, which will play an important role for determining the current density. 

We  consider the one-body correlation matrix in the eigenmode basis,
\be 
g_{\alpha\beta}(\kv,t) \equiv \Tr[\hat \psi_\alpha^\dagger(\kv,t)\hat \psi_\beta(\kv,t)\hat \rho(\kv,t)].
\ee 
The instantaneous occupancy of band $\alpha$  is given by the diagonal elements of this matrix, %$g_{\alpha\beta}(t)$:
$
\rho_\alpha(\kv,t) =g_{\alpha\alpha}(\kv,t).
$ 
However, we will also keep track of the off-diagonal elements of $g_{\alpha\beta}(\kv,t)$, since these are used to compute the current density in the next subsection. 
In the remainder of this step of the derivation, $\kv$ and $t$ are fixed parameters, and we therefore suppress them for brevity. 

Inserting the solution for $\hat \rho$ we obtained Eq.~\eqref{eqa:steady_state_solution}, we find 
\be 
g_{\alpha\beta}
    = 
\sum_{\bs n}f_{\bs n} \langle \bs n|\hat U \hat \psi^\dagger_{\alpha} \hat \psi_{\beta}\hat U^\dagger|\bs n\rangle +\mathcal O (\lambda^2),
\ee
where  $\hat U=\hat V\hat Q_2 \hat Q_1$.
Since $|\bs n\rangle$ is an eigenstate of $\hat V$, and $\hat Q_1^\dagger \hat \psi_\alpha\hat Q_1 = \hat c_\alpha$, we find 
\be 
g_{\alpha\beta}
    = 
\sum_{\bs n}f_{\bs n} \langle \bs n|\hat Q_2 \hat c^\dagger_{\alpha} \hat c_{\beta}\hat Q_2^\dagger|\bs n\rangle +\mathcal O (\lambda^2)
\ee
Since $\hat Q_2$ is a product of  exponentials of a quadratic operators, 
%and $\hat \psi_\alpha$ is a superposition of the orbital annihlation operators $\{\hat c_i\}$, we have that  
$\hat Q_2^\dagger\hat c_\alpha \hat Q_2$ must be  a linear combination of the operators $\{\hat c_i\}$; i.e.,  %it takes the form 
\be 
\hat Q_2^\dagger \hat c_\alpha \hat Q_2
    = 
    \sum_i Q_{\alpha i }\hat c_i
\label{eqa:u_sp}
\ee
for some unitary matrix $Q_{\alpha i}$ which we obtain below. 
Using $ \langle \bs n|\hat c^\dagger_{i} \hat c_{j}|\bs n\rangle=\delta_{ij}n_i$, we thus find 
\be 
g_{\alpha\beta} =\sum_{\bs n,i}f_{\bs n} n_i Q_{\alpha i}^*Q_{\beta i} + \mathcal O(\lambda^2),
\label{eqa:g_expr_123}
\ee
% We identify this as 
% \be 
% g_{\alpha\beta} =\sum_{i}\rho_i U_{\alpha i}^*U_{\beta i} + \mathcal O(\lambda^2),
% \label{eqa:g_expr_123}
% \ee
% %where we used \langle \bs n|\hat c^\dagger_{i} \hat c_{j}|\bs n\rangle=\delta_{ij}n_i$. 

To compute $Q_{\alpha i}$, we use  Eqs.~\eqref{eqa:f1_expr}-\eqref{eqa:q2_def} to obtain
\be 
\hat Q_2 \hat c_\alpha \hat Q_2^\dagger 
    = 
    \hat c_\alpha 
    +  \sum_{\beta \neq \alpha} \frac{\mathcal  M_{\alpha\beta}}{\varepsilon_\alpha-\varepsilon_\beta}\hat c_\beta 
    + \mathcal O(\lambda^2)
\ee
Combining this with Eq.~\eqref{eqa:u_sp}, we conclude  
\be 
Q_{\alpha\beta} 
    =
    \delta_{\alpha\beta} 
    + \frac{\mathcal M_{\alpha\beta}}{\varepsilon_\alpha-\varepsilon_\beta}(1-\delta_{\alpha\beta})
    + \mathcal O(\lambda^2)
\label{eqa:u_sp_result}
\ee

Inserting this result into Eq.~\eqref{eqa:g_expr_123}, we obtain 
\be 
g_{\alpha\beta} = \sum_{\bs n}f_{\bs n}\left( \delta_{\alpha\beta}n_\alpha +\mathcal M_{\beta\alpha}(1-\delta_{\alpha\beta}) \frac{n_\alpha-n_\beta }{\varepsilon_\alpha-\varepsilon_\beta}\right) + \mathcal O(\lambda^2)
\label{eqa:g_expr}\ee
where we used that $\frac{\mathcal M_{\beta\alpha}}{\varepsilon_\alpha-\varepsilon_\beta}\sim \mathcal O(\lambda)$.
Setting $\alpha=\beta$ and using $\rho_\alpha=g_{\alpha\alpha}$, we hence find 
%thus identify 
\be 
\rho_\alpha = \sum_{\bs n}f_{\bs n}n_{\alpha} + \mathcal O(\lambda^2).
\label{eqa:rho_alpha_result}
\ee

\subsection{Explicit solution for Boltzmann-form dissipator}
\label{seca:boltzmann}
We finally demonstrate our solution above for the  case where $\mathcal D(\kv,t)$ is given by the Boltzmann-type dissipator in Eq.~\eqref{eq:boltzmann}, $\mathcal D(\kv,t)\circ\hat{\mathcal O} = -\frac{1}{\tau}[\hat{\mathcal O}-\hat \rho_{\alpha}^{\rm eq}(\kv,t)]$. 
Here $\rho_{\alpha}^{\rm eq}(\kv,t)$ denotes the equilibrium state of the instantaneous Hamiltonian $\hat H(\kv,t)$ with temperature $1/\beta$ and chemical potential $\mu$. 
We treat $\kv$ as a fixed parameter and suppress it below. 

First, we obtain the coefficients $f_{\bs n}(t)$, which determine the band occupancies $\rho_\alpha(t)$.
Recall that $f_{\bs n}$ are given as the steady-state solution to the  the differential equation in Eq.~\eqref{eqa:f_expr}, $\partial_t f_{\bs n}(t) = \sum_{\bs m}R_{\bs n\bs n}^{\bs m\bs m}(t)f_{\bs m}(t)$.
Using Eq.~\eqref{eqa:r_matrix_simplified}  to find the coefficients $\{R_{\bs n\bs n}^{\bs m\bs m}(t)\}$, a straightforward computation yields  %Eq.~\eqref{eqa:simplified_r} yields 
\be 
 R_{\bs n\bs n}^{\bs m\bs m}(t) =-\frac{1}{\tau}\delta_{\bs n\bs m} + \frac{1}{\tau}\Tr\big[\hat \rho_{\rm eq}(t)|\Psi_{\bs n}(t)\langle\Psi_{\bs n}(t)|\big]
\ee
Thus 
\be 
\partial_t f_{\bs n}(t) = -\frac{1}{\tau}f_{\bs n}(t) + \frac{1}{\tau}\Tr\big[\hat \rho_{\rm eq}(t)|\Psi_{\bs n}(t)\rangle\langle\Psi_{\bs n}(t)|\big]
\ee 
where we used $\sum_{\bs m}f_{\bs m}(t)=1$.
This first-order inhomogeneous differential equation  has steady-state solution 
\be 
f_{\bs n}(t)  =\frac{1}{\tau }\int_0^t ds e^{-(t-s)/\tau}
\Tr\big[\hat \rho_{\rm eq}(s)|\Psi_{\bs n}(s)\rangle\langle\Psi_{\bs n}(s)|\big] 
\label{eqa:boltzmann_solution}
\ee

Next, we obtain the band occupancies, $\{\rho_\alpha(t)\}$, using 
\be 
\rho_\alpha(t) = \sum_{\bs n} f_{\bs n}(t)n_\alpha
%=\frac{1}{\tau }\int_0^t ds e^{-(t-s)/\tau}
%\langle\Psi_{\bs n}(s)|\hat \rho_{\rm eq}(s)|\Psi_{\bs n}(s)\rangle. 
\ee
Substituting in Eq.~\eqref{eqa:boltzmann_solution} and identifying  $\sum_{\bs n}|\Psi_{\bs n}(t)\rangle\langle\Psi_{\bs n}(t)|n_{ \alpha} = \hat \psi^\dagger_\alpha(t)\hat \psi_\alpha(t)$, we obtain 
\be 
\rho_{\alpha}(t)  =\frac{1}{\tau }\int_0^t ds e^{-(t-s)/\tau}\Tr[\hat \rho_{\rm eq}(s) \hat \psi^\dagger_\alpha(s)\hat \psi_\alpha(s)]
\ee

Next, we note that $
\Tr[\hat \psi^\dagger_\alpha(s)\hat \psi_\alpha(s)\hat \rho_{\rm eq}(s)]$ 
gives occupation probability of the  $\alpha$th band of the Hamiltonian $\hat H(t)$ in equilibrium at temperature $1/\beta$ and chemical potential $\mu$. 
We recognize this probability as $ f _\beta(\varepsilon_\alpha(s) - \mu)
$
where $f_{\beta}(\varepsilon)$ denotes the Fermi-Dirac distribution at temperature $1/\beta$. 
Thus 
\be 
\rho_{\alpha}(t)  =\frac{1}{\tau }\int_0^t ds e^{-(t-s)/\tau} f _\beta(\varepsilon_\alpha(s) - \mu).
\ee
%For this dissipator, we identify $\Gamma = \tau^{-1}$. 
Note that $\rho_\alpha(t)$ converges to its time-average in the limit $\tau^{-1} \ll \Omega$, consistently with what we claimed in Eq.~\eqref{eqa:rho_nearly_stationary}.
%f _\beta(\varepsilon_\alpha(t) - \mu)$, as we clained in Eq.~\eqref{eq:...}, as we claimed.

We finally compute the  time-average of $\rho_\alpha(t)$.
A straightforward computation shows
\be 
\bar \rho_\alpha= \lim_{t\to \infty}\frac{1}{t}\int_0^{s} dt f _\beta(\varepsilon_\alpha(s) - \mu).
\ee 
which was what we quoted in Eq.~\eqref{eq:boltzmann_result} in the main text.

\subsection{Derivation of current density}
\label{seca:current_result}
We finally obtain the expression for the current density  in  Eq.~\eqref{eqa:current_result}, i.e., we seek to show that 
\be 
\frac{1}{\hbar}\Tr[\nabla \hat H \hat \rho] = \sum_\alpha \rho_\alpha\dot{\vec r}_\alpha
+\mathcal O\left(\lambda^2  v_{\rm F}\right),
\label{eqa:current_goal}
\ee 
where $\dot{\vec r}_\alpha= \frac{1}{\hbar}\nabla \varepsilon_\alpha -\frac{e}{\hbar}\vec E \times \vec \Omega_\alpha$.
Here and below, we take both $\kv$ and $t$ to be implicit parameters. 

As our first step, we combine 
$
\Tr[\hat \rho \nabla \hat H] 
    = 
    \sum_{ij}\langle i|\nabla H|j\rangle  \Tr[\hat \rho \hat c^\dagger_i \hat c_j].
$
with 
$\hat c_i^\dagger=\sum_\alpha\langle \psi_\alpha|i\rangle  \hat \psi_\alpha^\dagger$ to write 
\be 
\Tr[\hat \rho \nabla \hat H] 
    = 
    \sum_{\alpha\beta}\langle \psi_\alpha|\nabla H|\psi_\beta\rangle  g_{\alpha\beta},
    \label{eqa:h_integrand}
\ee
where 
$g_{\alpha\beta}\equiv \Tr[\hat \psi^\dagger_\alpha \hat \psi_\beta]$ and is computed in Sec.~\ref{seca:rho_alpha_prescription}.
Combining Eqs.~\eqref{eqa:g_expr}-\eqref{eqa:rho_alpha_result} we can  express $g_{\alpha\beta}$ in terms of the band occupancies $\rho_\alpha$:
\be 
g_{\alpha\beta} 
    = \delta_{\alpha\beta}\rho_\alpha 
    +\mathcal M_{\beta\alpha}(1-\delta_{\alpha\beta}) \frac{\rho_\alpha-\rho_\beta }{\varepsilon_\alpha-\varepsilon_\beta} 
    + \mathcal O(\lambda^2)
\label{eqa:g_expr_42}\ee
Next, we use the spectral decomposition  $H=\sum_\alpha|\psi_\alpha\rangle\langle \psi_\alpha|\varepsilon_\alpha$ to find 
\be 
\langle \psi_\alpha|\nabla H|\psi_\beta\rangle 
    = \delta_{\alpha\beta}  \nabla \varepsilon_\alpha 
    + i {\boldsymbol{\mathcal A}}_{\alpha\beta}(\varepsilon_\alpha-\varepsilon_\beta),
\label{eqa:trace_result}
\ee
where 
$
    {\boldsymbol{ \mathcal A}}_{\alpha\beta }\equiv  i \langle  \psi_\alpha|\nabla \psi_\beta\rangle 
$.
Combining Eqs.~\eqref{eqa:h_integrand}-\eqref{eqa:trace_result}, we  hence find  
\be 
\Tr[\nabla \hat H \hat \rho] 
    = \sum_\alpha \rho_\alpha \nabla  \varepsilon_\alpha
    +i\sum_{\alpha\beta}(\rho_\alpha-\rho_\beta)   {\boldsymbol {\mathcal A}}_{\alpha\beta}\mathcal M_{\beta\alpha} +\mathcal O\left(\lambda^2 v_{\rm F}\right) %\frac{\Omega^2}{\delta \varepsilon^2}v_{\rm F},v_{\rm F}\Gamma/\delta\varepsilon\right)
\label{eqa:almost_there}
\ee 
We identify the first term in the right-hand side above as the contribution arising from the group velocity. 
%where $v_{\rm F}$ denotes the characteristic magnitude of $\norm{\nabla \hat H} $. 

% Next,  we show that the second term above gives the contribution from the anomalous velocity in Eq.~\eqref{seca:current_result}, i.e., \fn{Revit result afterwards}
% %a brief derivation, which is given in Appendix~\ref{X} shows that 
% \be 
% \sum_{\alpha\beta}
% (\rho_\alpha-\rho_\beta)   {\boldsymbol {\mathcal A}}_{\alpha\beta}\mathcal M_{\beta\alpha} 
% = -e\sum_\alpha \rho_\alpha \vec E \times \vec \Omega_\alpha
% \ee 
To rewrite the second term, we use that $|\partial_t \psi_\alpha\rangle =\frac{e}{\hbar}  \vec E\cdot| \nabla \psi_\alpha\rangle$, implying
$ 
\mathcal M_{\alpha\beta} =  \frac{e}{\hbar } \vec E \cdot \bs{\mathcal A}_{\alpha\beta}
$~\footnote
{
    The freedom in choosing a $\kv$-dependent gauge for  $|\psi_\alpha\rangle$  does not affect our discussion, since diagonal elements of $\mathcal M$ do not enter in our derivation.
}.
Thus 
\be 
\sum_{\alpha\beta}(\rho_\alpha-\rho_\beta)     {\mathcal M}_{\alpha\beta}{ \mathcal A}^j_{\beta\alpha} = 
\frac{e}{\hbar}\sum_{\alpha\beta,i}(\rho_\alpha-\rho_\beta)  E_i   { \mathcal A}^i_{\alpha\beta}{ \mathcal A}^j _{\beta\alpha} 
\label{eqa:berry_tensor}
\ee 
where $\mathcal A^i_{\alpha\beta}$ and $E_i$ denotes the $i$th vector component of $\bs{\mathcal A}_{\alpha\beta}$ and $\vec E$, respectively.
Next, we note 
\be 
\sum_{\alpha\beta}(\rho_\alpha-\rho_\beta)   \mathcal A^{i}_{\alpha\beta}\mathcal A^{j}_{\beta\alpha} = \sum_{\alpha\beta}\rho_\alpha ( \mathcal A^{i}_{\alpha\beta}\mathcal A^j_{\beta\alpha} -   \mathcal A^{j}_{\alpha\beta}\mathcal A^{i}_{\beta\alpha})% - \epsilon_{\muj\lambda}\Omega_\alpha^\lambda \rho_\alpha
\ee
Using the definition of $\mathcal A_{\alpha\beta}^i$ along with 
% $
%   \mathcal A^{i}_{\alpha\beta}\mathcal A^{j}_{\beta\alpha}
% = -\langle \psi_\alpha|\partial_i \psi_\beta\rangle\langle \psi_\beta|\partial_j \psi_\alpha\rangle
% $ and 
$\langle \psi_\alpha|\partial_i\psi_\beta \rangle = -\langle \partial_i \psi_\alpha|\psi_\beta\rangle$, we  find 
$
\sum_\beta    \mathcal A^{i}_{\alpha\beta}\mathcal A^{j}_{\beta\alpha} = \langle \partial_i \psi_\alpha|\partial_j \psi_\alpha\rangle
$.
Hence
\be 
\sum_{\alpha\beta}(\rho_\alpha-\rho_\beta)   \mathcal A^{i}_{\alpha\beta}\mathcal A^{j}_{\beta\alpha} = \sum_\alpha\rho_\alpha (\langle \partial_i \psi_\alpha|\partial_j \psi_\alpha\rangle
-\langle \partial_j \psi_\alpha|\partial_i \psi_\alpha\rangle
)
\ee
We identify the right-hand side as $ -i \sum_{\alpha,k}\rho_{\alpha}\epsilon_{ijk}\Omega_\alpha^{k}$, where $\epsilon_{ijk}$ denotes the Levi-Civita tensor and $\Omega_\alpha^k$ denotes the $k$th component of $\vec \Omega_\alpha$.
Thus, 
\be 
\sum_{\alpha\beta}
(\rho_\alpha-\rho_\beta)   {\boldsymbol {\mathcal A}}_{\alpha\beta}\mathcal M_{\beta\alpha} 
= -i\frac{e}{\hbar}\sum_\alpha \rho_\alpha \vec E \times \vec \Omega_\alpha
\label{eqa:vanom_result}
\ee 
Hence the second term in Eq.~\eqref{eqa:almost_there} gives the contribution to the particle velocity from the  anomalous velocity.
In particular, by inserting the above result into Eq.~\eqref{eqa:almost_there}, and dividing through with $\hbar$, we establish Eq.~\eqref{eqa:current_goal}, which was the goal of this subsection.

\subsection{Derivation of auxiliary results}
%Eqs.~\eqref{eq:...}-\eqref{eq:...}.}
\label{seca:eqs_derivations}
In this subsection we derive the auxiliary results which we quoted in the subsections above. 
Specifically, we derive Eqs.~\eqref{eqa:rho_nearly_stationary},~\eqref{eqa:group_velocity_result}, and 
\eqref{eqa:q1_exp}.
These results are established in Secs.~\ref{app:unitary_proof},~\ref{seca:near_stationary},~and~\ref{seca:group_velocity_bounds}, respectively. 

\subsubsection{Near-stationarity of $\rho_\alpha$ [Eq.~\eqref{eqa:rho_nearly_stationary}]}
\label{seca:near_stationary}
We first show that $\rho_\alpha$ is nearly stationary.

Our starting point is the equation of motion for the diagonal matrix elements of $\hat \rho(\kv,t)$ in the orbital basis, $\{f_{\bs n}(\kv,t)\}$,
$\partial_t f_{\bs n}(t) = -  \sum_k  R_{\bs n\bs n}^{\bs m\bs m}(t)f_{\bs m}(t)$. 
We note that  $ R_{\bs n\bs n}^{\bs m\bs m}(\kv,t)$ is of order $\Gamma$, but oscillates with characteristic frequency $\Omega$.
As a result, we expect $f_{\bs n}(\kv,t)$ to deviate from its time-average, $\bar f_{\bs n}(\kv)$, 
%\equiv\lim_{t\to \infty}\frac{1}{t}\int_0^t dt' f_{\bs n}(\kv,t')$ 
by a correction of order $\Gamma/\Omega$:
\be 
f_i(\kv,t) = \bar f_i(\kv) + \mathcal O (\lambda).
\label{eqa:occupancy_steady_state}
\ee 
Inserting this into Eq.~\eqref{eqa:rho_alpha_result}, we thus find 
\be 
\rho_\alpha = \bar \rho_\alpha + \mathcal O (\lambda). %+\mathcal O(\lambda^2). 
\ee 
where $\bar \rho_\alpha(\kv)$ denotes the time-average of $\rho_\alpha(\kv,t)$, and we neglected a correction of order $\lambda^2$, since it is subleading relative to $\Gamma/\Omega$.

\subsubsection{Bounds on group velocity [Eq.~\eqref{eqa:group_velocity_result}]}
\label{seca:group_velocity_bounds}
We next establish the bounds on the group velocity in Eq.~\eqref{eqa:group_velocity_result}.
To obtain this result,  we note that $\frac{1}{\hbar}|\nabla \varepsilon_\alpha|= \frac{1}{\hbar}|\langle \psi_\alpha|\nabla H|\psi_\alpha\rangle|\leq v_{\rm F}$. 
This establishes the first condition in Eq.~\eqref{eqa:group_velocity_result}. 

To establish the second condition in  Eq.~\eqref{eqa:group_velocity_result}, we use that 
\be 
\frac{e}{\hbar}\vec E(t) \times \vec \Omega_\alpha(t)=
i\sum_{\beta}(  {\boldsymbol {\mathcal A}}_{\alpha\beta}\mathcal M_{\beta\alpha} -{\boldsymbol {\mathcal A}}_{\beta\alpha}\mathcal M_{\alpha\beta})
\ee 
This follows from Eq.~\eqref{eqa:vanom_result} after setting $\rho_{\alpha}$ equal to $1$ for one particular choice of $\alpha$ and $0$ for all other choices. 
Next, we use  that $|\bs{\mathcal A}_{\alpha\beta}| =|\langle \psi_\alpha|\nabla H|\psi_\beta\rangle/(\varepsilon_\alpha-\varepsilon_\beta)|\leq v_{\rm F}/\delta \varepsilon$. 
Since $|\mathcal M_{\alpha\beta}|\lesssim \lambda \delta \varepsilon$ [see Eq.~\eqref{eqa:m_bound}],  we thus conclude that 
$\frac{e}{\hbar}|\vec E(t) \times \vec \Omega_\alpha(t)|\lesssim \lambda v_{\rm F}$. This was what we aimed to show.

\subsubsection{Expression for $\hat Q_1(t)$ [Eq.~(\ref{eqa:q1_exp})]}
\label{app:unitary_proof}
We finally prove that, for each $\alpha$, the unitary operator in Eq.~\eqref{eqa:q1_exp},
\be 
    \hat Q_1(t) 
    =
    \mathcal T e^{-i\int_0^t dt' \sum_{\alpha\beta} \mathcal M_{\alpha\beta}(t')\hat \psi^\dagger_\alpha(t')\hat \psi_\beta(t')}
    \hat V_1,
\ee
 transforms the eigenmode of the Hamiltonian, $\hat \psi_\alpha(\kv,t)$ into the orbital annihilation operator $\hat c_\alpha$:
  \be 
\hat Q_1^\dagger(t) \hat \psi_\alpha(t) \hat Q_1(t) =  \hat c_\alpha,
\label{eqa:q_result_proof}
\ee 
%where $\mathcal M_{\alpha\beta}(t) =- i \langle \psi_\alpha(t)|\partial_t \psi_\beta (t)\rangle$. 

We first note that $\hat Q_1(t)$ is quadratic and conserves the number of fermions. 
% fermion number. 
Hence, $\hat Q^\dagger_1(t)\hat \psi_\alpha(t) \hat Q_1(t) $ must be a linear combination of the orbital annihilation operators:
\be 
\hat Q^\dagger_1(t)\hat \psi_\alpha(t) \hat Q_1(t) = \sum_i K_{\alpha i}(t) \hat c_i
\label{eqa:q_superposition}
\ee
for some matrix $K_{\alpha i}(t)$.
 Eq.~\eqref{eqa:q_result_proof} is satisfied if $K_{\alpha i}(t) = \delta_{\alpha i}$. 

We can find $K_{\alpha i}(t)$ from the single-particle evolution of the system, using $|i\rangle = \hat c^\dagger_i|0\rangle$ and $|\psi_\alpha(t)\rangle = \hat \psi_\alpha^\dagger(t)|0\rangle$:
\be
K_{\alpha i}(t)=\langle \psi_\alpha(t)| Q_1(t)|i\rangle,
\ee
where $Q_1(t)$ is the operator $\hat Q_1(t)$ projected into the single-particle space:
\be 
    Q_1(t) 
    = 
    \mathcal T e^{-i\int_0^t ds \sum_{\alpha\beta}|\psi_\alpha(s)\rangle\langle \psi_\beta(s)|\mathcal M_{\alpha\beta}(s)}e^{\sum_{ij}|i\rangle \langle j|\log(M)_{ij}}.
\ee
Since $M_{ij}=  \langle i|\psi_j(0)\rangle$, we find 
\be 
e^{\sum_{ij}|i\rangle\langle   j|\log(M)_{ij}} = \sum_{ij}|i\rangle\langle j |M_{ij}
\ee
Using $\sum_{ij}|i\rangle\langle j |M_{ij} = \sum_i|\psi_i(0)\rangle \langle i|$, we find 
\be 
K_{\alpha i}(t) = \langle \psi_\alpha(t)| \mathcal T e^{-i\int_0^t ds \sum_{ij} \mathcal M_{ij}(s)|\psi_i(s)\rangle\langle\psi_j(s)|
}|\psi_i(0)\rangle,
\ee
implying
\be 
K_{\alpha i}(0) = \delta_{\alpha i}.
\ee 
To see that $K_{\alpha i}(t)$ also equals $\delta_{\alpha i}$ at later times, we take the time-derivative  above:
\be 
\partial_t K_{\alpha i}(t) = \Big(\langle\partial_t \psi_\alpha(t)| -i \sum_\beta \mathcal A_{\alpha\beta}(t)\langle \psi_\beta(t) |\Big)Q_1(t)|i\rangle
%\hat \psi^\dagger_\alpha(t')\hat \psi_\beta(t') \mathcal T e^{-i\int_0^t dt' \sum_{\alpha\beta} \mathcal M_{\alpha\beta}(t')\hat \psi^\dagger_\alpha(t')\hat \psi_\beta(t')
%}|\psi_i(0)\rangle
\label{eqa:q_derivative}
\ee
Since %$\langle \psi_\alpha(t)|\psi_\beta(t)\rangle = \delta_{\alpha\beta}$ implies that  
$\langle  \psi_\alpha(t)|\partial_t \psi_\beta(t)\rangle = - \langle \partial_t \psi_\alpha(t)| \psi_\beta(t)\rangle$, $\mathcal M_{\alpha\beta}(t) =- i\langle \partial_t \psi_\alpha(t) |\psi_\beta(t)\rangle$.
This result, along with $\sum_\beta |\psi_\beta(t) \rangle\langle \psi_\beta(t)|=1$, implies
\be 
 \sum_\beta \mathcal M_{\alpha\beta}(t)\langle \psi_\beta(t) | =-i\langle\partial_t \psi_\alpha(t)| .
\ee 
Using this in Eq.~\eqref{eqa:q_derivative}, we conclude that
$ 
\partial_t K_{\alpha i}(t) = 0.
$ 
Since we found above that $K_{\alpha \beta}(0) = \delta_{\alpha \beta}$, it hence follows that 
\be
K_{\alpha i}(t) = \delta_{\alpha i}
\ee
 at all times  $t$. 
Using this result in Eq.~\eqref{eqa:q_superposition}, we conclude that Eq.~\eqref{eqa:q_result_proof} holds. 
This was what we wanted to show, and concludes this appendix.

\section{Derivation of bound on $d_0$}
\label{app:d0_bound}
Here we derive the condition for adiabaticity which we quote  above Eq.~\eqref{eq:d0_result} and the text above.
I.e., we seek to establish that the time-dependence of $H(\kv,t)$ can be considered adiabatic for $\kv$-points where 
\be 
\min_t |\kv+e\vec A(t)/\hbar| \gtrsim \sqrt{\frac{\norm{R}eE}{\hbar v_0^2}}.
\label{eqa:d0_goal}\ee 
See main text for definition of quantities and notation.

Our starting point is Eq.~\eqref{eq:adiabaticity_condition}, which states that  the dynamics of the system are adiabatic for $\kv$-points where 
\be 
\hbar\norm{ \partial_t H(\kv+e\vec A(t)/\hbar)}\ll \delta \varepsilon^2(\kv+e\vec A(t)/\hbar)
\label{eqa:adiabaticity_condition_0}
\ee 
for all $t$.

We consider the dynamics near a Weyl point,  where the Hamiltonian takes the linearized form $H(\kv)={\bs \sigma}\cdot R \kv + \vec V \cdot \kv$ [see Eq.~\eqref{eq:weyl_hamiltonian}].   
We ignore the second term arising from the Weyl cone tilt $\vec V$, since it only affects the time evolution  through an overall phase factor. 
With this linearized form we find 
\begin{eqnarray} 
\delta \varepsilon(\kv) &=& |R\kv|
,\\
\hbar \norm{\partial_t H(\kv+e\vec A(t)/\hbar)} &=& e |R\vec E(t)|
\end{eqnarray}
Thus dynamics in the system are adiabatic if 
\be 
e |R\vec E(t)|\ll |R(\kv+e\vec A(t)/\hbar)|^2
\label{eqa:weyl_adiabaticity_condition}
\ee 

We now use that  $|R\vec E(t)|\lesssim \norm{R}E$, where $E$ denotes the characteristic magnitude of the driving-induced electric field. Moreover, 
$ 
|R\vec v| \geq v_0|\vec v|
$
where $v_0$ denotes the smallest eigenvalue of $R$. 
Combining these two inequalities with Condition~\eqref{eqa:weyl_adiabaticity_condition},
we conclude that the time-dependence of the Hamiltonian is adiabatic if 
\be 
\norm{R} \frac{e E}{\hbar } \lesssim v_0^2  |\kv+e\vec A(t)/\hbar|^2
\ee 
for all $t$. 

Rearranging the factors above, we conclude that the  dynamics of the system are adiabatic if Condition~\eqref{eqa:d0_goal} is satisfies. 
This was what we wanted to show.

\end{document}